\documentclass{aa}

\usepackage{multirow}
\usepackage{amsmath}
\usepackage{amssymb}
\usepackage[colorlinks,linkcolor=blue,anchorcolor=blue,citecolor=blue]{hyperref}
\hypersetup{
  colorlinks=true,
  allcolors=[rgb]{0,0,0.8},
  pdftitle={The SRG/eROSITA All-Sky Survey: Optical Identification and Properties of Galaxy Clusters and Groups in the Western Galactic Hemisphere},
  pdfauthor={Kluge et al.},
  }
\usepackage{graphicx,subfig} 
\usepackage{lscape}
\usepackage{longtable}
\usepackage{txfonts}
\usepackage{natbib}
\usepackage{color}
\usepackage{booktabs}
\usepackage{stfloats}

\newcommand{\erosita}{eROSITA\xspace}
\newcommand{\xmm}{\textit{XMM-Newton}\xspace}

\newcommand{\planck}{\textit{Planck}\xspace}
\newcommand{\rosat}{ROSAT\xspace}

\newcommand{\erass}{eRASS1\xspace}
\newcommand{\esass}{\texttt{eSASS}\xspace}
\newcommand{\redmapper}{\texttt{redMaPPer}\xspace}
\newcommand{\eromapper}{\texttt{eROMaPPer}\xspace}

\usepackage{natbib,twoopt}
\usepackage{xcolor}

\makeatletter
\newcommandtwoopt{\citeads}[3][][]{\href{http://adsabs.harvard.edu/abs/#3}%
    {\def\hyper@linkstart##1##2{}%
     \let\hyper@linkend\@empty\citealp[#1][#2]{#3}}}
  \newcommandtwoopt{\citepads}[3][][]{\href{http://adsabs.harvard.edu/abs/#3}%
    {\def\hyper@linkstart##1##2{}%
     \let\hyper@linkend\@empty\citep[#1][#2]{#3}}}
  \newcommandtwoopt{\citetads}[3][][]{\href{http://adsabs.harvard.edu/abs/#3}%
    {\def\hyper@linkstart##1##2{}%
     \let\hyper@linkend\@empty\citet[#1][#2]{#3}}}
  \newcommandtwoopt{\citeyearads}[3][][]%
    {\href{http://adsabs.harvard.edu/abs/#3}
    {\def\hyper@linkstart##1##2{}%
     \let\hyper@linkend\@empty\citeyear[#1][#2]{#3}}}
\makeatother
\usepackage{graphicx}
\usepackage{multirow}
\usepackage{txfonts}

\begin{document}
\title{The SRG/eROSITA All-Sky Survey}
\subtitle{Optical identification and properties of galaxy clusters and groups in the western galactic hemisphere}

\author{M. Kluge\inst{1}\thanks{E-mail: mkluge@mpe.mpg.de},
J. Comparat\inst{1},
A. Liu\inst{1},
F. Balzer\inst{1},
E. Bulbul\inst{1},
J. Ider Chitham\inst{1},
V. Ghirardini\inst{1},
C. Garrel\inst{1},
Y. E. Bahar\inst{1},
E. Artis\inst{1},
R. Bender\inst{1,2},
N. Clerc\inst{3},
T. Dwelly\inst{1},
M. H. Fabricius\inst{1},
S. Grandis\inst{2,4},
D. Hernández-Lang\inst{2,5}
G. J. Hill\inst{6,7},
J. Joshi\inst{8},
G. Lamer\inst{9},
A. Merloni\inst{1},
K. Nandra\inst{1},
F. Pacaud\inst{8},
P. Predehl\inst{1},
M. E. Ramos-Ceja\inst{1},
T. H. Reiprich\inst{8},
M. Salvato\inst{1},
J. S. Sanders\inst{1}, 
T. Schrabback\inst{4,8}
R. Seppi\inst{1},
S. Zelmer\inst{1},
A. Zenteno\inst{10},
X. Zhang\inst{1}}

\institute{
Max Planck Institute for Extraterrestrial Physics, Giessenbachstrasse 1, 85748 Garching, Germany 
\and
University-Observatory, Ludwig-Maximilians-University, Scheinerstrasse 1, D-81679 Munich, Germany
\and
IRAP, Université de Toulouse, CNRS, UPS, CNES, F-31028 Toulouse, France
\and
Universit\"at Innsbruck,  Institut f\"ur Astro- und Teilchenphysik, Technikerstr. 25/8, 6020 Innsbruck, Austria
\and
Excellence Cluster Origins, Boltzmannstr. 2, 85748 Garching, Germany
\and
McDonald Observatory, University of Texas at Austin, 2515 Speedway, Stop C1402, Austin, TX 78712, USA
\and
Department of Astronomy, University of Texas at Austin, 2515 Speedway, Stop C1400, Austin, TX 78712, USA
\and
Argelander-Institut f\"ur Astronomie (AIfA), Universit\"at Bonn, Auf dem H\"ugel 71, 53121 Bonn, Germany
\and
Leibniz-Institut f\"ur Astrophysik Potsdam (AIP), An der Sternwarte 16, 14482 Potsdam, Germany
\and
Cerro Tololo Inter-American Observatory, NSF’s National Optical-Infrared Astronomy Research Laboratory, Casilla 603, La Serena, Chile
}

\date{Received 20 December 2023 / Accepted 30 April 2024}
\titlerunning{\erass\ Catalog of Galaxy Clusters and Groups: Optical Identification}
\authorrunning{Kluge et al.}

\abstract
{
The first SRG/\erosita All-Sky Survey (\erass) provides the largest intracluster medium-selected galaxy cluster and group catalog covering the western Galactic hemisphere.
Compared to samples selected purely on X-ray extent, the sample purity can be enhanced by identifying cluster candidates using optical and near-infrared data from the DESI Legacy Imaging Surveys. Using the red-sequence-based cluster finder \eromapper, we measured individual photometric properties (redshift $z_\lambda$, richness $\lambda$, optical center, and BCG position) for 12\,000 \erass clusters over a sky area of 13\,116\,deg$^2$, augmented by 247 cases identified by matching the candidates with known clusters from the literature.
The median redshift of the identified \erass sample is $z=0.31$, with 10\% of the clusters at $z>0.72$. The photometric redshifts have an accuracy of $\delta z/(1+z)\lesssim0.005$ for $0.05<z<0.9$.
Spectroscopic cluster properties (redshift $z_{\rm spec}$ and velocity dispersion $\sigma$) were measured a posteriori for a subsample of 3210 and 1499 \erass clusters, respectively, using an extensive compilation of spectroscopic redshifts of galaxies from the literature.
We infer that the primary \erass sample has a purity of 86\% and optical completeness >95\% for $z>0.05$.
For these and further quality assessments of the \erass identified catalog, we applied our identification method to a collection of galaxy cluster catalogs in the literature, as well as blindly on the full Legacy Surveys covering 24\,069\,deg$^2$.
Using a combination of these cluster samples, we investigated the velocity dispersion-richness relation, finding that it scales with richness as $\log(\lambda_{\rm norm})=2.401\times\log(\sigma)-5.074$ with an intrinsic scatter of $\delta_{\rm in}=0.10\pm0.01$\,dex.
The primary product of our work is the identified \erass cluster catalog with high purity and a well-defined X-ray selection process, opening the path for precise cosmological analyses presented in companion papers.
}


\maketitle

\section{Introduction} \label{sec:introduction}

In the last decades, two outstanding cosmological questions have been raised. What is the nature of dark matter? What drives the accelerated expansion in the late-time Universe? 
These puzzles can be addressed using observations of the large-scale structure (LSS). There exist two sets of LSS-based cosmological observables \citep[e.g., see the reviews from ][]{Weinberg2013aa,Will2014,Ishak2019aa}. 
The first set is connected to the homogeneous cosmological background. 
These probes use standard rulers such as baryon acoustic oscillations and are sensitive to the geometry and the expansion history of the Universe. 
The second and complementary set of observables relates to the inhomogeneous universe and how the large-scale structure has grown with time: cluster counts, redshift-space distortions in galaxy clustering, and cosmic shear.
The growth of the structures in the cosmic web is mainly related to the cosmological model through the $n$-point statistical functions of the dark matter halos (halo mass function, power spectrum, bi-spectrum, etc.). 
Provided that the cluster scaling relation or the galaxy bias function is under control, one can constrain a set of cosmological parameters from the cluster number counts or the galaxy clustering. 
Currently, the largest samples of intra-cluster medium (ICM) selected clusters of galaxies considered in cosmological analysis are in the regime of few hundreds \citep{Vikhlinin2009aa, Vikhlinin2009ab, Pacaud2016aa, Pacaud2018aa, PlanckCollaboration2016ab, Zubeldia2019aa, de-Haan2016aa, Bocquet2019aa}. 
Their constraint on the cosmological model alone is not stringent and depends strongly on the calibration of the mass-observable relation. 

\erosita, on board the Spektrum Roentgen Gamma (SRG) orbital observatory launched in 2019, is a sensitive, wide-field X-ray telescope which has performed an all-sky survey of unprecedented depth \citep{Merloni2012aa,Predehl2021aa}. 
The sensitivity of \erosita extends to 10\,keV on the high end, but is at its highest in the soft X-ray band, specifically in the 0.2-2.3\,keV range, which makes it particularly suitable for detecting and studying emission from the hot gas in galaxy clusters. 
\erosita is now delivering an outstanding sample of all the most massive clusters up to redshift $z=1.32$ (\citealt{Bulbul2023}, and this analysis) to constrain cosmological parameters with percent-level precision \citep{Ghirardini2023} using a mass observable relation calibrated using weak gravitational lensing (\citealt{Grandis2023,Kleinebreil2023}; Pacaud et al. in prep.). 

To make the most of the unprecedentedly large X-ray-selected galaxy cluster sample provided by \erosita, it is vital to measure accurately the redshifts of the galaxy clusters. 
Measuring redshifts based on X-ray data is only possible from the emission lines found in the X-ray spectra of the ICM \citep[e.g.,][]{Hashimoto2004aa, Lamer2008aa, Lloyd-Davies2011aa, Yu2011aa}.
For \erosita, this venue is available only for a small subset of the clusters and at a relatively low precision ($\sim$10\%) \citep{Borm2014aa}. 
Indeed, since the redshift precision depends strongly on the ICM gas temperature and signal-to-noise ratio of the X-ray spectra, precise measurements are only feasible for very bright clusters observed with long exposure times (e.g., at the ecliptic poles of the \erass). 

Galaxy clusters contain over-densities of early-type galaxies relative to the field \citep{Dressler1980aa}. 
Up to date, the most accurate and precise cluster redshifts are obtained from an ensemble of galaxy redshifts \citep[e.g.,][]{Beers1990aa, Bohringer2004aa, Clerc2016aa, IderChitham2020aa} measured via optical spectroscopy \citep[e.g.,][]{Szokoly2004aa, Koulouridis2016aa, Clerc2020aa}. 
Dedicated spectroscopic observations of the galaxies in \erosita clusters are ongoing with the SDSS-V \citep{Kollmeier2017aa} and soon to be started with 4MOST \citep{Finoguenov2019ab}, so that by 2029 (planned end of 4MOST) essentially all \erosita clusters will have measured spectroscopic redshifts. 
Until then, we estimate photometric redshifts using multi-band photometric data to sample the spectral energy distribution of the sources of interest \citep[e.g., review from ][]{Salvato2009aa}. 
The performance of photometric redshift techniques depends on the quality of the photometry, how well the desired spectral features are sampled by the subset of the spectrum encompassed by the photometric filters, the robustness of the calibration methods, and how representative the spectroscopic training datasets are. 
Estimating the cluster photometric redshift consists of two steps: identifying cluster-bound galaxies in the optical and infrared data and estimating their redshift via ensemble averaging. 
Optical cluster finding algorithms are classified by their methodology: matched-filter based algorithms \citep[e.g.,][]{Postman1996aa, Olsen2007aa, Szabo2011aa}, Voronoi tessellation methods \citep[e.g.,][]{Ramella2001aa, Soares-Santos2011aa, Murphy2012aa}, friends-of-friends \citep[e.g.,][]{Wen2012aa} and percolation algorithms \citep[e.g.,][]{Dalton1997aa, Rykoff2014aa}. 
Photometric redshift estimation methods are classified by the information they use: red sequence \citep[e.g.,][]{Gladders2000aa,Gladders2007aa,Oguri2014aa}, color overdensities \citep[e.g.,][]{Miller2005aa}, brightest cluster galaxies (BCG) \citep[e.g.,][]{Koester2007aa, Hao2010aa}, photometric redshifts of all galaxies \citep[e.g.,][]{Wen2012aa, Tempel2018aa, Bellagamba2018aa, Aguena2021aa}, or spectroscopic galaxy surveys \citep[e.g.,][]{Duarte2015aa, Old2015aa}.

In this analysis, we consider a sample of candidate clusters detected as extended X-ray sources in the first eROSITA all-sky survey \citep[\erass,][]{Merloni2023,Bulbul2023}.
We measure each cluster's photometric redshift and optical richness following the method of \citet{Rykoff2014aa}. 
The basis for these measurements is the optical and near-infrared inference models of the extragalactic sky published in the 9th and 10th data releases of the DESI Legacy Imaging Surveys
\citep{Dey2019aa}\footnote{\url{https://www.legacysurvey.org}}. 
It covers almost all the extragalactic sky with a footprint extending over 24\,069~deg$^2$. 
In the event of sufficient coverage with galaxy spectra, we measure the cluster spectroscopic redshift and cluster line-of-sight-velocity dispersion using the method from \citet{Clerc2020aa} and \citet{Kirkpatrick2021aa}. 

The article is organized as follows. 
In Section \ref{sec:data}, we introduce the optical and X-ray data. 
Section \ref{sec:rm} describes the optical cluster finding method and defines how the different redshift types are measured. 
Our main result, the \erass identified cluster and group catalog, is presented in Section \ref{sec:erass}. We evaluate its properties in the following sections. With the help of consistently reanalyzed catalogs of known clusters, we estimate the optical completeness of the \erass catalog in Section \ref{sec:quality}. In Section \ref{sec:numberdensities}, we compute cluster number densities and examine their dependence on the optical survey depth. 
We discuss the purity and the properties of the remaining contaminants in the \erass catalog in Section \ref{sec:pcont}.
The accuracy and precision of the photometric redshifts are quantified in Section \ref{sec:redshift_accuracy}.
Finally, in Section \ref{sec:richness_vdisp}, we combine all cluster catalogs to tightly constrain the richness--velocity dispersion relation and measure its intrinsic scatter.
Our results are summarized in Section \ref{sec:summary}.

Throughout the paper, we assume a flat cosmology with $H_0=70$ km s$^{-1}$ Mpc$^{-1}$, $\Omega_{\rm m}=0.3$, and $\sigma_8=0.8$. Redshifts are given in the heliocentric reference frame. No corrections for Virgo infall or the CMB dipole moment have been applied.

\section{X-ray, optical, and near-infrared data} \label{sec:data}

In this section, we introduce the data considered in the X-ray wavelength domain (Sec. \ref{subsec:erosita}) and in the optical and near-infrared range (Sec. \ref{subsec:data_legacy}). 
The observed X-ray radiation of galaxy clusters stems from thermal bremsstrahlung and line emission by the hot Intra-Cluster Medium (ICM). 
In the optical and near-infrared, old stellar populations emit most of the received light. 
The different wavelengths thus trace distinct parts of the galaxy clusters. 

\subsection{The X-ray \erass cluster catalog}
\label{subsec:erosita}
Galaxy cluster candidates are selected from the first \erosita All-Sky Survey catalog \citep[\erass,][]{Merloni2023}. 
It consists of X-ray sources in half of the sky at Galactic longitude $179.9442\degr < l < 359.9442\degr$. 
Galaxy clusters emit X-ray radiation from their diffuse hot gas. 
They appear as extended sources in the \erosita observations as opposed to point sources such as Active Galactic Nuclei (AGNs) or stars, although there can be misclassifications.
Therefore, we consider extended \erass sources as galaxy cluster candidates. 
This sample is described in detail in \cite{Bulbul2023} and forms the basis for the \erass cluster candidate catalog. 

\begin{figure}
    \includegraphics[width=\linewidth]{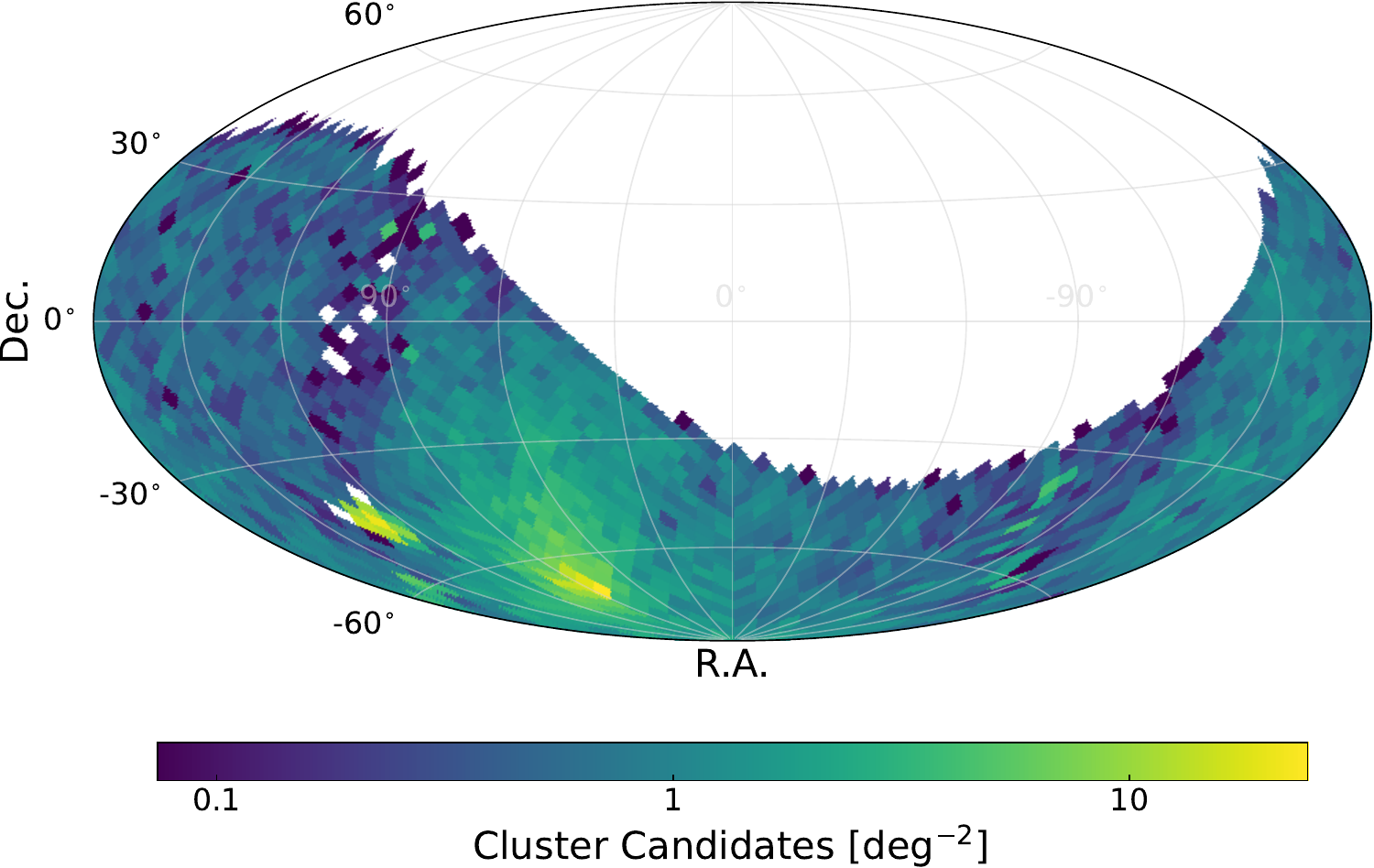}
    \caption{Pixelized number density of 22\,718 cleaned \erass extended sources, that is, cluster candidates in the German part of the \erosita survey. The HEALPix map has a resolution of ${\rm NSIDE}=16~\hat{=}~13.4$\,deg$^2$. The highest number density occurs near the southern ecliptic pole around $\rm{R.A.}=90\degr$ and $\rm{Dec.}\approx-67\degr$ where the exposure time is largest. Another overdensity of extended sources near the Galactic plane around $\rm{R.A.}\approx135\degr$ and $\rm{Dec.}\approx-45\degr$ is associated with the Vela supernova remnant. It lies outside the common footprint with the Legacy Surveys (see Figure \ref{fig:footprint}).
    \label{fig:footprint_erass}}
\end{figure}

The survey area of \erosita used here is that to which the German consortium holds data rights and covers half the sky: 20\,626\,deg$^2$ (see Figure \ref{fig:footprint_erass}). The joint coverage of this with the DESI Legacy Imaging Surveys \citep[LS,][]{Dey2019aa} (see Section \ref{subsec:data_legacy}) is 13\,178\,deg$^2$, primarily because the LS avoids the area around the Galactic plane and the Magellanic clouds. 
As described in \cite{Bulbul2023}, we further masked regions of known non-cluster extended X-ray sources (using the flag IN\_XGOOD=False), in total 62\,deg$^2$. This reduces the \erass cluster candidate catalog area to 13\,116\,deg$^2$. 

A fraction of the extended sources are cataloged multiple times. 
Most of these split sources were removed by applying the SPLIT\_CLEANED flag \citep{Bulbul2023}.
In total, the \erass cluster candidate catalog contains 14\,818 cleaned sources in the common region between the German part of the \erosita survey and the LS.
Figure \ref{fig:footprint_erass} shows the sky density of cleaned galaxy cluster candidates. 
The deeper exposures towards the ecliptic poles cause the gradient of source density. 
This is the main cluster catalog analyzed in the article. Other cluster catalogs analyzed are presented in Section \ref{sec:quality}.

\begin{figure}
    \includegraphics[width=\linewidth]{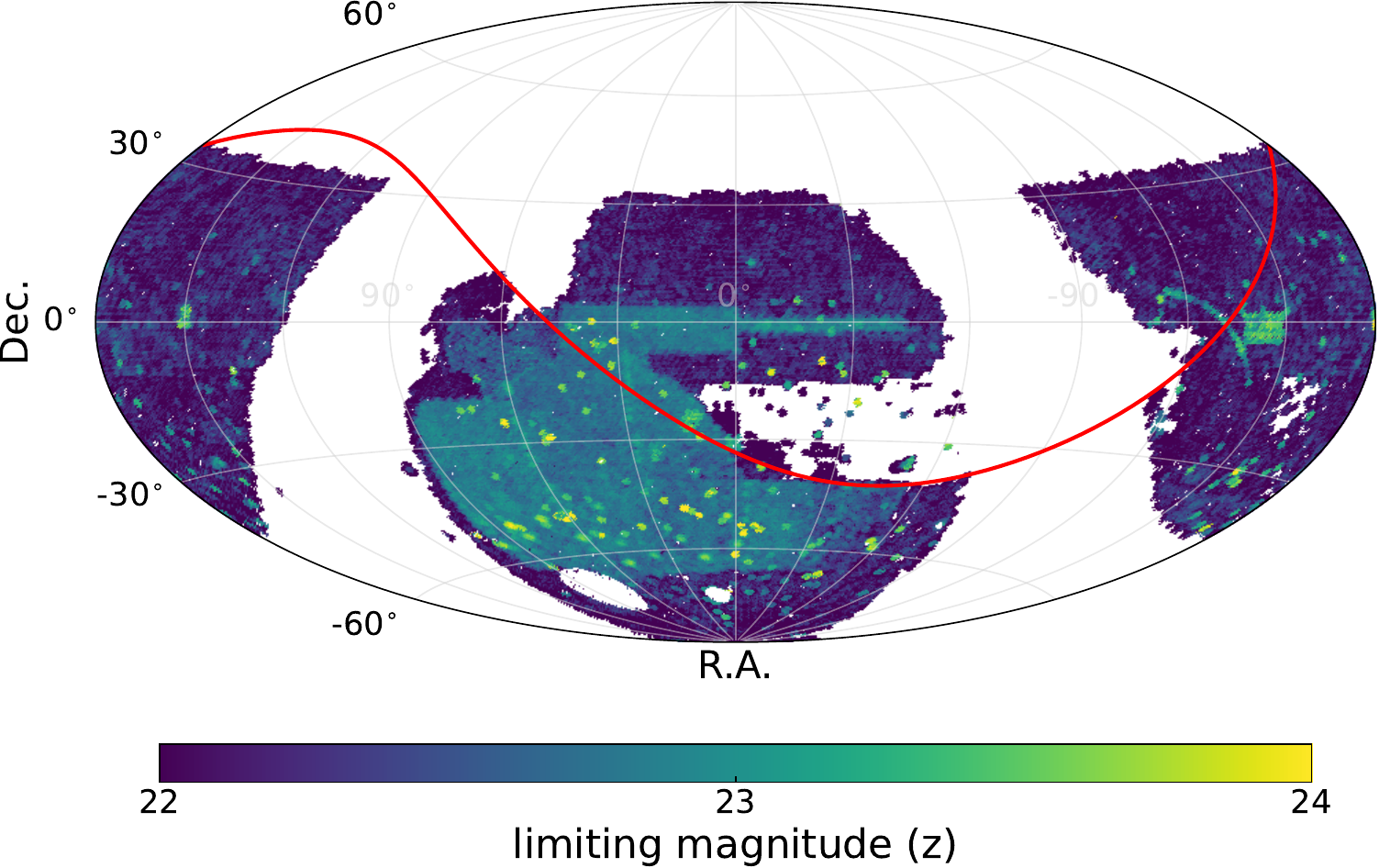}\\\\
    \includegraphics[width=\linewidth]{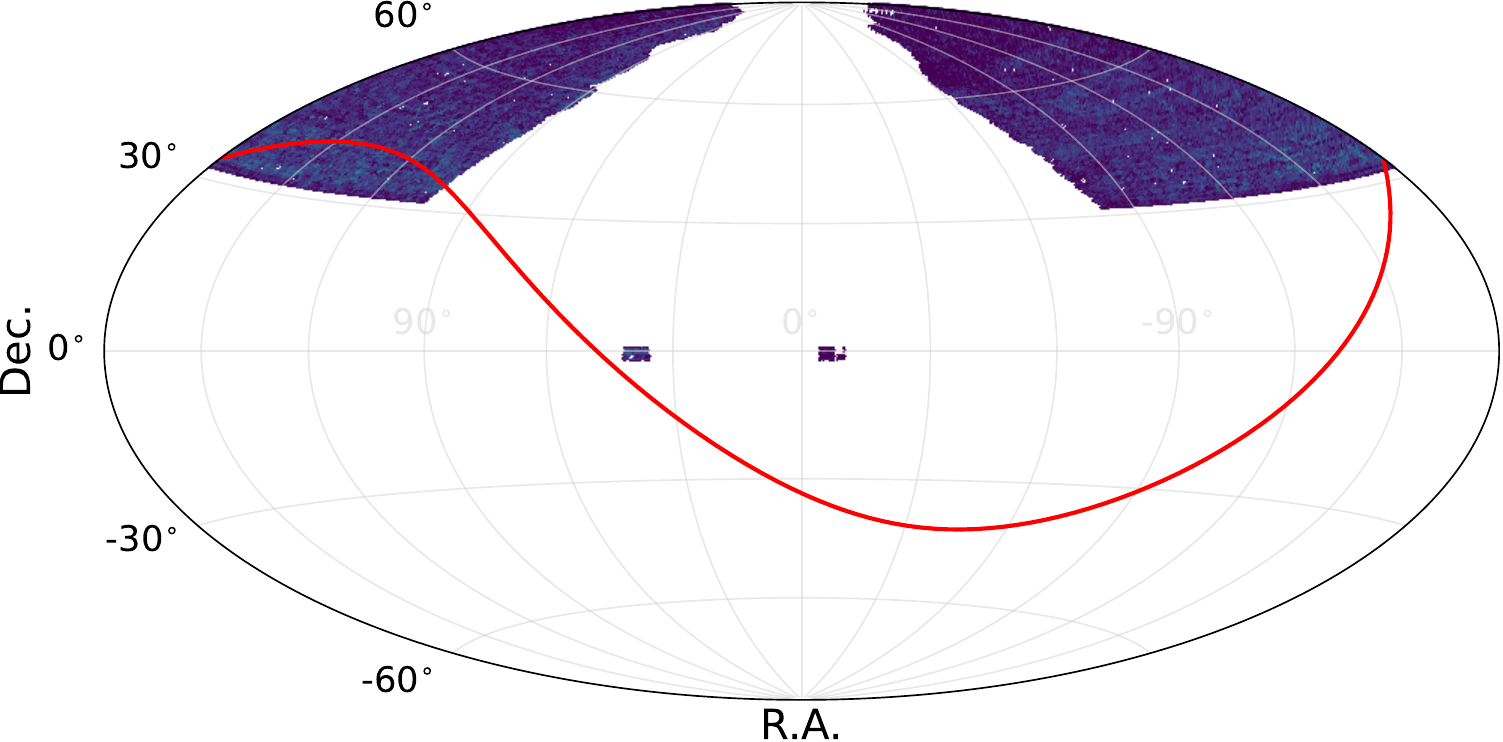}
    \caption{Maps of the limiting magnitude in the $z$ band for sources detected with 10$\sigma$ significance. The footprint maps show the coverage for the Legacy Surveys DR10 south (top panel) and DR9 north (bottom panel). Only regions with data in all the $g$-, $r$-, and $z$-bands were considered. The half sky that is covered by \erass corresponds to the region below the red line. The limiting magnitudes in the $z$ band directly translate to the limiting redshift $z_{\rm{vlim}}$ (see Appendix \ref{sec:zvlim} for details). \label{fig:footprint}}
\end{figure}

\subsection{DESI Legacy Imaging Surveys, 9th and 10th releases} \label{subsec:data_legacy}

We combined the optical and near-infrared inference model data from the 9th and 10th release of the DESI Legacy Imaging Surveys \citep[LS,][]{Dey2019aa} to obtain the largest coverage of the extragalactic sky (Galactic latitude $|b|\gtrsim20\degr$). The Galactic plane was not covered by the observations and was thereby excluded from our analysis.
Observations at optical wavelengths were carried out with three different telescopes as described in Sections \ref{sec:lsdr10south} and \ref{sec:lsdr9north}. Consequently, the survey area was split at a declination of $\approx$32.375$\degr$. We refer to the southern part as LS DR10 South and the northern part as LS DR9 North.

To increase the wavelength coverage, the LS utilized 7 years (LS DR10) or 6 years (LS DR9) of infrared imaging data from the Near-Earth Object Wide-field Infrared Survey Explorer \citep[NEOWISE,][]{Lang2014,Mainzer2014,Meisner2017a,Meisner2017b}.
The limitation of blended sources due to the much lower spatial resolution in the NEOWISE data ($\approx 6\arcsec$) compared to the LS data ($\lesssim 1\arcsec$) was partly overcome by applying forced photometry on the 
"un"-blurred WISE maps \citep[unWISE,][]{Lang2014} at the locations of the sources that were detected in the LS. Including the 3.4$\mu$m $W1$ band in our analysis allowed us to extend the galaxy cluster samples to higher redshift ($z\gtrsim0.8$).

Photometry was measured consistently for all surveys using \textit{The Tractor} algorithm \citep{Lang2016} based on seeing-convolved PSF, de Vaucouleurs, exponential disk, or composite de Vaucouleurs + exponential disk models. All magnitudes are given in the AB system.
We applied the correction for Galactic extinction provided in the LS catalogs. Those corrections were derived using the maps from \cite{Schlegel1998aa} with updated extinction coefficients for the DECam.

\begin{figure}
    \includegraphics[width=\linewidth]{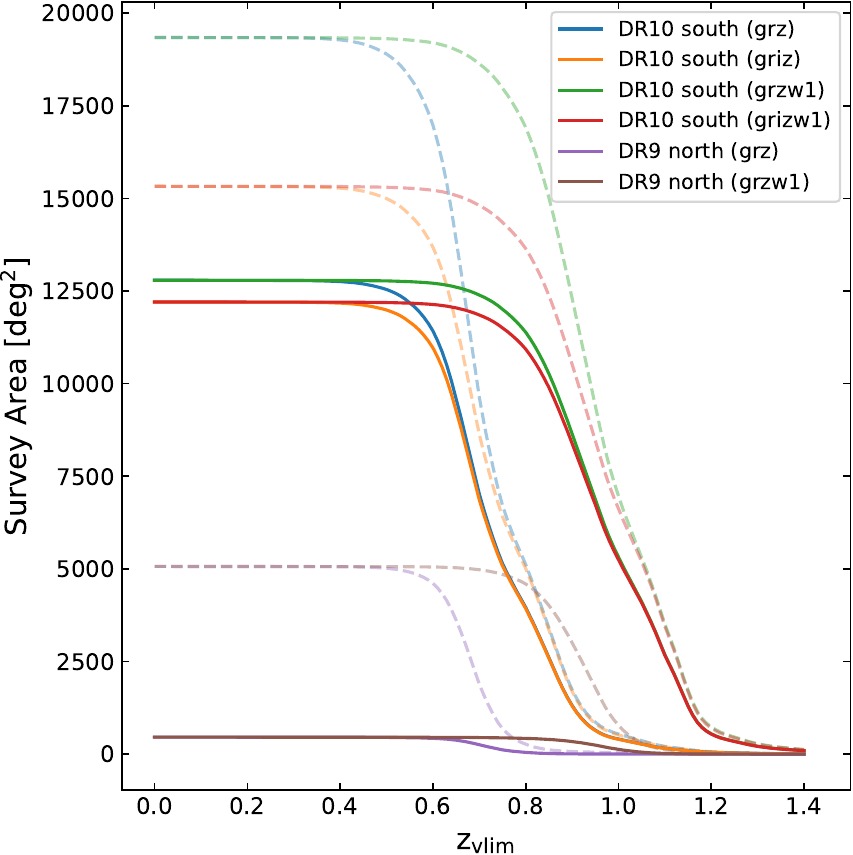}
    \caption{Effective survey area for the Legacy Surveys DR10 south and DR9 north depending on redshift. At the limiting redshift $z_{\rm vlim}$, the faintest considered cluster member galaxy was observed with 10$\sigma$ confidence in the $z$ band. Details are given in Appendix \ref{sec:zvlim}. The luminosity of that galaxy is $L=0.2$\,L$_*$ for the \eromapper\ (\erosita Matched-filter Probabilistic Percolation cluster finder) runs which used the $grz$ and $griz$ filter bands and $L=0.4$\,L$_*$ for the runs that also included the $w1$ band. The dashed lines correspond to the full LS footprint while the continuous lines show the area that is reduced to the overlapping region with \erass (see Figure \ref{fig:footprint}). \label{fig:surveyarea}}
\end{figure}

\subsubsection{LS DR10 south} \label{sec:lsdr10south}

The catalogs were obtained by processing CTIO/DECam observations \citep{Flaugher2015} from the DECam Legacy Survey \citep[DECaLS,][]{Dey2019aa}, the Dark Energy Survey \citep[DES,][]{TheDarkEnergySurveyCollaboration2005aa, DarkEnergySurveyCollaboration2016aa}, and publicly released DECam imaging (NOIRLab Archive) from other projects, including the DECam \erosita survey (DeROSITAs; PI: A. Zenteno).
In the DES area ($\sim5000$ deg$^2$), the depth reached is higher than elsewhere in the footprint, see Figure \ref{fig:footprint}. 

The LS DR10 south covers 19\,342\,deg$^2$ in the $g$, $r$, and $z$-bands (see also Figure \ref{fig:surveyarea}). When the $i$-band is included, the survey area is 15\,326\,deg$^2$. The overlap with \erass is 12\,791 (12\,205) deg$^2$ when the $i$-band is excluded (included). We describe the calculation of the survey area in Appendix \ref{sec:zvlim}. 

\subsubsection{LS DR9 north} \label{sec:lsdr9north}

In the north (Decl. > 32.375$\degr$), LS uses the Beijing-Arizona Sky Survey \citep[BASS,][]{Zou2017aa} for $g$- and $r$-band coverage, and The Mayall $z$-band Legacy Survey \citep[MzLS,][]{Silva2016aa} for the $z$-band coverage. 
Since no additional observations were obtained between DR9 and DR10, the inference catalogs for this part of the sky are that of DR9. 
It covers 5068\,deg$^2$ (see Figure \ref{fig:footprint}). The overlap with \erass is 462\,deg$^2$.

\subsubsection{Processing of released catalogues} \label{sec:catprocessing}
We processed the LS DR10 south and LS DR9 north data independently because the filter responses of the DECam differ from those of the cameras used by the MzLS and BASS. 
Fortunately, there is a common area of 341\,deg$^2$ between the two survey parts, which we used to perform consistency checks (see Section \ref{sec:richness}).

Each source in the LS has MASKBITS and FITBITS flags. 
They encode possible issues encountered in the data (mask) or during the catalog inference process (fit).\footnote{\url{http://legacysurvey.org/dr10/bitmasks}} 
We discarded sources that have set any of the MASKBITS=[0,1,4,7,10,11,12,13] when FITBITS!=9. 
This mainly affects regions around bright and medium bright stars, around bright galaxies in the Siena Galaxy Atlas \citep{Moustakas2021aa}, and around globular clusters. The bright galaxies in the Siena Galaxy Atlas were kept even when the listed MASKBITS had been set.

\subsection{Spectroscopic galaxy redshift compilation} \label{sec:spec_compilation}

We have compiled 4\,882\,137 spectroscopic galaxy redshifts from the literature. 
This compilation serves two goals: calibrating the red-sequence models (Appendix \ref{sec:redseq}; $\sim$90\,000 redshifts) and calculating spectroscopic cluster redshifts and velocity dispersions.
The references considered to create the compilation are listed in Appendix \ref{sec:spec_compilation_refs}.
There, we list the selection criteria applied to the published catalogs to retrieve high-quality redshifts and to avoid stars or quasi-stellar objects. Dedicated spectroscopic follow-up programs are being undertaken for \erass clusters. We included 154 unpublished galaxy redshifts from Balzer et al. (in prep), who utilize the VIRUS instrument \citep{Hill2021} on the \textit{Hobby-Eberly Telescope} \citep{Ramsey1998,Hill2021}.

All galaxies in the spectroscopic compilation were matched to sources in the LS catalogs using a search radius of 1$\arcsec$. If multiple redshifts were available for a source, we chose the redshift from the closest match.

\section{Measurement of cluster candidate counterparts in the optical and near-infrared} \label{sec:rm}
Star formation in galaxies is quickly quenched after the galaxy reaches the first peri-center on its orbit in the cluster \citep[e.g.,][]{Oman2016,Lotz2019}. 
Optical and near-infrared colors of quenched galaxies are relatively insensitive to stellar ages older than a few Gyrs \citep[e.g.,][]{Bruzual2003aa}. 
Hence, cluster members can be identified by relatively uniform optical and near-infrared colors. 
These colors strongly depend on redshift, enabling the measurement of photometric cluster redshifts.
We identify an X-ray extended source (cluster candidate) as a genuine galaxy cluster by associating it with a coincident overdensity of red-sequence galaxies. The reliability of the identification depends on several factors that we investigate and quantify in the following sections.

We introduce the cluster finder \redmapper in Section \ref{sec:redmapper} and our enhancements to it in the following subsections. We refer to the enhanced \redmapper package as \eromapper\ (\erosita Matched-filter Probabilistic Percolation cluster finder).

\subsection{\redmapper} \label{sec:redmapper}

We applied the well-tested and publicly available red-sequence Matched-filter Probabilistic Percolation cluster finder \citep[\redmapper\footnote{\url{https://github.com/erykoff/redmapper}},][]{Rykoff2012aa,Rykoff2014aa,Rykoff2016aa}. 
Its usage is flexible and multi-purpose \citep[e.g.,][]{Rykoff2012aa,Rozo2014aa, Rykoff2016aa,Bleem2020aa,Finoguenov2020aa}.
The \redmapper algorithm applies three filters in luminosity, color, and sky position. 
It finds clusters and creates samples thereof. 
Spectroscopic verification demonstrates the high purity of the cluster samples obtained: $\gtrsim 90\%$ of the \redmapper-identified systems are condensations in redshift space, even at the lowest values of optical richness in the HectoMAP redshift survey down to $r = 21.3$ \citep{Sohn2018aa,Sohn2020aa}.
Further tests by reshuffling members of true clusters and re-detecting those clusters confirm a high purity of $\gtrsim 95\%$ in the SDSS when $z \in [0.1, 0.6)$ \citep{Rykoff2014aa}, meaning these clusters are not significantly affected by projection effects. In this paper, we define purity as the probability that an extended \erass X-ray source that was optically identified using \eromapper is neither an X-ray background fluctuation nor an AGN.
The completeness of the returned samples has been estimated for a similar red-sequence-based cluster finder \citep[RedGOLD,][]{Licitra2017aa,Licitra2016} to be $\sim$100\% ($\sim$70\%) at $z<0.6$ ($z<1.0$) for galaxy clusters with $M>10^{14}$\,M$_\odot$ \citep{Euclid-Collaboration2019aa}. The decreasing completeness at high redshift relative to results from other cluster finders like AMICO \citep{Bellagamba2018aa} is likely attributed to the increasing fraction of blue star-forming galaxies \citep{Nishizawa2018aa} to which \redmapper is insensitive.

Many spectroscopic observing campaigns use a target selection based on \redmapper catalogs \citep[e.g.,][]{Clerc2016aa,Rykoff2016aa,Rines2018aa,Sohn2018aa,Clerc2020aa,Kirkpatrick2021aa}. It has also been used for numerous cosmological experiments \citep[e.g.,][]{Costanzi2019ab,Kirby2019aa,IderChitham2020aa,Costanzi2021aa}, as well as mass calibration analyses \citep[e.g.,][]{Saro2015aa,Baxter2016aa,Farahi2016aa,Melchior2017aa,Jimeno2018aa,Murata2018aa,Capasso2019ab,McClintock2019aa,Palmese2020aa,Raghunathan2019aa}. The impact of projection effects \citep[e.g.,][]{Costanzi2019aa,Myles2020aa}, centering \citep[e.g.,][]{Rozo2014aa,Hoshino2015aa,Hikage2018aa,Hollowood2019aa,Zhang2019aa} and intrinsic alignment \citep{Huang2018aa} have also been studied in detail.

\redmapper can be configured in several different modes, two of which are used in this paper. When configured in blind (cluster-finding) mode, only the optical and near-infrared galaxy catalogs from the LS are used to identify clusters. When configured in scanning mode, \redmapper also considers a positional prior from an input cluster catalog. The search radius for cluster members around the fixed input coordinates is equivalent to the cluster radius $R_\lambda$ \citep[see Equation (4) in][]{Rykoff2014aa}

\begin{equation}
    R_\lambda = 1.0 h^{-1}\,{\rm Mpc}~(\lambda/100)^{0.2}. \label{eq:clusterradius}
\end{equation}

To identify a cluster at a fixed location, \redmapper evaluates a likelihood function $\mathcal{L_\lambda}$ on a redshift grid (see Equation (76) in \citealt{Rykoff2014aa}):

\begin{equation}
    \ln\mathcal{L}_\lambda = -\frac{\lambda}{S}-\sum_i\ln(1-p_{{\rm mem},i}), \label{eq:lmax}
\end{equation}

where $\lambda$ denotes the optical richness (see Equation (\ref{eq:richness}). The likelihood depends only on the membership probabilities $p_{\rm mem}$, which in turn depend on the color distance from the red-sequence model in all considered filter bands, galaxy luminosity, galaxy spatial distribution, and global background galaxy density \citep{Rykoff2014aa,Rykoff2016aa}. The richness $\lambda$ is defined as the sum of the membership probabilities multiplied by a scaling factor $S$, which depends amongst other properties on the masked fraction \citep[][see also Section \ref{fig:richness_des_ero}]{Rykoff2014aa}:

\begin{equation}
    \lambda = S \cdot \sum_i p_{{\rm mem},i}. \label{eq:richness}
\end{equation}

The masked fraction of a cluster is calculated from the number of sources inside $R_\lambda$ with nonzero MASKBITS and FITBITS (see Section \ref{subsec:data_legacy}) over the total number of sources within $R_\lambda$.
Using simulations, it was estimated that the richness errors quoted in the \redmapper catalogs underestimate the observational uncertainties by $\sim$40--70\% due to observational noise in the membership probabilities \citep{Costanzi2019aa}. A cluster with richness $\lambda=20$ has a typical signal-to-noise ratio of 3.

In the case of a free cluster center (\redmapper blind mode), a centering likelihood is logarithmically added to $\ln\mathcal{L_\lambda}(z)$. The maximum $\ln\mathcal{L}_{\rm max}$ of the resulting likelihood $\ln\mathcal{L}(z)$ is determined, and the corresponding redshift is taken as the photometric cluster redshift $z_\lambda$. All other optical cluster properties are evaluated at this redshift.
A minimum of two initial member galaxies is required to start the algorithm. A richness cut of $\lambda>3$ was applied at a later stage.

\subsubsection*{Enhancements to \redmapper}
First, we promoted \redmapper to a strongly parallel application using the {\tt pyspark}\footnote{\url{https://spark.apache.org/docs/3.4.1/api/python}} programming interface. 
This makes it feasible to run in blind mode on the full LS DR9 and DR10 galaxy catalogs (almost all of the extragalactic sky) within a few days.
Second, we adopted the spectroscopic post-processing from SPIDERS \citep{Clerc2020aa, Kirkpatrick2021aa}; see Section \ref{sec:zspec}. 
Finally, we wrote a module to select and rank targets for dedicated spectroscopic follow-up programs from SDSS-V (BHM-clusters) and 4MOST (S5).
In the following sections, the package encapsulating these features will be referred to as \eromapper.

\subsection{Cluster counterpart association, photometric redshift, and richness measurements} \label{sec:zlambda2}

The optical cluster-finding process and the determination of the cluster optical properties (photometric redshift, richness, etc.) are simultaneous and interdependent. This can result in three distinct outcomes.

\subsubsection{Cases with a single optical cluster counterpart} \label{sec:eromapper_singleclusters}

We determined the photometric cluster redshift, denoted $z_\lambda$, by the maximum value of a parabola that is fitted to the $\mathcal{L_\lambda}(z)$ curve. It is close to the highest peak of the blue curve in the example in Figure \ref{fig:zspec_example}, bottom-right panel. 
The likelihood includes via the membership probabilities the color distance to our red-sequence model, which is described in Appendix \ref{sec:redseq}. 
We analyze the accuracy and precision of the photometric redshifts in Section \ref{sec:redshift_accuracy}. 

\subsubsection{Ambiguous cases with multiple optical clusters along the line of sight} \label{sec:multipleclusters}

We identified ambiguous cases with multiple distinct peaks in $\ln\mathcal{L}(z)$. 
These correspond to multiple red-sequence overdensities that overlap along the line of sight without necessarily being physically connected. 
We define the two Gaussians as distinct if their peaks are farther apart than $|z_{\lambda,1}-z_{\lambda,2}|>0.05$, the logarithmic peak ratio is at least $\ln\mathcal{L}(z_{\lambda,2}) / \ln\mathcal{L}(z_{\lambda,1}) > 0.2$, and both cluster likelihoods are above $\ln\mathcal{L}(z_{\lambda,i})>10$ (which roughly corresponds to a richness $\lambda>14$). In this consideration, $z_{\lambda,1}$ is the redshift for the peak of the Gaussian that is closest to the photometric cluster redshift $z_\lambda$, and $z_{\lambda,2}$ is the redshift for the peak of the secondary Gaussian.

\subsubsection{Cases with no optical cluster counterpart}

Cluster candidates (extended X-ray sources) are considered unidentified by \eromapper if 1) it is unable to identify at least two red-sequence galaxies at any redshift, 2) the richness of the cluster is $\lambda<3$, or 3) the cluster candidate is sufficiently far outside of the LS footprint. Clusters near the edge of the footprint can still be detected when the circle with radius $R_\lambda$ partly overlaps with the footprint. More information for these cases is given in Appendix \ref{sec:infootprint}.

\subsection{Spectroscopic cluster redshifts} \label{sec:zspec}

Spectroscopic cluster redshifts $z_{\rm spec}$ provide $\sim$10 times more accurate estimates of the true cluster redshifts compared to our photometric redshifts ($z_\lambda$; see Section \ref{sec:redshift_accuracy}). They are on the other hand more expensive to obtain in terms of telescope time and, therefore, are currently only available for a subset of the clusters. Moreover, they provide a means to estimate the bias and uncertainties of the photometric redshifts (see Section \ref{sec:redshift_accuracy}) and allow us to calculate cluster velocity dispersions $\sigma$.

The automated algorithm to obtain the spectroscopic cluster redshifts has been adopted from the work of \cite{Clerc2016aa} and \cite{Ferragamo2020aa} and is the basis of the automatic spectroscopic redshift pipeline used within the SPIDERS cluster program \citep{IderChitham2020aa, Clerc2020aa, Kirkpatrick2021aa}. The algorithm is iterative, with the subscript $k$ referring to the $k^{\rm th}$ iteration of the procedure.

First, we matched the cluster members selected by \eromapper to our spectroscopic galaxy compilation (see Section \ref{sec:spec_compilation}). The number of matches is $N_{\rm spec,0}$. Figure \ref{fig:zspec_example}, bottom panel, shows the member redshift distribution for the example cluster 1eRASS J085401.2+290316 by the gray bars. The initial spectroscopic cluster redshift was estimated by the bi-weight location estimate \citep{Beers1990aa}

\begin{equation}
\label{eq:zbiwt}
z_{\rm bwt} = \hat{z} + \frac{\sum_{|u_j|<1} \ (z_j - \hat{z}) [1 - u_j^2(z, z_j, 6)]^2} {\sum_{|u_j|<1} \ [1 - u_j^2(z, z_j, 6)]^2},
\end{equation}
where $z$ is a vector of galaxy redshifts, $\hat{z}$ is the sample median, $j$ is the index of the spectroscopic member galaxy and $u_j$ is given by

\begin{equation}
u_j(z, z_j, a) = \frac{(z_j - \hat{z})}{a \times {\rm MAD}(z)}. \label{eq:u}
\end{equation}
 
Here, $a=6$ is the tuning constant that regulates the weighting and corresponds to the clipping threshold in units of the median absolute deviation ${\rm MAD}(z)$ of the member galaxy redshifts.

The proper line-of-sight velocity offset \citep{Danese1980aa} of all member galaxies was then computed relative to the estimate of the cluster redshift;
\begin{equation}
\label{eq:proper_v}
\frac{v_j}{c} = \frac{z_j - z_{\rm bwt}}{1+z_{\rm bwt}}.
\end{equation}

Figure \ref{fig:zspec_example}, bottom panel, illustrates for the example cluster 1eRASS J041610.3-240351 the necessity to clip outliers. While the bulk of member redshifts is concentrated around $z=0.40$, ten photometrically selected member galaxies have significantly lower or higher redshifts. Therefore, we performed an initial velocity clipping of members with $|v_j| > 5000$~km\,s$^{-1}$ to reject them from the spectroscopic sample of member galaxies. 
This results in $N_{\rm spec,1}$ spectroscopic members for the first iteration ($k=1$), which are used to recompute the bi-weight cluster redshift using Equation (\ref{eq:zbiwt}). This procedure is iterated until $N_{\rm spec,k}$ converges or a maximum of $k=20$ iterations is reached. In each iteration $k>1$, the clipping velocity is recalculated as $v_j > 3\sigma_k$, where $\sigma_k$ is the cluster velocity dispersion for the $k$-th iteration (see Section \ref{sec:vdisp}). There are several possible outcomes of the clipping procedure:

\begin{enumerate}
    \item[1)] $N_{\rm spec,1} = 0$: the initial $5000$\,km\,s$^{-1}$ clipping rejected all members: the procedure cannot proceed and a flag is issued to indicate that convergence failed. This can occur for true systems when several distinct structures along the line of sight are far apart in redshift space.
    \item[2)] $0 < N_{{\rm spec},k} < 3$: There is an insufficient number of spectroscopic members left after $k$ steps: it is not possible to estimate the bi-weight redshift estimate (Equation (\ref{eq:zbiwt})) or the velocity dispersion (Equation (\ref{eq:siggap})) and a flag is issued accordingly.
    \item[3)] $N_{{\rm spec},k} \geq 3$: the process successfully converged, and the cluster redshift and velocity dispersion are estimated.
\end{enumerate}

In the last case, the remaining objects are called spectroscopic members.
Even when the iterative procedure converges, it is essential to check the final clipping velocity. If it is larger than the initial $5000$\,km\,s$^{-1}$, there is likely a substructure that biases the spectroscopic cluster redshift. A flag was assigned in this case.
The spectroscopic cluster redshift $z_{\rm spec}$ was calculated by taking the mean of all $z_{\rm bwt}$ values, calculated after bootstrapping the clipped spectroscopic member redshifts 64 times. The standard deviation of these $z_{\rm bwt}$ values was adopted as the spectroscopic cluster redshift uncertainty $\delta z_{\rm spec}$.

In cases 1) and 2), it can still be possible to assign a spectroscopic redshift to a cluster. 
If there was a spectroscopic redshift available for the identified central galaxy (CG), it was taken as the cluster redshift $z_{\rm spec,cg}$. 
The uncertainty of the CG redshift was used as the cluster redshift uncertainty. 
This underestimates the cluster redshift uncertainty because the central galaxy can have a non-negligible line-of-sight velocity with respect to the cluster as a whole \citep{Lauer2014}. We quantify this effect for the \erass clusters in Section \ref{sec:bestz}.
For an example cluster in Figure \ref{fig:zspec_example}, we mark $z_{\rm spec,cg}$ by the green line. 
The darker blue line shows the confidence interval $\delta z_{\rm spec}$ of the more robust $z_{\rm spec}$.

\subsection{Literature cluster redshifts} \label{sec:litz}

\begin{table}[]
    \caption{Literature catalogs matched to the \erass extended X-ray sources.}
    \centering
    \small
    \begin{tabular}{ll}
\hline\hline
Survey & Source\\
\hline
ABELL & \cite{Andernach1991} \\
ACTDR5 & \cite{Hilton2020aa}  \\
CODEX & \cite{Finoguenov2020aa}  \\
eFEDS & \cite{Liu2022}  \\
GALWEIGHT & \cite{Abdullah2020aa}  \\
GOGREEN-GCLASS & \cite{Balogh2020aa}  \\
KIM-HIGHZ-LENSING & \cite{Kim2021}  \\
LP15 & \cite{Aguado-Barahona2019aa}  \\
MADCOWS & \cite{Gonzalez2019aa}  \\
MARD-Y3 & \cite{Klein2019aa}  \\
MCXC & \cite{Piffaretti2011aa}  \\
NEURALENS & \cite{Huang2021aa}  \\
NORAS & \cite{Bohringer2017aa}  \\
PSZ1 & \cite{Planck-Collaboration2015aa}  \\
PSZ2 & \cite{Planck-Collaboration2016aa}  \\
REDMAPPER-DES-SVA & \cite{Rykoff2016aa}  \\
REDMAPPER-DESY1 & \cite{McClintock2019aa}  \\
REDMAPPER-SDSS-DR8 & \cite{Rykoff2016aa}  \\
RXGCC & \cite{Xu2022}  \\
SPT2500D & \cite{Bocquet2019aa}  \\
SPTECS & \cite{Bleem2020aa}  \\
SPTPOL100D & \cite{Huang2020ab}  \\
WENHAN-HIGHZ-SPECZ & \cite{Wen2018aa}  \\
XCLASS & \cite{Koulouridis2021}  \\
XCSDR1 & \cite{Mehrtens2012}  \\
XXL365 & \cite{Adami2018aa}  \\

\hline\hline
    \end{tabular}
    \tablefoot{Further details are given in \cite{Bulbul2023}.}
    \label{tab:litz_match_clusters}
\end{table}

We matched all X-ray cluster candidates (including those outside the LS footprint) with the public catalogs listed in Table \ref{tab:litz_match_clusters}. The matching radius was 2\arcmin and if multiple matches were found, we selected the cluster nearest to the X-ray centroid. This enabled us to assign redshifts to \erass
clusters that were not found by \eromapper or even choose the literature redshift $z_{\rm lit}$ as our best redshift $z_{\rm best}$ if an incorrect optical counterpart was selected (see Section \ref{sec:bestz}). 
The literature redshift can either be photometric or spectroscopic. We did not distinguish these two cases.

\subsection{Assigning the best redshift type} \label{sec:bestzmethods}

We selected the best redshift $z_{\rm best}$ from the available spectroscopic $z_{\rm spec}$ or $z_{\rm spec,cg}$, photometric $z_\lambda$, and literature redshifts $z_{\rm lit}$ by assigning priorities to them in the following order:

\begin{enumerate}
    \item $z_{\rm spec}$ if the cluster has at least three spectroscopic members and the final velocity clipping converged,
    \item $z_{\rm spec,cg}$ if there is a spectroscopic redshift for the galaxy at the optical center,
    \item $z_\lambda$ if a photometric redshift is available and it is within the calibrated redshift range (see below),
    \item $z_{\rm lit}$ otherwise.
\end{enumerate}

\begin{figure*}
    \centering
    \includegraphics[width=0.4\linewidth]{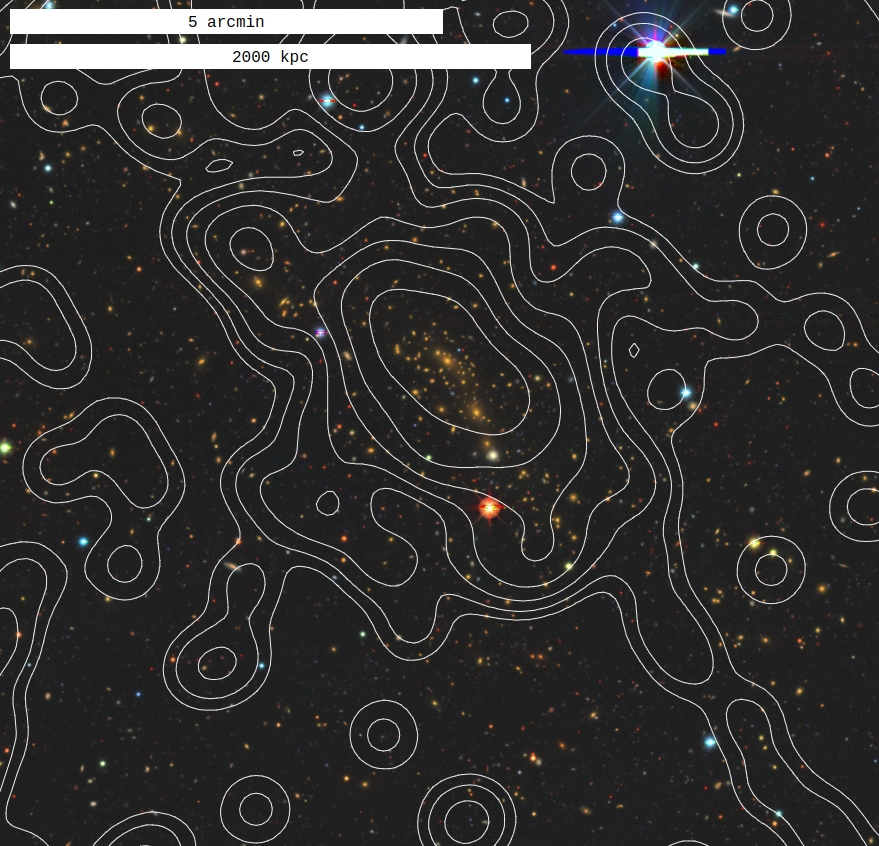}
    \includegraphics[width=0.4\linewidth]{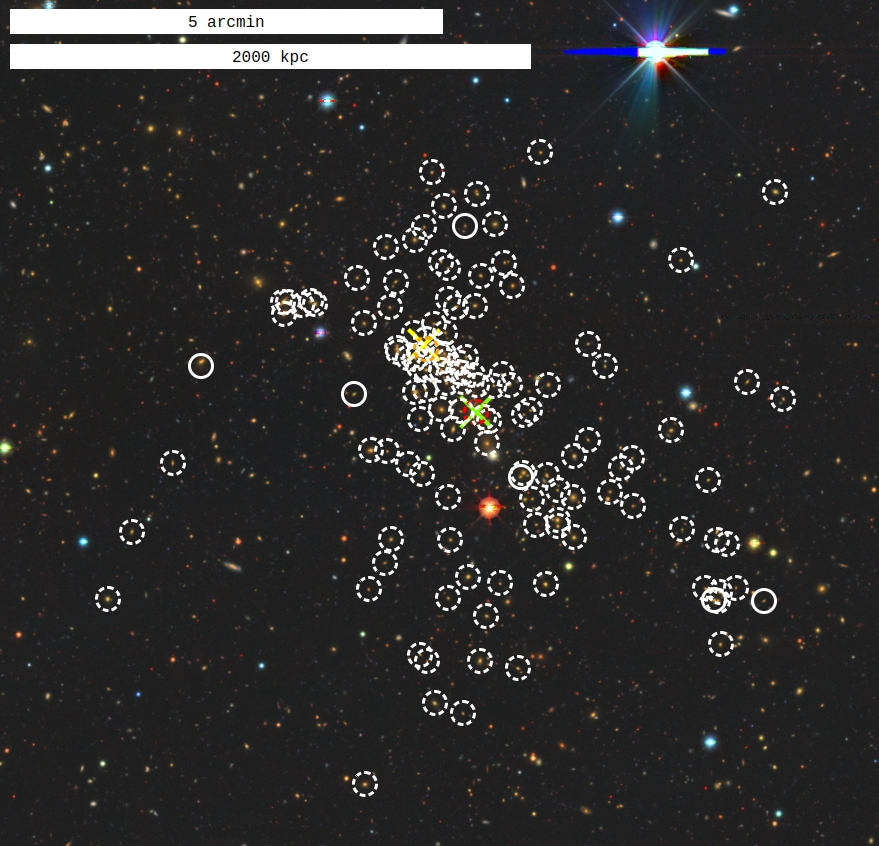}\\
    \vspace{0.1cm}
    \includegraphics[width=0.4\linewidth]{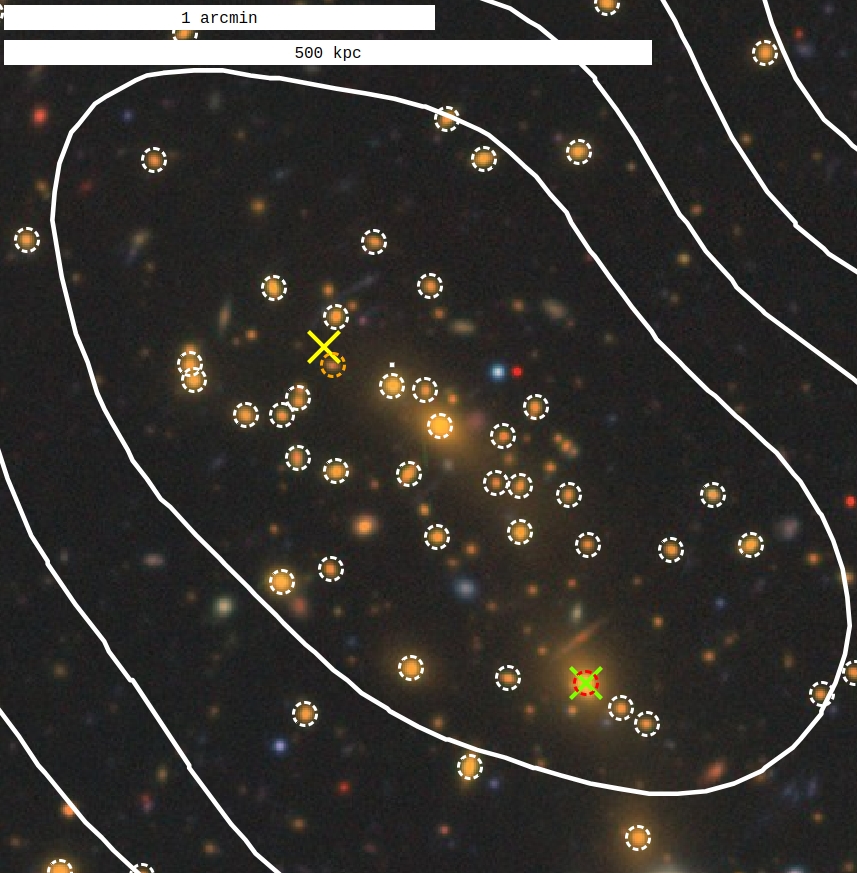}
    \includegraphics[width=0.4\linewidth]{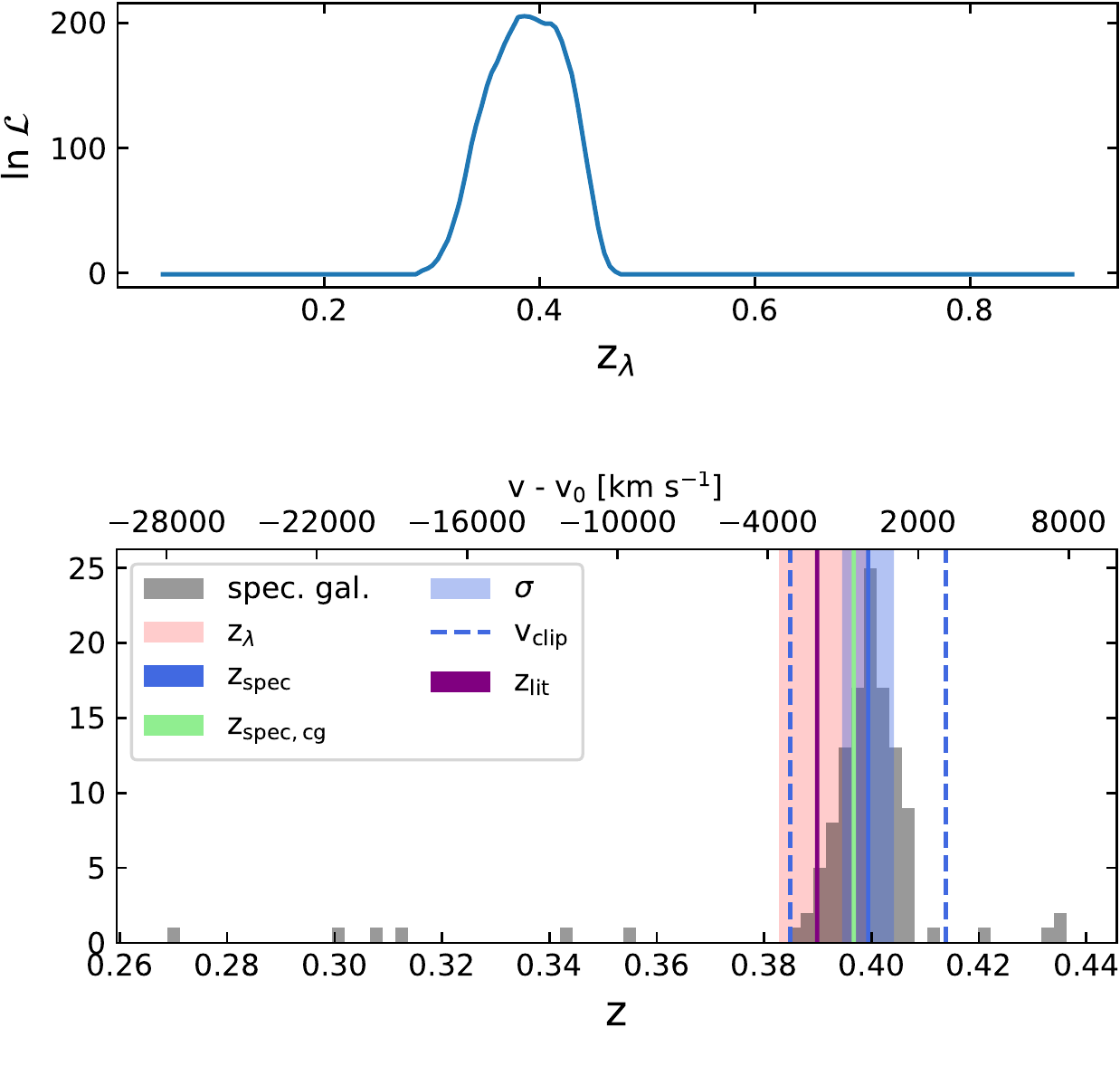}
    \caption{Illustrative example 1eRASS J041610.4-240351 from the \erass cluster catalog. The panels in the top row show an LS $grz$ image of the full cluster overlaid with \erass X-ray contours (left) and cluster members detected in this work (circles, right). Spectroscopic information is available for all cluster members marked by dashed circles. The bottom-left panel shows a zoom-in. The red circle marks the BCG, which agrees with the optical cluster center (green cross).
    The orange circle is located at the position of the central galaxy, which is the one closest to the X-ray center (yellow cross). North is up; east is left.
    In the bottom-right panel, upper subpanel, the cluster likelihood $\ln\mathcal{L}(z)$ shows one peak at $z=0.385$ which is close to the cluster photometric redshift $z_\lambda=0.3910\pm0.0082$.
    The bottom-right panel, lower subpanel, demonstrates the measurements of the spectroscopic redshift and the velocity dispersion for the shown cluster. The cluster has a richness of $\lambda=128.7\pm8.5$. The 127 photometric members are marked by circles. Of those, 121 have spectroscopic redshifts and are marked by dashed circles. 
    The black histogram in the bottom panel shows the spectroscopic redshifts of the photometric members. Ten outlier galaxies with $z\gtrless z_{\rm spec}\pm3\sigma$ (blue dashed lines) are discarded, leaving $N_{\rm members}=111$ spectroscopic members. The cluster velocity dispersion is $\sigma=1034\pm67$\,km\,s$^{-1}$ (light blue line). The photometric cluster redshift (light red shades) agrees with the spectroscopic cluster redshift $z_{\rm spec}=0.39936\pm0.00039$ (darker blue line). The redshift of the galaxy at the optical center is $z_{\rm spec,cg}=0.3967\pm0.0004$ (green line). The literature redshift $z_{\rm lit}=0.38989$ (purple line) is a photometric redshift from MARDY3 \citep{Klein2019aa}.}
    \label{fig:zspec_example}
\end{figure*}

The spectroscopic redshift $z_{\rm spec}$ received the highest priority because it has a low uncertainty and is unbiased. The $z_{\rm spec,cg}$ is also unbiased. Still, it has a larger uncertainty because the central galaxy can have a non-zero line-of-sight velocity relative to the cluster (see example in Figure \ref{fig:zspec_example}). We perform a detailed analysis of the photometric redshift accuracy in Section \ref{sec:redshift_accuracy}.

The photometric redshifts were calibrated in a limited redshift range $0.05<z_\lambda<0.9$ (for the $grz$ \& $griz$ runs), or $0.05<z_\lambda<1.2$ (for the $grzw1$ \& $grizw1$ runs). Details are given in Appendix \ref{sec:redseq}. Only photometric redshifts inside that range are reliable. If a cluster photometric redshift is outside these limits and no spectroscopic redshift is available, we adopted the literature redshift as the best redshift if it was available.

\subsection{Velocity dispersion from spectroscopic member redshifts} \label{sec:vdisp}

Simultaneously with the spectroscopic cluster redshifts (see Section \ref{sec:zspec}), the velocity dispersion \citep{Beers1990aa} was calculated using the biweight scale estimator (Equation (\ref{eq:sigbwt}), if $N_{{\rm spec},k} \geq 15$) or the gapper estimator (Equation (\ref{eq:siggap}), if $3\leq N_{{\rm spec},k} < 15$).

The biweight scale estimator \citep{Tukey1958aa} is defined\footnote{Strictly the prefix is $\sqrt{\left(\frac{N_{\rm spec}^2}{N_{\rm spec} - 1}\right)}$ although $\sqrt{N_{\rm spec}}$ is an adequate approximation for large $N_{\rm spec}$.} as
\begin{equation}
\label{eq:sigbwt}
\sigma_{\rm bwt}(N_{\rm spec}) = \sqrt{N_{\rm spec}} \ \frac{\sqrt{\sum_{|u_j| < 1} \ (v_i - \hat{v})^2 (1 - u_j^2)^4}} {|(\sum_{|u_j| < 1} \ (1 - u_j^2) (1 - 5u_j^2))|},
\end{equation}
where $u_j = u_j(v, v_j, 9)$ (see Equation (\ref{eq:u})). 
The gapper estimator \citep{Wainer1976aa} is based on the gaps of an ordered statistic, $x_j,x_{i+1},\dots,x_n$. It is defined as a weighted average of gaps:
\begin{equation}
\label{eq:siggap}
\sigma_{\rm gap}(N_{\rm spec}) =\frac{\sqrt{\pi}}{N_{\rm spec} \,(N_{\rm spec}-1)}\sum_{j=1}^{N_{\rm spec}-1} w_j \,g_j,
\end{equation}
where weights and gaps are given by $w_j$ and $g_j$ respectively
\begin{align}
 w_j &= j\,(N_{\rm spec} -1),\\
 g_j &= x_{j+1} -x_j. 
\label{eq:gap_eq1}
\end{align}
Spectroscopic galaxies with $v_j > 3\sigma$ were rejected as outliers during a further $\sigma$ clipping process. This procedure was iterated until no outlier galaxies remained or up to a maximum of 20 iterations ($k\leq20$).

Analogous to the spectroscopic cluster redshift, the velocity dispersion was calculated 64 times after bootstrapping member galaxies. The mean of these results was adopted as the final value for the velocity dispersion, and the standard deviation was adopted as the uncertainty. The ratio of the numbers of flagged converged bootstrapping results to all converged bootstrapping results was stored as the velocity dispersion flag. The closer it is to 1, the less robust the velocity dispersion is.

\section{The \erass identified cluster and group catalog} \label{sec:erass}

The main outcome of this work is the optical properties (redshifts, richnesses, etc.; see Appendix \ref{sec:column_descriptions}) of the galaxy clusters and groups detected in \erass. 
Most extended X-ray sources are genuine galaxy clusters. 
However, AGN (or other point sources), as well as random background fluctuations and supernova remnants may have been detected as extended sources, inducing contamination among the cluster candidates. 
The contamination rate is anti-correlated with the brightness of the source and its likelihood to be extended $\mathcal{L}_{\rm ext}$. 
Using simulations, \citet{Seppi2022} predicted that the purity is $\sim$97\% ($\sim$50\%) when considering a sample with $\mathcal{L}_{\rm ext}>3$ and above an average flux limit of 8.0 (0.4) $\times10^{-13}$ ergs s$^{-1}$ cm$^{-2}$. 
The \erass cluster candidate catalog is not strictly flux-limited because of the spatially varying exposure time. The precise limits are described by the selection function \citep[][in particular their Figure E.1]{Clerc2023}. However, an approximate flux limit in an aperture enclosing 500 times the critical density of the Universe at the cluster's redshift is $F_{500}\approx0.4\times10^{-13}$ ergs s$^{-1}$ cm$^{-2}$ \citep{Bulbul2023}. Depending on the science case, more or less pure and complete samples can be selected by applying cuts in $F_{500}$ or $\mathcal{L}_{\rm ext}$.

For the same flux thresholds, the completeness is expected to be $\sim$90\% ($\sim$11.3\%).
Selecting cluster candidates with a higher threshold in $\mathcal{L}_{\rm ext}$ yields higher purity but at the cost of lower completeness for the same flux limit \cite[see also][]{Bulbul2023}. The threshold of $\mathcal{L}_{\rm ext}>3$ was applied in this work to prefer high completeness. The low purity for the low flux threshold was increased significantly in this work by requiring cluster candidates to be identified using optical and near-infrared imaging data.

In that process, we cleaned the \erass cluster candidate catalog of 2497 extended X-ray source detections for which we found no counterpart in the optical data, and of 458 detections which we classified visually as contamination. These results were achieved by running \eromapper in scan mode on the positions of all candidate galaxy clusters and groups in \erass.
We detail the procedure for constructing the catalog in Section \ref{sec:erass_construction} and describe its contents in Section \ref{sec:cluster_confirmation_stats}.

\subsection{Constructing the catalog} \label{sec:erass_construction}

In scan mode, we fixed the cluster coordinates to the cataloged X-ray coordinates (R.A., Decl.). The member search radius is given in Equation (\ref{eq:clusterradius}).
The \eromapper runs were done with different combinations of the $g$,$r$,$i$,$z$, and $w1$ filters. 
The optimal filter band combination changes with redshift. 
We find in Sections \ref{sec:redshift_bias} and \ref{sec:redshift_uncertainties} that $grz$ performed well at low and intermediate redshifts $z_\lambda\leq 0.8$ while $grizw1$ was best suited for high redshifts $z_\lambda>0.8$. 
For the catalogs obtained in \eromapper scan mode, we merged the different runs afterward using a priority scheme.

The richness $\lambda$ measurement varied from run to run. 
That is because membership probabilities change when more filter bands are included and different galaxy luminosity cuts are applied.
We calculate a normalized richness $\lambda_{\rm norm}$ where these systematic effects are corrected in Section \ref{sec:richness}.

\subsubsection{Catalog merging} \label{sec:merging}

Six different \eromapper runs were done on the \erass cluster candidate catalog: four of them for the southern LS in the filter band combinations $grz$, $griz$, $grizw1$, and $grzw1$, and two of them for the northern LS in the filter band combinations $grz$ and $grzw1$. No $i$-band data are available for the northern surveys. The resulting catalogs were merged using the following priority scheme:

\begin{align*}
    1. \quad & \text{LS DR10 south}  && grz    && \text{if $z_\lambda\leq0.8$}
    & (10,\!823), \\
    2. \quad & \text{LS DR10 south}  && griz   && \text{if $z_\lambda\leq0.8$}
    & (314), \\
    3. \quad & \text{LS DR9~~~north} && grz    && \text{if $z_\lambda\leq0.8$}
    & (202), \\
    4. \quad & \text{LS DR10 south}  && grizw1 && \text{if $z_\lambda>0.8$}
    & (485), \\
    5. \quad & \text{LS DR10 south}  && grzw1  && \text{if $z_\lambda>0.8$}
    & (113), \\
    6. \quad & \text{LS DR9~~~north} && grzw1  && \text{if $z_\lambda>0.8$}
    & (7), \\
    7. \quad & \text{other}            &&       &&   & (303), \\
    \cline{7-7}
       \quad &                         &&       &&   & (12,\!247).
\end{align*}

Most clusters were identified in more than one run. The priority scheme avoids multiple entries in the merged catalog for the same cluster by selecting the result only from the run with the highest priority.
In parenthesis is the number of clusters for each category that went into the merged \erass identified catalog.
For example, in run 1, we found 10\,823 clusters, while adding the $i$ band (run 2) complemented the catalog by an additional 314 clusters.\footnote{Vice versa, there were 446 out of 10\,320 clusters identified in the LS DR10 south $grz$ run that also have $i$-band coverage but were rejected when the $i$ band was included in the \eromapper run. The mean probability of being a contaminant is high for those clusters: $P_{\rm cont}=49\%$ (see Section \ref{sec:pcont}). Most of them (95\%) have low richnesses $\lambda<16$, for which the $i$-band might help to identify random line-of-sight projections.}
In the category ``other'' fall clusters that are either at $z>0.8$ and only detected in the $grz$ or $griz$ runs (114), or at $z\leq0.8$ and only detected in the $grizw1$ or $grzw1$ runs (35). If they were detected in more than one run and were always outside the constrained redshift range (7), we still followed the priority scheme but neglected the redshift constraint. Moreover, we included clusters with no counterpart in the LS but are matched with a cluster from the literature (147).
The result is a robustly constructed catalog of the optical properties for 12\,247 identified \erass galaxy clusters and groups.

\subsubsection{Richness normalization} \label{sec:richness}

The cluster richness $\lambda$ is the scaled sum of the membership probabilities $p_{{\rm mem},i}$ (see Equation (\ref{eq:richness})). 
Calculating the scaling factor ("SCALEVAL" in the catalog or $S$ in Equation (\ref{eq:richness})) is part of the \redmapper algorithm \citep{Rykoff2014aa} and includes corrections for the masked area and the limited depth in the optical images (see Appendix \ref{sec:zvlim}).
In this section, we homogenize the richness measurements depending on the filter band combination and the minimum galaxy luminosity.
For the $grz$ and $griz$ runs, we selected only galaxies with a minimum luminosity of $L>0.2$\,L$_*$, where L$_*$ is the break of the Schechter luminosity function \citep{Schechter1976aa}. This luminosity cut minimizes the scatter of the X-ray luminosity at fixed richness \citep{Rykoff2012aa}. Where the LS is sufficiently deep to reliably (with 10$\sigma$ significance) measure galaxy luminosities down to the minimum member galaxy luminosity (i.e., $z_\lambda<z_{\rm vlim}$), the luminosity cut ensures consistent richness measurements across redshift because L$_*$ is an intrinsic property of the galaxy populations. For the $grzw1$ and $grizw1$ runs, we applied a higher minimum luminosity of $L>0.4$\,L$_*$ to obtain (noisier but) unbiased richnesses at higher redshift $z>0.8$ (see Appendix \ref{sec:zvlim}).
When the luminosity threshold increases, the richness is systematically smaller because fewer galaxies are selected. To correct for this effect, we define a normalized richness;

\begin{equation}
  \lambda_{\rm norm}=S_{\rm norm}\lambda \label{eq:richnessnorm}
\end{equation}
and its uncertainty

\begin{equation}
  \delta\lambda_{\rm norm}=S_{\rm norm}\delta\lambda.
\end{equation}

It was determined by comparing the richness of the same clusters measured in different runs. The relations were fit by minimizing the uncertainty-weighted squared orthogonal distances to the best-fit line \citep{Boggs1989}. We also applied an orthogonal cut at low richness to not bias the slope. The resulting scaling factors are given in Table \ref{tab:lambdanorm}.

\begin{table}
    \caption{Richness scaling factors.}
    \centering
    \begin{tabular}{ll}
    \hline\hline
    Filter Band Combination & $S_{\rm norm}$ \\
    \hline
    $grz$ (DR10 south) & 1.000 \\
    $griz$ (DR10 south) & 1.039 \\
    $grz$ (DR9 north) & 1.020 \\
    $grizw1$ (DR10 south) & 2.451 \\
    $grzw1$ (DR10 south) & 2.336 \\
    $grzw1$ (DR9 north) & 2.550\\
    \hline\hline
    \end{tabular}
    \tablefoot{Richness scaling factors $S_{\rm norm}$ used to calculate the normalized richness $\lambda_{\rm norm}=S_{\rm norm}\lambda$. As the reference, we chose $S_{\rm norm}=1$ for the $grz$ filter bands in the LS DR10 south. The biggest impact on $S_{\rm norm}$ is the minimum galaxy luminosity of the cluster members. It is fainter for the $grz$ and $griz$ filter band combinations ($L>0.2$\,L$_*$) than for the runs that include the $w1$ band ($L>0.4$\,L$_*$). The richness increases when fainter members are included.
    } \label{tab:lambdanorm}
\end{table}

\subsubsection{Assigning the BCG}\label{sec:bcg:choice}

Near the center of a galaxy cluster often resides a disproportionately bright and extended early-type galaxy called the Brightest Cluster Galaxy (BCG). These galaxies are distinct from normal massive early-type galaxies due to their embedding in the faint Intracluster Light (ICL, see recent reviews by \citealt{Contini2021,Arnaboldi2022,Montes2022}). 

The choice of the BCG can be ambiguous in $\sim$20\% of the cases \citep{Kluge2020aa}. For simplicity, we defined the BCG as the brightest member galaxy in the $z$ band. We explore the justification for this definition in Section \ref{sec:bcg_analysis}.

\subsection{Properties of the \erass catalog of identified clusters and groups} \label{sec:cluster_confirmation_stats}

In this section, we detail the overall statistics of the results obtained with the \eromapper runs on the \erass cluster candidate catalog.

\begin{figure}
    \includegraphics[width=\linewidth]{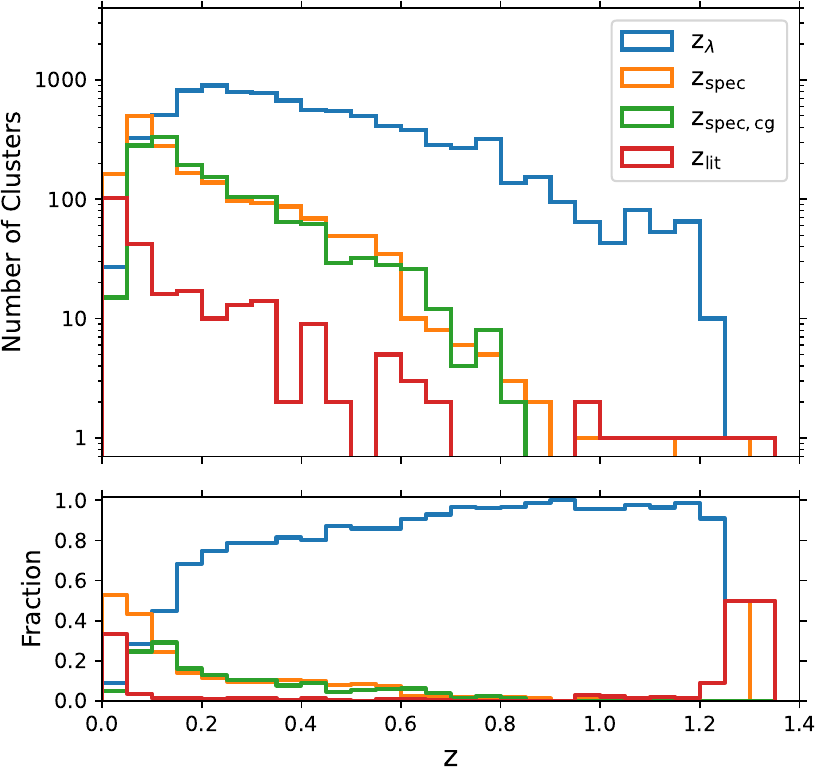}
    \caption{Number (top panel) and relative fraction (bottom panel) of \erass clusters per redshift bin depending on their best redshift type. The total numbers are 8790 photometric redshifts ($z_\lambda$, Section \ref{sec:zlambda2}), 1759 spectroscopic redshifts determined using the bootstrap method ($z_{\rm spec}$, Section \ref{sec:zspec}), 1451 spectroscopic redshifts of the galaxy at the optical center ($z_{\rm spec,cg}$, Section \ref{sec:zspec}), and 247 literature redshifts ($z_{\rm lit}$, Section \ref{sec:litz}). Clusters with unflagged velocity dispersion information (Section \ref{sec:vdisp}) have the same distribution as the $z_{\rm spec}$ sample.} \label{fig:bestztype}
\end{figure}

\subsubsection{Cluster identification statistics} \label{subsec:CTPs}

The \erass source catalog covers half of the sky. It contains 26\,682 extended sources \citep{Merloni2023}. After cleaning split sources and masking regions of known non-cluster extended X-ray sources \citep[see Section \ref{subsec:erosita} and][]{Bulbul2023}, we obtain a sample of 22\,718 cluster candidates with extent likelihood $\mathcal{L}_{\rm ext}>3$. We ran \eromapper in scan mode on this sample and obtained optical properties (redshifts, richnesses, optical centers, and BCG positions) for 12\,705 clusters.

Most of the rejected candidates (7516) are outside of the LS footprint (see Figure \ref{fig:footprint} and Appendix \ref{sec:infootprint}) and have no match to known clusters in the literature (see Section \ref{sec:litz}). A further 2497 rejected candidates have LS coverage, but no optical red sequence counterpart has been found with \eromapper. These X-ray sources are most likely AGN or random background fluctuations misidentified as extended sources \citep{Bulbul2022}. We discuss the properties of these contaminating sources in Section \ref{sec:pcont}.
It is also possible that some real clusters were not optically identified because their redshift is higher than the limiting redshift of the LS $z\gg z_{\rm vlim}$.
Finally, we performed a visual cleaning procedure as described in Section \ref{sec:bestz}. This further reduced the sample to a final size of 12\,247 \erass clusters in a sky area of 13\,116\,deg$^2$, so an average density of about one cluster per square degree.

The association of the X-ray signal with the optical cluster was straightforward in 97\% of the cases. For the remaining 332 \erass clusters (3\%) with $z_\lambda<z_{\rm vlim}$, we found overlapping structures in projection. A secondary photometric redshift $z_{\lambda,2}$ is provided in these cases. It is mentioned in \cite{Bulbul2023} that one eFEDS cluster \citep[eFEDSJ091509.5+051521,][]{Klein2021aa,Liu2022} has inconsistent redshift ($z=0.249$) to its counterpart 1eRASS J091510.8+051440 ($z_{\rm spec}=0.136$). We find that it agrees with $z_{\lambda,2}=0.269$.

Close but separate X-ray sources can have a large number of common members identified when their separation is smaller than the optical cluster radius $R_\lambda$. This leads to a small number of 413 (3\%) \erass clusters that share >70\% of their members with another cluster that has a higher likelihood $\mathcal{L}_{\rm max}$. We marked the cluster with lower $\mathcal{L}_{\rm max}$ in the \erass catalog by flagging it as {\tt SHARED\_MEMBERS}.

\begin{table}[]
    \caption{Redshift types and occurrences in the \erass catalog.}
    \centering
    \begin{tabular}{llcrr}
        \hline\hline
        Pri. & Redshift Type & Notation & Total & Best \\
        \hline
        1 & spectroscopic & $z_{\rm spec}$ & 1906 & 1759 \\
          & ($\geq$3 members) &  &  &  \\
        2 & spectroscopic & $z_{\rm spec,cg}$ & 2881 & 1451 \\
          & (gal. at opt. center) &  &  &  \\
        3 & photometric & $z_\lambda$ & 12\,100 & 8790 \\
          & & & & \\
        4 & literature & $z_{\rm lit}$ & 3886 & 247 \\
        \hline
        \\
        Total  & & & & 12\,247 \\
         \hline\hline
    \end{tabular}
    \tablefoot{The priorities in the first column and redshift types in the second and third columns are described in Section \ref{sec:bestzmethods}. The total number of available redshifts for each type is given in column four, and column five shows how often each redshift type is selected as the best redshift $z_{\rm best}$.}
    \label{tab:bestztype}
\end{table}

\subsubsection{Cluster redshifts} \label{sec:bestz}

Each of the 12\,247 \erass clusters has at least one redshift assigned. In 5428 cases where multiple redshifts are available, we followed a priority scheme to assign the best redshift $z_{\rm best}$ from up to four available redshift types (see Section \ref{sec:bestzmethods}). Table \ref{tab:bestztype} summarizes the results for the \erass identified catalog. Most values of $z_{\rm best}$ are photometric (72\%). We perform a detailed analysis of the accuracy of the photometric redshifts in Section \ref{sec:redshift_accuracy}.
Higher-quality spectroscopic redshifts are available for a significant fraction (26\%). Literature redshifts were adopted only in rare cases (2\%).
Figure \ref{fig:bestztype} shows the number of clusters per redshift bin with associated redshifts. At low redshifts $z<0.15$, more than half of the \erass clusters have a reliable spectroscopic redshift. This number falls below 10\% above $z>0.6$.

The spectroscopic redshifts are calculated using the robust bootstrapped bi-weight method (15\%; see Section \ref{sec:zspec}) or adopted from the spectroscopic redshift of the central galaxy (12\%).
We note that the formal uncertainties for the latter method underestimate the real cluster redshift uncertainties because individual cluster member galaxies can have a non-negligible line-of-sight velocity with respect to the cluster rest frame (see Section \ref{sec:zspec} and Figure \ref{fig:zspec_example}). For the \erass clusters, the mean offset velocity is 80\% of the cluster velocity dispersion. This corresponds to a mean uncertainty of $\delta z=0.0013$.

Literature redshifts, when available, were adopted for clusters (a) outside of the LS footprint (126 cases), (b) within the LS footprint but without an optical detection (21 cases), (c) where we decided after visual inspection that the optical detection is in projection to the X-ray source (95 cases), or (d) at low or high redshifts where $z_{\lambda}$ is outside of the calibration range for the red sequence ($z<0.05$ and $z>1.2$, see Appendix \ref{sec:redseq}) and no spectroscopic redshift was available (5 cases). In total, 247 \erass clusters have $z_{\rm best}=z_{\rm lit}$.
The 21 matched clusters within the LS footprint but without optical detection by us, are
(a) at very low redshift $z_{\rm lit}<0.05$ (14 cases), (b) at higher redshift than the limiting redshift of the LS (2 cases), (c) a very poor group (1eRASS J022946.2-293740), (d) consist only of late-type galaxies (1eRASS J091433.5+063417), or not detected for unknown reasons (1eRASS J050808.8-525124, 1eRASS J055136.2-532733, 1eRASS J102144.9+235552). 

Furthermore, there are cases in which more than one cluster is projected near the X-ray emission. In these cases, \eromapper chooses the cluster with the highest likelihood $\ln\mathcal{L}_{\rm max}$, which correlates with the richness by definition. This is generally a good strategy when the galaxy sample is complete. However, we show in Section \ref{sec:lowzcompleteness} that the optical completeness drops below $z\lesssim0.05$ because we discarded galaxies near bright and extended galaxies.
Consequently, if the cluster is at low redshift, \eromapper often chooses a background cluster. We identified these cases by 
\begin{itemize}
    \item matching all clusters with the NGC catalog \citep{Dreyer1888} using a 30\arcsec~search radius,
    \item looking for a secondary low-$z_{\lambda,2}$ peak in the $\ln\mathcal{L}(z)$ distributions (see Section \ref{sec:zlambda2}),
    \item selecting clusters with low $z_{\rm lit}$,
    \item searching for close pairs in the X-ray images
\end{itemize}
and inspecting their X-ray images, optical images, and member galaxy distributions visually. For the cases that we judged to be misidentified, we set $z_{\rm best}=z_{\rm lit}$ if a literature redshift was available (87 cases). If no $z_{\rm lit}$ was available or we judged that the X-ray signal is not spatially associated with the cluster's ICM emission, we discarded that cluster (458 cases).
This includes 31 cases where the X-ray signal is associated with the central emission of a nearby late-type or strongly distorted galaxy at $z_{\rm lit}<0.01$.

\subsubsection{\erass clusters in the large-scale structure}\label{subsec:LSS:clusters}

\begin{figure}
    \centering
    \includegraphics{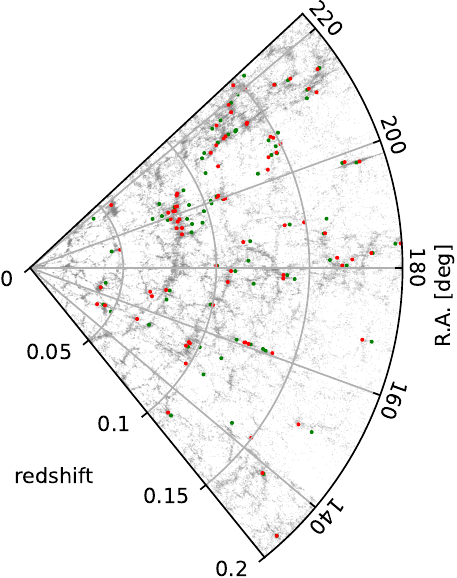}
    \caption{Slice of the cosmic volume with \erass clusters with available spectroscopic redshifts $z_{\rm spec}$ overplotted onto our spectroscopic galaxy compilation (gray points). The slice has a thickness of $\pm2\degr$ around ${\rm Decl.} = 0\degr$. Red points mark the positions of \erass clusters when spectroscopic redshifts were used. They trace the nodes of the cosmic web. Green points mark the same clusters, but this time, we used the photometric redshifts. They scatter around the nodes of the cosmic web because of the higher redshift uncertainty.}
    \label{fig:lss}
\end{figure}

Massive galaxy clusters are located at the nodes of the cosmic web. 
Galaxies trace this web and can be visualized by selecting a thin slice limited in declination around $\pm2\degr$.
Figure \ref{fig:lss} shows a zoom-in on the redshift range $0\leq z\leq0.2$.
We also restricted the range in right ascension ($125\degr < {\rm R.A.} < 223\degr$) to the overlapping area in the western Galactic hemisphere that is well covered by \erosita (see Figure \ref{fig:footprint}).
Gray data points correspond to the redshifts of the galaxies in our spectroscopic compilation (see Sec. \ref{sec:spec_compilation} and Appendix \ref{sec:spec_compilation_refs}). Three slices in R.A. have a high galaxy density. They correspond to the GAMA fields G09, G12, and G15 \citep{Driver2022}.
Overplotted in green are the positions of the \erass clusters when using their photometric redshifts. We show in Section \ref{sec:redshift_uncertainties}
that these redshifts have an uncertainty of $\delta_z\approx0.005-0.015$.
This is sufficiently large to cause an apparent displacement of the \erass clusters from their true position in the cosmic web. 
The red points correspond to the same clusters but are located at their spectroscopic redshifts, with a $\sim$10 times higher precision (see Section \ref{sec:redshift_accuracy}). In these cases, the \erass clusters trace the cosmic web well, as can be seen for example around $z\approx0.08$ and ${\rm R.A.}\approx200\degr$.

\subsubsection{Consistency with known clusters}

\begin{figure}
    \centering
    \includegraphics[width=\linewidth]{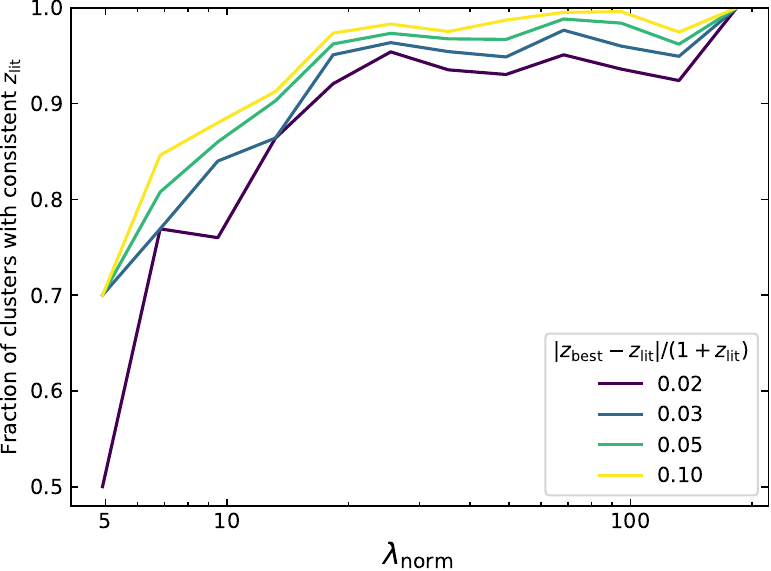}
    \caption{Fraction of \erass clusters with consistent best redshift $z_{\rm best}$ and literature redshift $z_{\rm lit}$ depending on richness $\lambda_{\rm norm}$. Four different redshift tolerances are applied.}
    \label{fig:lmax_deltaz}
\end{figure}

We matched all \erass cluster candidates with the public catalogs in Table \ref{tab:litz_match_clusters} as described in Section \ref{sec:litz} and found 3886 pairs. The consistency of $z_{\rm best}$ with the matched $z_{\rm lit}$ is shown in Figure \ref{fig:lmax_deltaz}. We only considered clusters where the best redshift type is not the literature redshift and the masking fraction is below $<10\%$. 
The four colored lines refer to different allowed redshift deviations. 
Above a richness of $\lambda_{\rm norm}=20$, 94\% of the matched clusters have consistent redshifts within $|z_{\rm best}-z_{\rm lit}|/(1+z_{\rm lit})<0.02$. 
The fraction increases to $>98\%$ with larger allowed redshift deviations $|z_{\rm best}-z_{\rm lit}|/(1+z_{\rm lit})<0.10$. The remaining inconsistencies are explained by ambiguous cluster choices in the presence of multiple structures along the line of sight (see Section \ref{sec:multipleclusters}).
At low richness ($\lambda_{\rm norm}<20$), these ambiguous choices become more frequent because the purity decreases as we show in Section \ref{sec:pcont}. Here, about 80\% (92\%) of the matched clusters have consistent redshifts within $|z_{\rm best}-z_{\rm lit}|/(1+z_{\rm lit})<0.02$~~~($<0.10$).

Surprisingly, two matched clusters with large richnesses $\lambda_{\rm norm}>100$ are inconsistent with the literature redshift.
The first outlier, 1eRASS~J094813.0+290709, is a rich cluster with $\lambda=133\pm7$ at high redshift $z_\lambda=0.766\pm0.013$. The \erass X-ray signal is well centered on that cluster. 
Its matched counterpart in the CODEX catalog has a significantly lower redshift $z_{\rm lit}=0.253$ and lower richness $\lambda_{\rm lit}=19\pm2$. 
We identify another cluster in close proximity (2.4\arcmin~east) in our blind-mode \eromapper catalog, which we describe in Section \ref{sec:blind_construction}. Its redshift $z_{\rm spec}=0.254$ is consistent with the CODEX redshift. One likely reason for the different choices is the limited depth. CODEX was limited to a lower redshift range $z\lesssim0.6$ because it relied on shallower SDSS data where the high-$z$ cluster is not visible. Another possibility is miscentering. CODEX relied on ROSAT data with a much larger PSF. This could have played a role in determining the cluster center in the CODEX catalog.

The second outlier, 1eRASS~J020628.4-145358 (Abell 305), has a robust redshift of $z_\lambda=0.298\pm0.006$. Its matched counterpart MCXC J0206.4-1453 is at $z_{\rm lit}=0.153$. The literature redshift was measured using spectroscopic redshifts of two galaxies \citep{Romer1994}. We identified one late-type galaxy LEDA 918533 close ($\approx20\arcsec$) to the \erass X-ray emission peak, which could be a foreground galaxy that was misclassified as a cluster member by \cite{Romer1994}.

\subsubsection{Cluster velocity dispersions}

For all \erass clusters with at least three spectroscopic member galaxies, we calculated the cluster velocity dispersion $\sigma$ as the line-of-sight velocity dispersion of those galaxies. The total number of \erass clusters with an estimated velocity dispersion is identical to the number of 1906 clusters with available spectroscopic redshift $z_{\rm spec}$ (see Table \ref{tab:bestztype}). However, the flagging is stricter. For $z_{\rm spec}$, we required the velocity clipping of the full sample of spectroscopic members to converge. For $\sigma$, we required the velocity clipping for all bootstrapped realizations to converge. This gave more robust results by reducing the number of clusters with substructure. Of the 1906 spectroscopic clusters, 1499 were not flagged and, hence, have robust $\sigma$ estimates. 

The clusters with robust $\sigma$ estimates against redshift are similar to the distribution shown for $z_{\rm spec}$ in Figure \ref{fig:bestztype} by the orange line. We do not show the histogram for $\sigma$ explicitly.

\begin{figure*}
    \includegraphics[width=\linewidth]{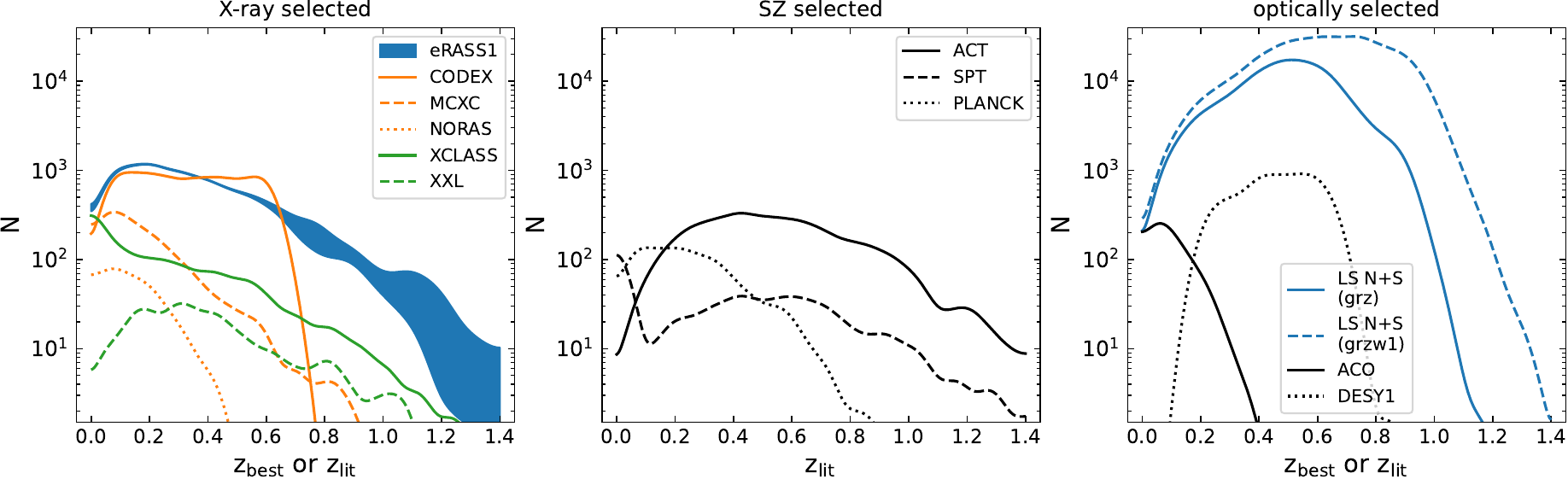}
    \caption{Number of clusters per redshift interval $\Delta z_\lambda=0.05$. For catalogs created in this work (blue), we used the best redshift $z_{\rm best}$, and for literature catalogs, we used the published redshift $z_{\rm lit}$. Each panel refers to a different cluster selection method. The smoothed histograms were obtained by summing over the redshift posteriors of all clusters. For better visibility, the distributions were additionally smoothed with a Gaussian kernel with a standard deviation $\Delta z_\lambda=0.04$. For the \erass catalog, the upper border of the filled region refers to all 12\,247 clusters, and the lower border refers to the subset of 10\,959 clusters with $z_\lambda<z_{\rm vlim}$. A richness cut $\lambda_{\rm norm}>16$ and a limiting redshift cut $z_\lambda<z_{\rm vlim}$ were only applied to the optically selected combined LS DR10 south and LS DR9 north catalogs (LS N+S). That makes them comparable to the DES year 1 catalog, which has a consistent richness cut of $\lambda>20$ applied (see Section \ref{sec:desrichnesses}).
    \label{fig:zscanhisto}}
\end{figure*}

Of those 1499 clusters, 358 have high-quality velocity dispersion values calculated with the bi-weight scale estimator (see Section \ref{sec:vdisp}). As we required a large number of $N_{\rm members}\geq15$ spectroscopic members for this method, these values have a low relative uncertainty: $\delta\sigma/\sigma\sim20\%$. 
The remaining 1141 velocity dispersions were calculated with the gapper estimator. The relative uncertainties are on the order of $\delta\sigma/\sigma\sim40\%$.
In total, 6.5\% of the photometric members with spectroscopic redshift information were discarded by the velocity clipping procedure.

\section{Quality assessments of the \erass catalog} \label{sec:quality}

In addition to the \erass catalog, we ran \eromapper in scan mode on various catalogs of candidate clusters from the literature, selected using different methodologies.
By doing so, we can compare our results to the redshifts and richnesses measured in previous works.
In addition to this, we ran \eromapper in blind mode on the full LS DR10 south and LS DR9 north. This enabled the analysis of selection effects that arose purely from processing the optical and near-infrared data. 
These effects can manifest as a depth-dependent contamination rate and possible deviations of the cluster number density from a theoretical halo-mass function.

\begin{figure*}
    \includegraphics[width=0.24\linewidth]{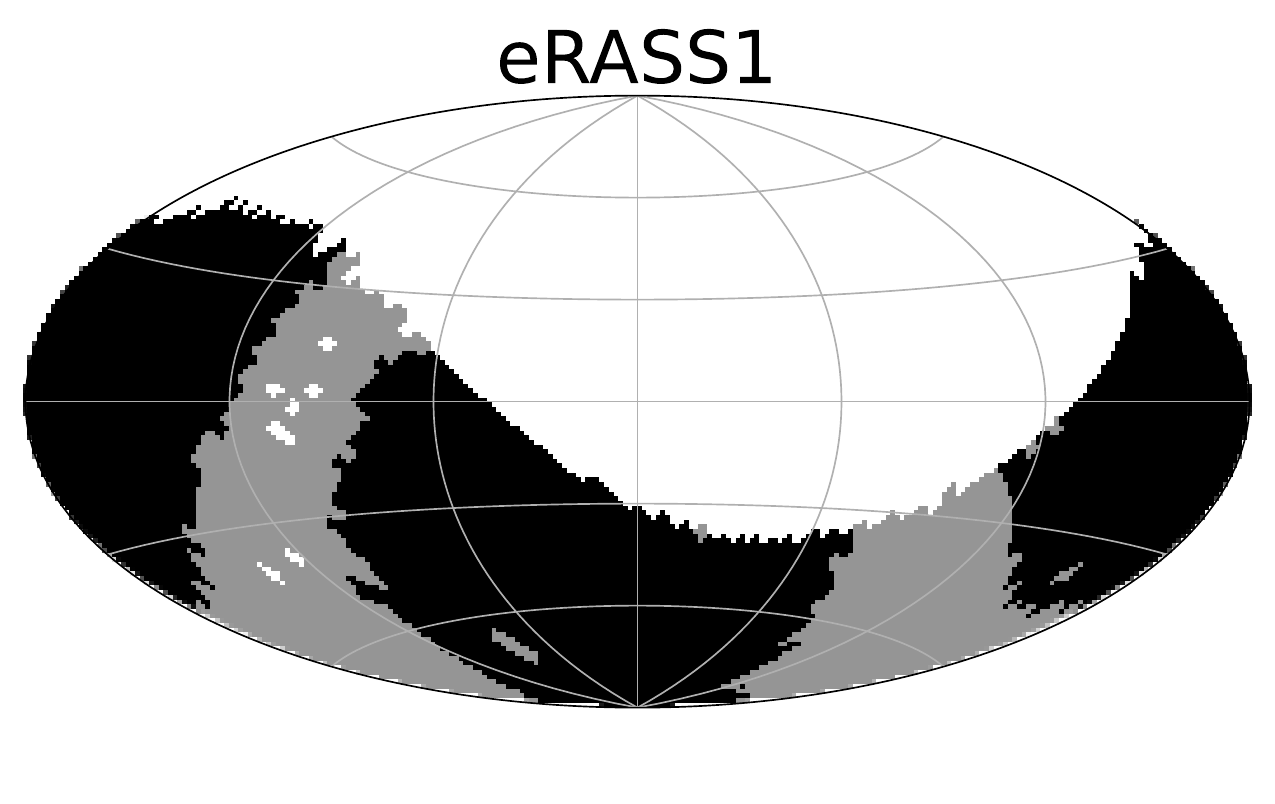}
    \includegraphics[width=0.24\linewidth]{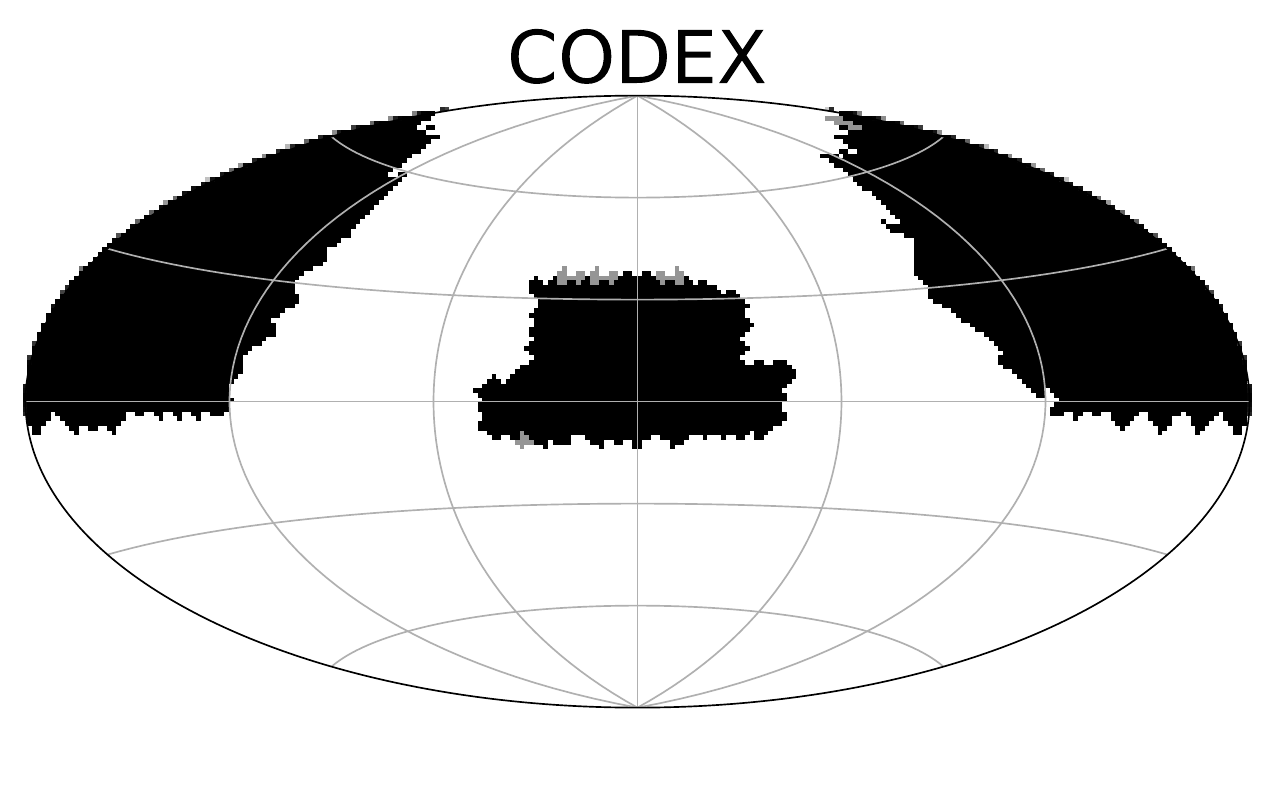}
    \includegraphics[width=0.24\linewidth]{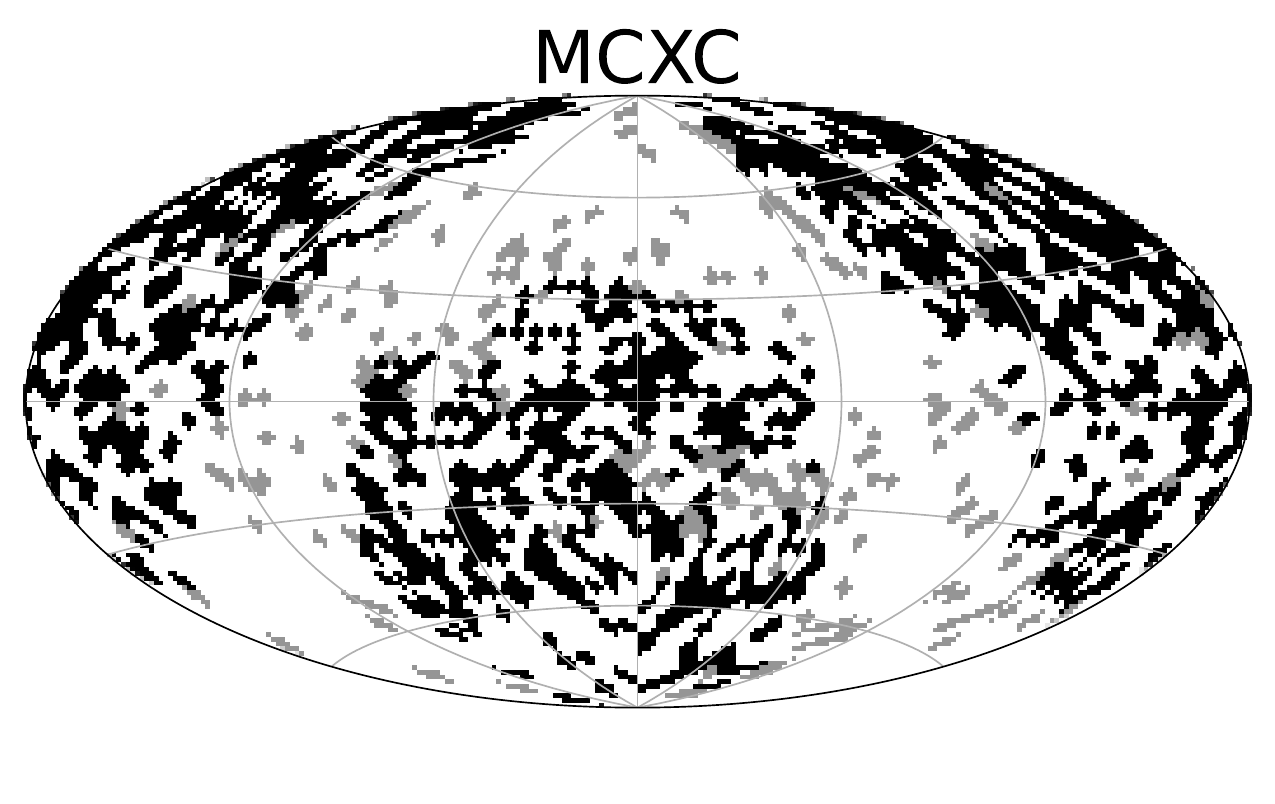}
    \includegraphics[width=0.24\linewidth]{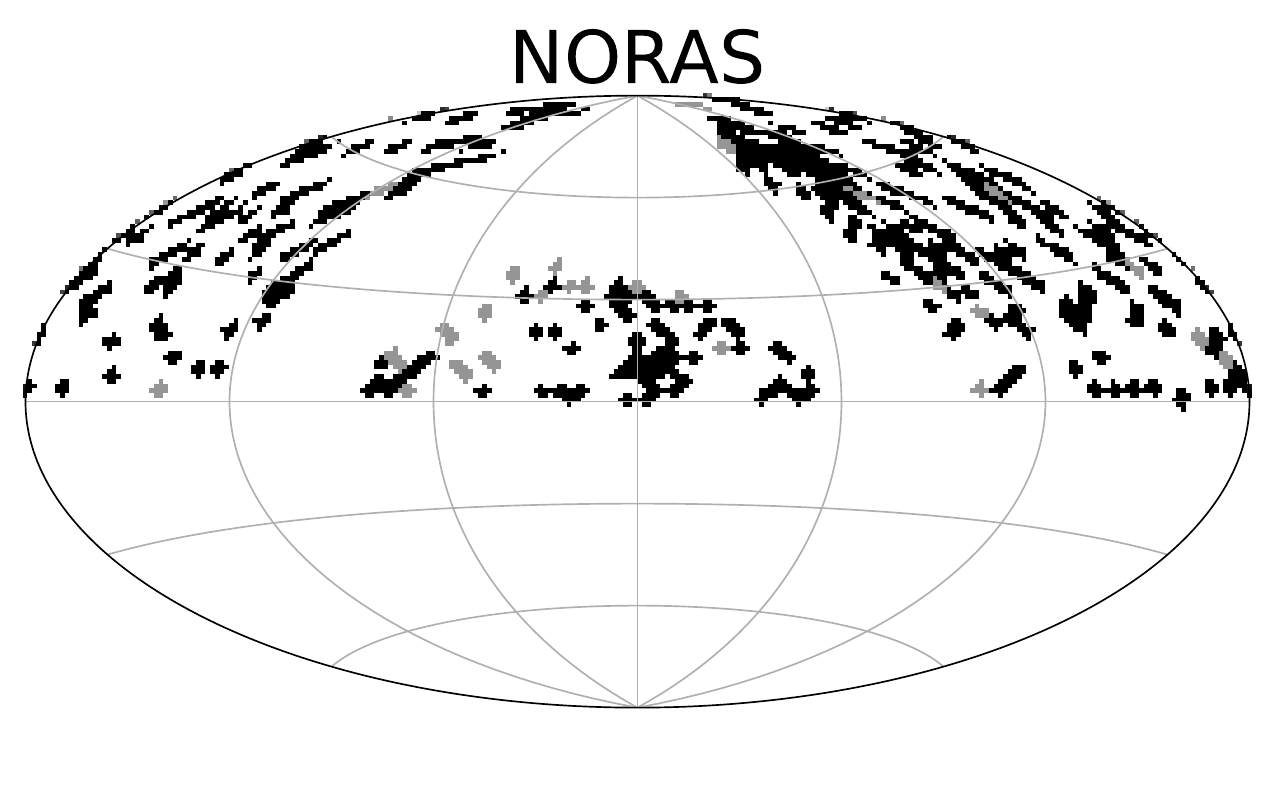}\\
    \includegraphics[width=0.24\linewidth]{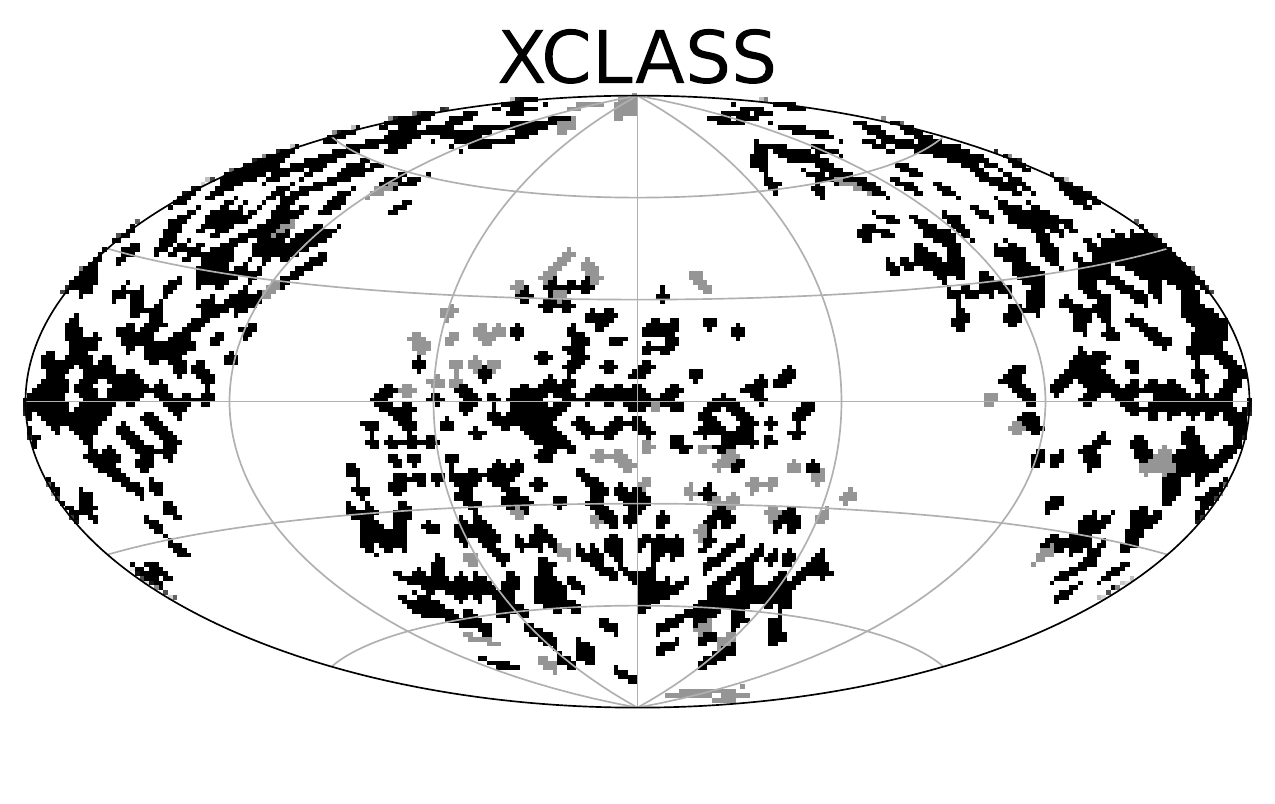}
    \includegraphics[width=0.24\linewidth]{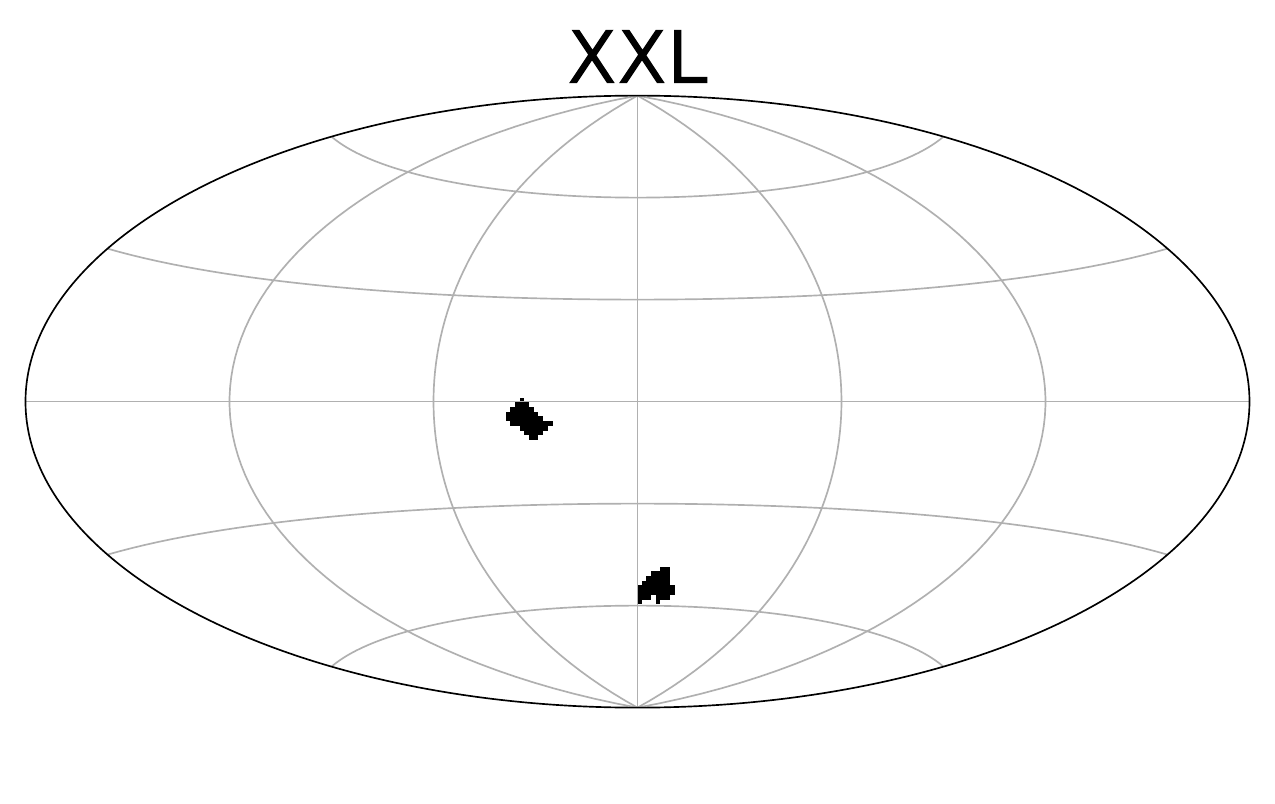}
    \includegraphics[width=0.24\linewidth]{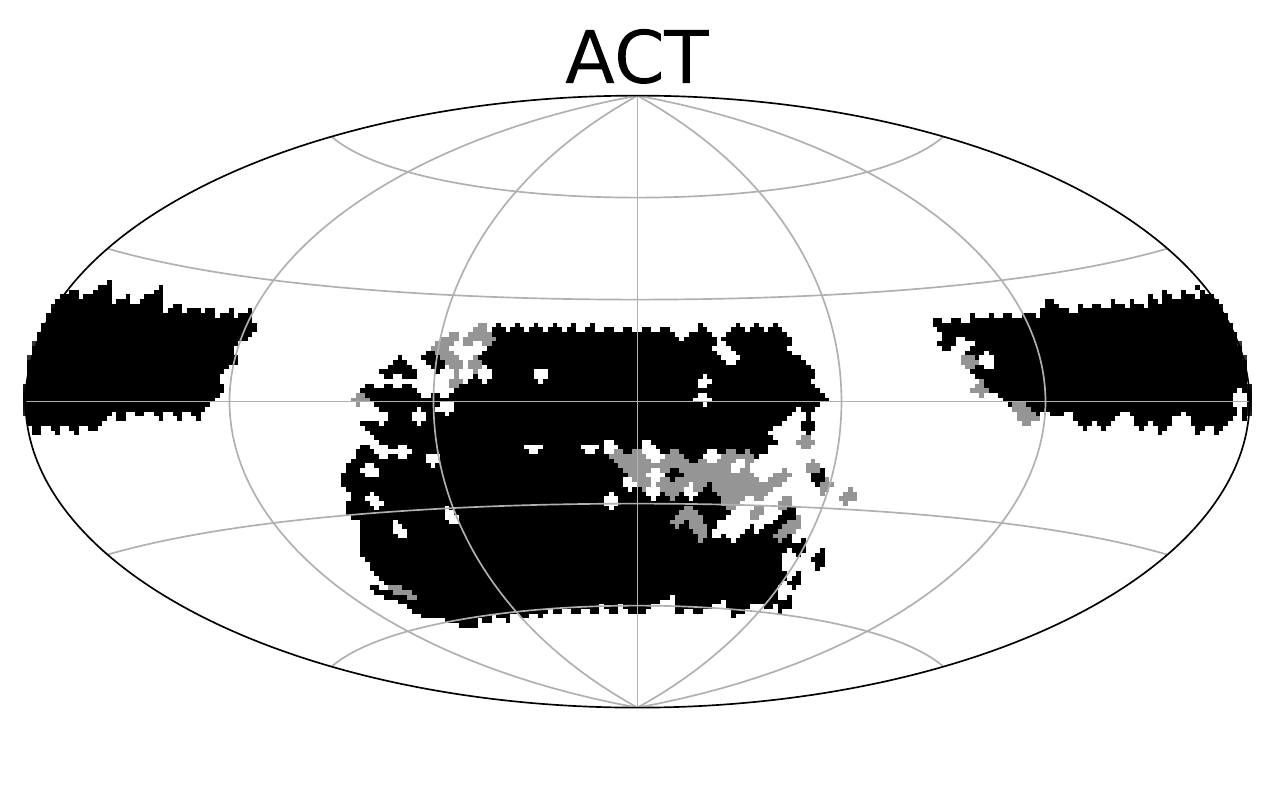}
    \includegraphics[width=0.24\linewidth]{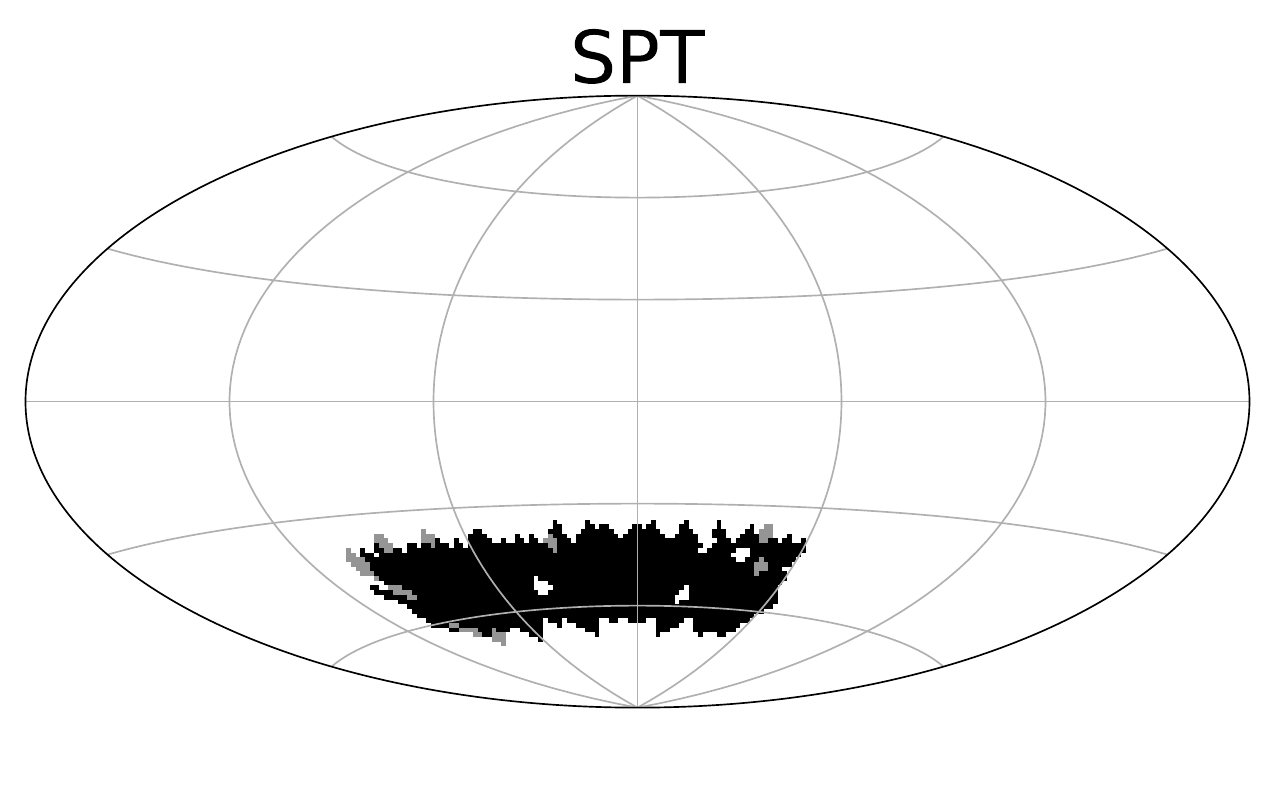}\\
    \includegraphics[width=0.24\linewidth]{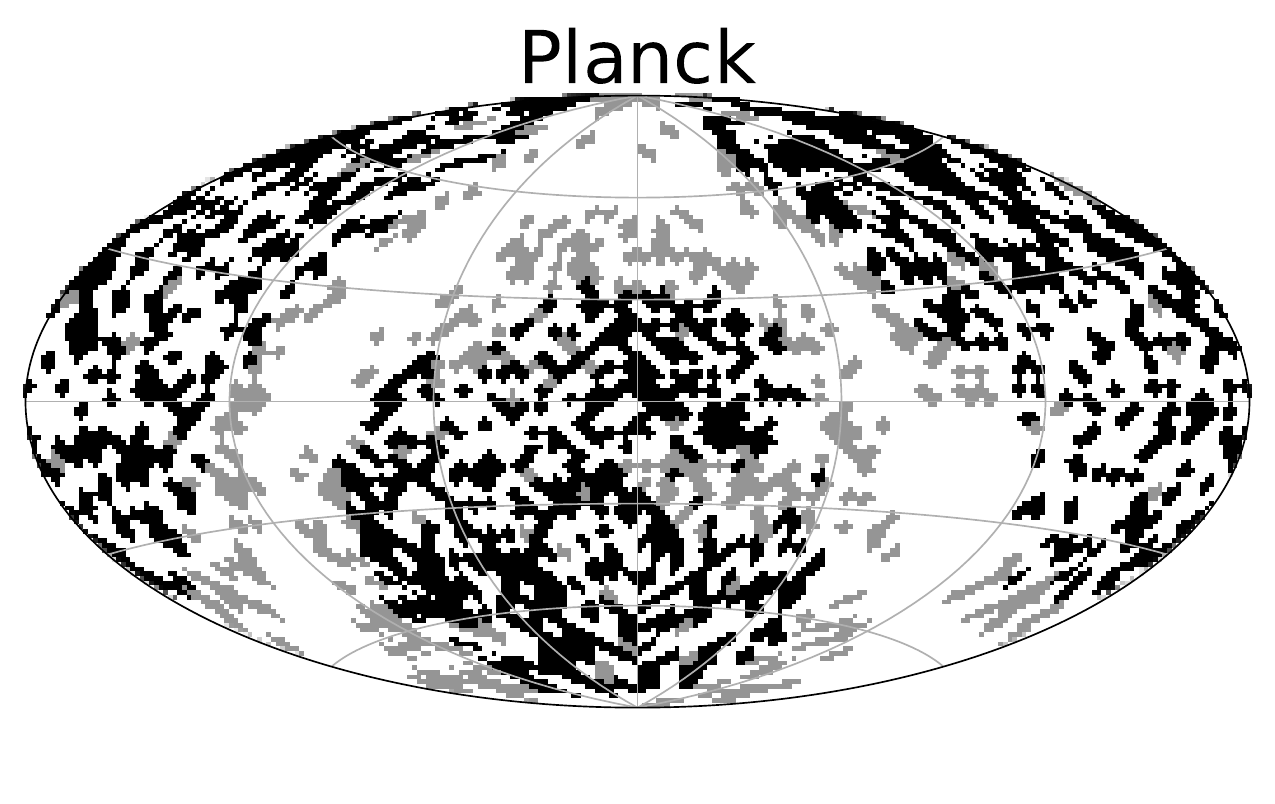}
    \includegraphics[width=0.24\linewidth]{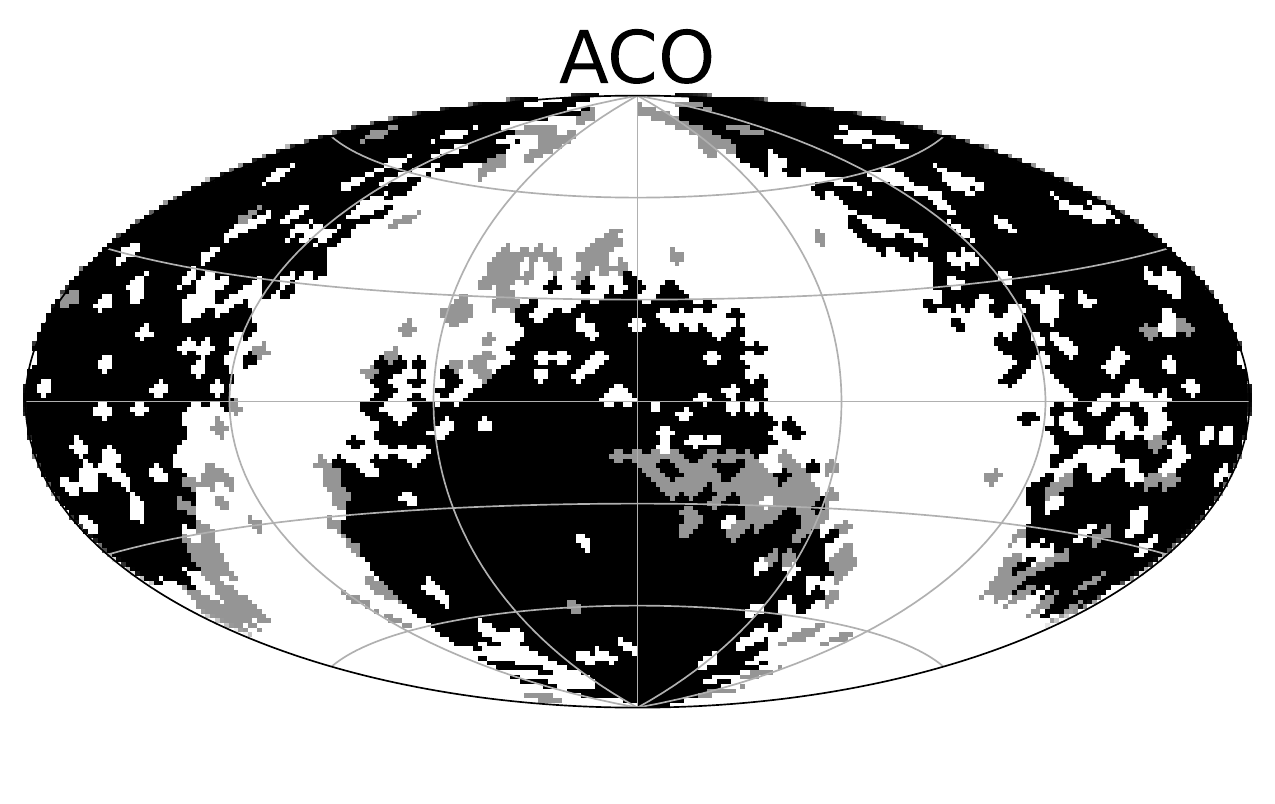}
    \includegraphics[width=0.24\linewidth]{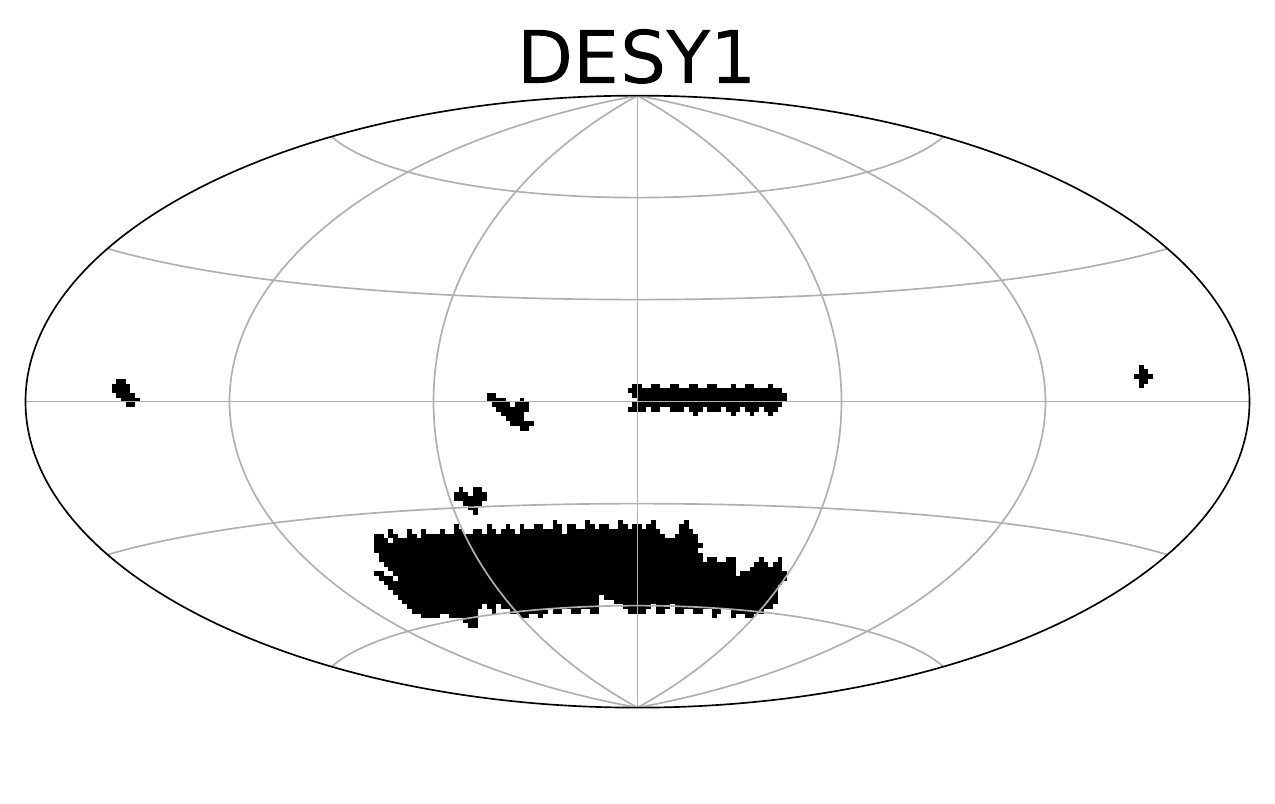}
    \includegraphics[width=0.24\linewidth]{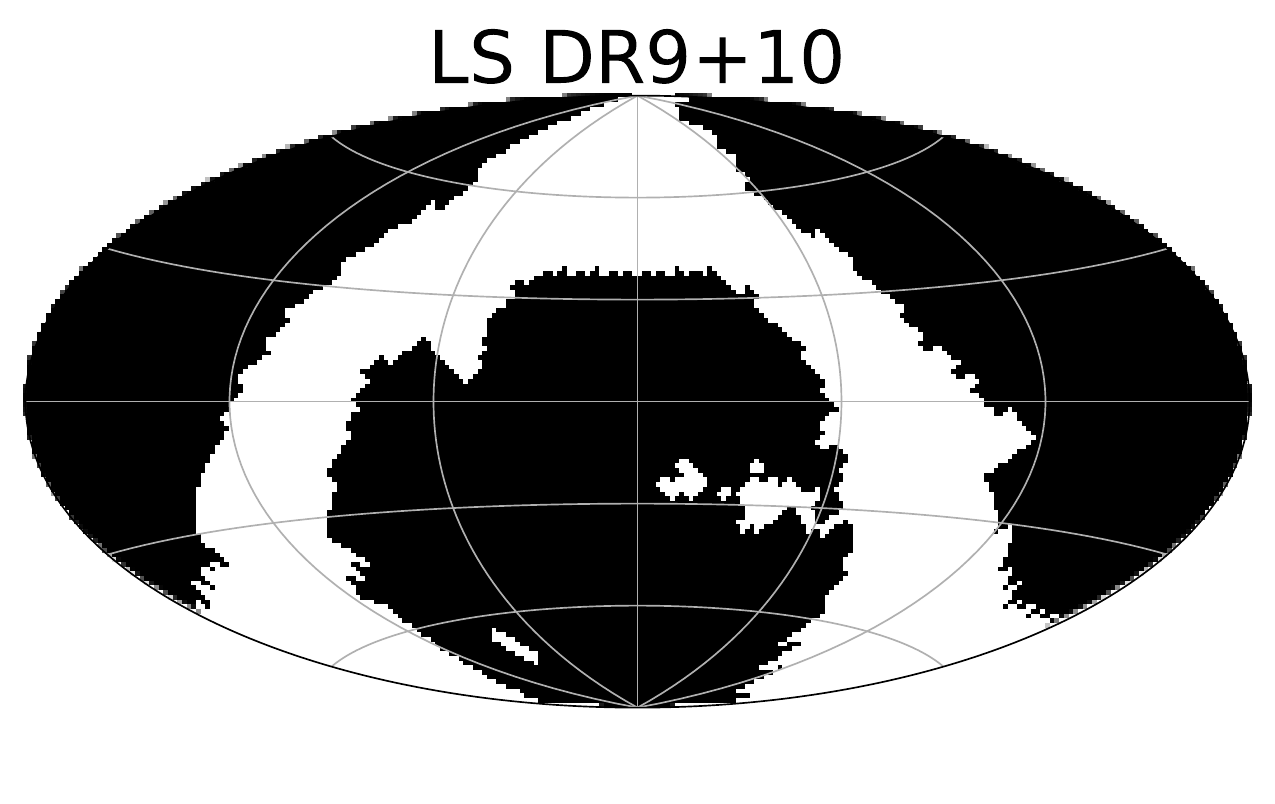}
    \caption{Footprint maps for the cluster catalogs used in this work. The HEALPix resolution is ${\rm NSIDE}=16~\hat{=}~13.4$\,deg$^2$. Black regions overlap with the LS footprint and gray regions are outside of it. White holes within the footprint indicate that the local cluster number density in a catalog is less than 1 per heal pixel.
    \label{fig:footprints}}
\end{figure*}

The input catalogs used are listed in Table \ref{tab:scancat}. These include clusters that are selected using their X-ray signal: CODEX \citep{Finoguenov2020aa}, MCXC \citep{Piffaretti2011aa}, NORAS \citep{Bohringer2017aa}, XXL \citep{Adami2018aa}, XCLASS \citep{Koulouridis2021}, using the Sunyaev Zeldovich (SZ) effect: ACT \citep{Hilton2020aa}, SPT \citep{Bleem2015aa}, \planck \citep{Planck-Collaboration2016aa}, or via optical overdensities of galaxies: ACO \citep{Abell1989}, DES \citep{Abbott2020aa}. The number of clusters contained in these catalogs against redshift is shown in Figure \ref{fig:zscanhisto}. Approximate footprints are shown in Figure \ref{fig:footprints}. Black heal pixels, each covering an area of 13.4\,deg$^2$, are within the footprint of the LS, and gray heal pixels are outside of them. Patchy maps indicate a low number density of clusters.

\begin{table*}
  \caption{Overview of the catalogs analyzed with \eromapper in this work.}
  \centering
  \small
  \begin{tabular}{ll|rrrrr}
  \hline\hline
     Survey & Source & Cataloged & In LS     & Survey Area \\
            &        & Clusters  & Footprint & [deg$^2$] \\
     \hline
     \textit{X-ray selected} & & & & & \\
     ~~~~\erass & \cite{Bulbul2023} \& this work & 12\,247 & 12\,247 & 13\,116 \\
     ~~~~CODEX & \cite{Finoguenov2020aa} & 10\,382 & 10\,079 & 10\,269 \\
     ~~~~MCXC & \cite{Piffaretti2011aa}  & 1743 & 1477 & $\sim$all sky \\
     ~~~~NORAS & \cite{Bohringer2017aa} & 378  & 344  & 13\,519 \\
     ~~~~XXL & \cite{Adami2018aa}   & 302 & 299 & 50 \\
     ~~~~XCLASS & \cite{Koulouridis2021} & 1559 & 1433 & 269 \\
     \hline
     \textit{SZ selected} & & & & & \\
     ~~~~ACT & \cite{Hilton2020aa}   & 4195 & 4044 & 13\,211 \\
     ~~~~SPT & \cite{Bleem2015aa}   & 677  & 656  & 2500 \\
     ~~~~PSZ2 & \cite{Planck-Collaboration2016aa} & 1653 & 1207 & $\sim$all sky \\
     \hline
     \textit{optically selected} & & & & & \\
     ~~~~LS DR10 south $grz$    & this work & 112\,609 & 112\,609 & 19\,342 \\
     ~~~~LS DR10 south $grzw1$  & this work & 273\,150 & 273\,150 & 19\,342 \\
     ~~~~LS DR10 south $griz$   & this work &  91\,790 &  91\,790 & 15\,326 \\
     ~~~~LS DR10 south $grizw1$ & this work & 242\,044 & 242\,044 & 15\,326 \\
     ~~~~LS DR9  north $grz$    & this work &  27\,516 &  27\,516 & 5068 \\
     ~~~~LS DR9  north $grzw1$  & this work &  93\,227 &  93\,227 & 5068 \\
     ~~~~ACO & \cite{Abell1989}   & 5250 & 4635 & $\sim$all sky \\
     ~~~~DESY1 & \cite{Abbott2020aa} & 6729 & 6724 & 1437 \\
     \hline\hline
  \end{tabular}
  \tablefoot{The survey area was calculated from the high-resolution HEALPix maps for \erass and the optically selected LS catalogs or taken from the catalog paper in the source column. Three catalogs have survey areas that cover approximately the full sky but (partly) exclude the Galactic plane, as shown in Figure \ref{fig:footprints}.
  A richness cut of $\lambda_{\rm norm}>16$ and a redshift constraint $z_\lambda<z_{\rm vlim}$ were applied to the LS DR9 north and LS DR10 south catalogs. Although the number of detections is huge, the contamination fraction in these catalogs is not well known.
  \label{tab:scancat}}
\end{table*}

\subsection{Comparison to other X-ray-selected catalogs}

We show the number of \erass clusters against redshift $z_{\rm best}$ as the blue line in Figure \ref{fig:zscanhisto}, left panel. The median redshift of the \erass sample is $z_{\rm best}=0.31$, and the highest is $z_{\rm best}=1.32$ (1eRASS J020547.4-582902). The line width increases toward higher redshift. Its upper border corresponds to the full sample of 12\,247 clusters. The lower border corresponds to the subsample of 10\,959 clusters with $z_\lambda<z_{\rm vlim}$; that is, the photometric redshift must be below the local limiting redshift of the LS (see Appendix \ref{sec:zvlim}). 
We explore the increasing contamination associated with exceeding the limiting redshift in Section \ref{sec:blind}.

The \erass cluster catalog contains the largest number of identified clusters of all considered literature catalogs (see Table \ref{tab:scancat}). 
Compared to other X-ray-selected catalogs, the advantage is the larger survey area than \xmm (green lines in Figure \ref{fig:zscanhisto}, see also Figure \ref{fig:footprints} and Table \ref{tab:scancat}) and the better sensitivity and smaller PSF than ROSAT (orange lines in Figure \ref{fig:zscanhisto}).

For CODEX, the source detection on RASS data had been improved, reaching a similar depth to \erass. The contamination in the cluster candidate sample is high but the published catalog had already been cleaned by identifying the candidates using \redmapper and SDSS data. A great advantage of \erosita over \rosat is that the sharper PSF is better suited to filter sources at high redshift by their extent, leading to purer cluster samples. Moreover, the CODEX sample was limited in redshift to $z\lesssim0.6$ by the usage of shallower SDSS imaging data.

\subsection{High-$z$ optical completeness} \label{sec:highzcompleteness}

We estimated the high-$z$ completeness of the \erass identified catalog using SZ-selected catalogs. The SZ signal is independent of redshift\footnote{While the SZ signal is theoretically independent of redshift, in practice, the detection efficiency, particularly for observations made by the \planck satellite, was affected by the angular resolution of the instrument's beam. This can result in a slightly higher detectability of the SZ effect for galaxy clusters at lower redshifts.}, which has been exploited to produce highly complete catalogs with 90\% completeness for $M_{500\rm{c}} > 3.8 \times 10^{14}$\,M$_\odot$ at \citep[ACT,][]{Hilton2020aa} or nearly 100\% completeness for $M_{500\rm{c}} > 7 \times 10^{14} h_{70}^{-1}$\,M$_\odot$ at $z>0.25$ \citep[SPT,][]{Bleem2015aa}.
Figure \ref{fig:zscanhisto}, middle panel, confirms that the number of clusters remains large at high literature redshift in the ACT, SPT, and \planck catalogs. 
Hence, they are well suited for testing the optical completeness of high-$z$ clusters in the \erass catalog.
For this, we ran \eromapper in scan mode on the locations of the literature clusters.
We define the optical completeness $C$ as the number of \eromapper-identified clusters $N_{\rm lit\&ero}$ to the total number of clusters $N_{\rm lit}$ in the SZ-selected literature catalogs

\begin{equation}
    C(z) = \frac{N_{\rm lit\&ero}(z)}{N_{\rm lit}(z)}.
\end{equation}

\begin{figure*}
    \centering
    \includegraphics[width=0.49\linewidth]{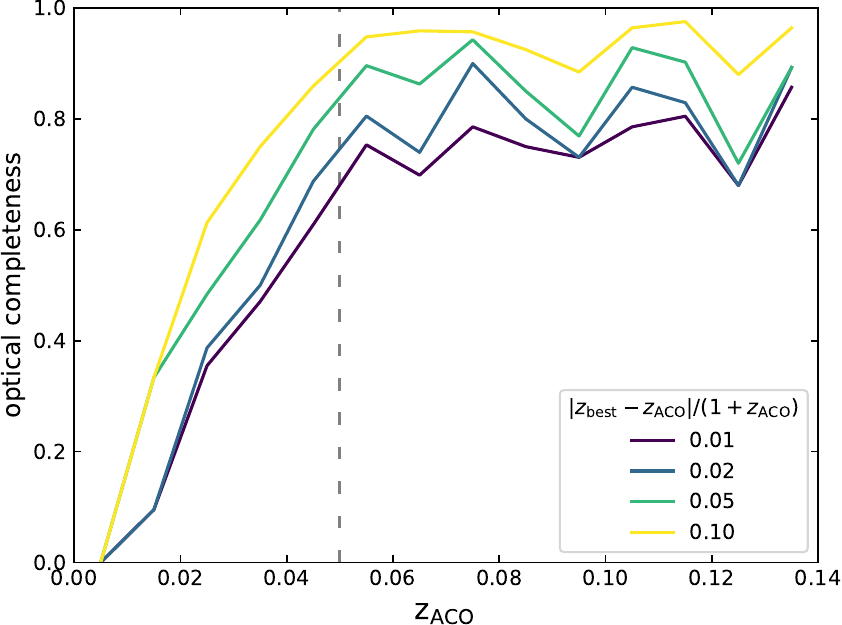}
    \includegraphics[width=0.49\linewidth]{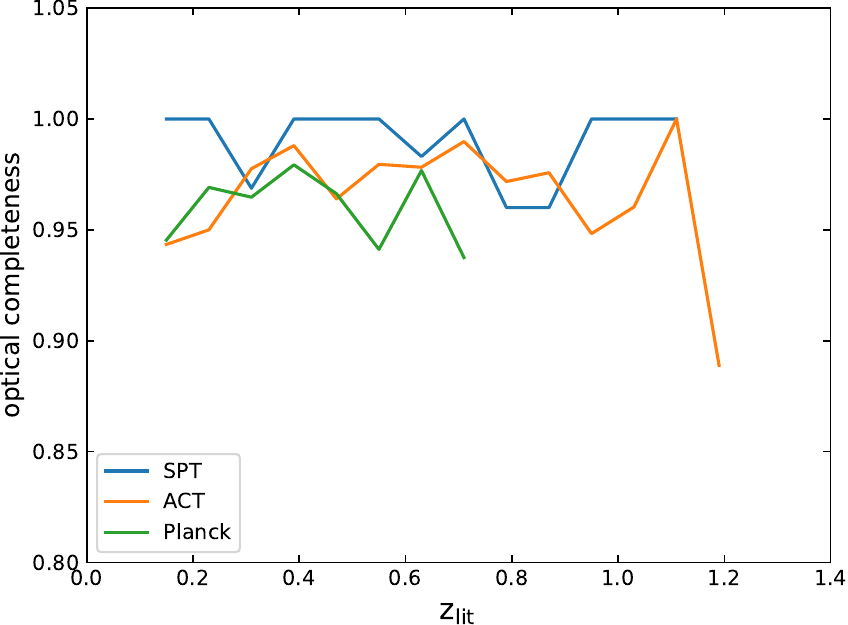}
    \caption{Optical completeness defined as the ratio of clusters confirmed after running \eromapper in scan mode on the cluster coordinates in various catalogs from the literature. At low redshifts, we used the ACO catalog and quantified the completeness for four different redshift tolerances. At high redshifts, we used the SPT, ACT, and \planck catalogs and considered clusters as confirmed when the redshifts agreed within $|z_{\rm best}-z_{\rm lit}|/(1+z_{\rm lit})<0.1$. We restricted the samples to clusters where the LS are sufficiently deep, that is, $z_{\rm best}<z_{\rm vlim}$. A minimum of five clusters was required per redshift bin.
    }
    \label{fig:completeness}
\end{figure*}

Thereby, we only considered clusters that lie within the LS footprint. Furthermore, the literature redshift must be lower than the limiting redshift of the LS $z_{\rm lit} < z_{\rm vlim}$. To avoid mismatches in the cases of overlapping clusters, we required $|z_{\rm best}-z_{\rm lit}|<0.1$ for a cluster to be considered a match. Figure \ref{fig:completeness}, right panel, confirms a high optical completeness $C>0.95$ for $z_\lambda<1.1$.
We emphasize that the optical completeness does not take the completeness of the X-ray source catalog into account (see Section \ref{sec:erass}). It is solely sensitive to the processing of the optical and near-infrared data.

The median richness is $\lambda_{\rm norm}=50$ for the ACT and SPT catalogs. For \planck, it is $\lambda_{\rm norm}=70$, which explains the lower number of clusters even though it is an all-sky survey. In the \erass catalog, the median richness is lower with $\lambda_{\rm norm}=25$ because \erosita is more sensitive to galaxy groups, especially at low redshift \citep{Bahar2024}.

\subsection{Low-$z$ optical completeness} \label{sec:lowzcompleteness}

We estimated the low-$z$ completeness of the \erass identified catalog with the help of the optically selected ACO catalog \citep{Abell1989}. The ACO catalog had been compiled by visually inspecting photographic plates. It is more sensitive to lower redshifts.
For the following investigation, we selected 577 ACO clusters with a cataloged redshift $z_{\rm ACO}<0.14$ that lie within the LS footprint. Additionally, we included 16 ACO clusters that are formally outside the footprint. However, they were only classified as such because the clean region around the BCG was masked due to the assigned MASKBITS (see Sections \ref{sec:catprocessing} and \ref{sec:redmapper}).

Figure \ref{fig:completeness}, left panel, shows the ratio of ACO clusters confirmed after running \eromapper in scan mode on the ACO coordinates over total ACO clusters. The redshift $z_{\rm ACO}$ was taken from the ACO catalog. The yellow curve demonstrates a high optical completeness of $C\approx95\%$ beyond $z_{\rm ACO}>0.05$.
Below $z_{\rm ACO}<0.05$, the optical completeness drops steeply: of 185 ACO clusters, we confirmed 132 ($C=71\%$).
The optical limitation stemmed from the LS photometry around bright extended galaxies. While we kept galaxies cataloged in the Siena Galaxy Atlas, we excluded those that overlap with them. These optical detections are often spurious. The consequence of the removal is that a background cluster was frequently preferred over a low-$z$ cluster.

\begin{figure}
    \centering
    \includegraphics[width=\linewidth]{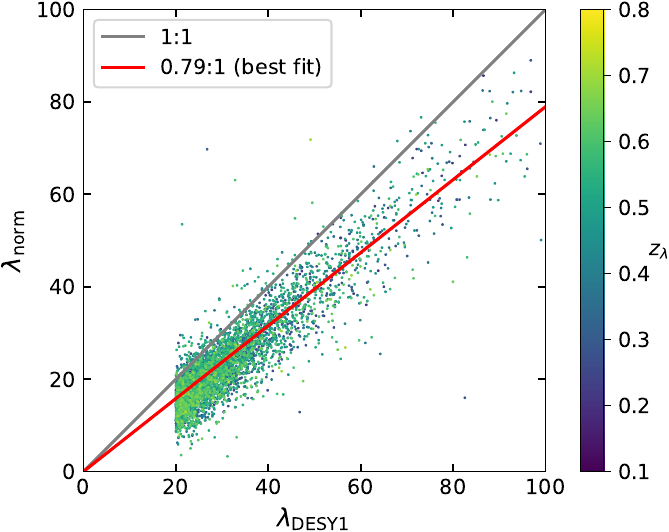}
    \caption{Comparison between the richnesses $\lambda$ measured using \eromapper and the published richnesses for the same DES year 1 clusters color-coded by their photometric redshift in the \eromapper catalog. 
    The 1:1 relation is shown in gray and the red line is the best-fit linear relation with a slope of $\alpha=0.79$.}
    \label{fig:richness_des_ero}
\end{figure}

\subsection{Richness comparison} \label{sec:desrichnesses}

The DES year 1 catalog was created using the cluster finder \redmapper, similar to this work \citep{Abbott2020aa}. This allowed a direct comparison of the measured richnesses. The catalog was obtained in \redmapper blind mode, that is, without positional priors. The redshift range was limited to $0.2<z_\lambda<0.65$. 
Figure \ref{fig:richness_des_ero} compares the richnesses computed in this work $\lambda_{\rm norm}$ against the published richnesses $\lambda_{\rm DESY1}$ \citep{Abbott2020aa}. We used 4716 clusters with good redshift agreement $\Delta z/(1+z_\lambda)<0.01$, a low masking fraction (<10\%), and which were identified in the LS DR10 south using the $grz$ filter band combination, for which $\lambda=\lambda_{\rm norm}$ (see Section \ref{sec:richness} and Table \ref{tab:lambdanorm}).

To compute the systematic scaling factor, we fitted a linear relation to the data points by minimizing the orthogonal residuals. To avoid biases due to the sharp richness cut at $\lambda_{\rm DESY1}=20$, we made an additional orthogonal cut below the line defined by $\lambda_{\rm norm} < -1 / \alpha \cdot \lambda_{\rm DESY1} + 50$, where $\alpha=0.79$ is the best-fit slope. This procedure rejected 365 additional data points. We obtained the best-fit relation;
\begin{equation}
    \lambda_{\rm norm} = 0.79 \cdot \lambda_{\rm DESY1}.
\end{equation}
The richnesses measured in this work are, on average, 21\% lower for identical DES year 1 clusters. A possible explanation is the different handling of galaxy outliers with respect to the red-sequence model in earlier versions of \redmapper.
We did not correct for this effect.
Moreover, the data points in Figure \ref{fig:richness_des_ero} were color-coded by the cluster redshift $z_\lambda$. We did not see any obvious dependency of the scaling factor on redshift.

\subsection{BCG selection} \label{sec:bcg_analysis}

A BCG is, by definition, the brightest member galaxy. Ambiguity arises from using BCGs as proxies for the cluster center and the association with the intracluster light that surrounds BCGs. The three commonly used selection criteria (being the brightest, most extended, and most central member) resemble the historical definition of a cD galaxy \citep{Matthews1964,Morgan1965,Morgan1975,Albert1977}. These qualities can conflict in up to 20\% of the cases, especially when shallow photometry underestimates the faint light in the outskirts. \citep{vonderLinden2007,Kluge2020aa}. Moreover, merging or unrelaxed clusters can have a significantly miscentered BCG or two similarly bright BCGs \citep[in 7\% of the cases in the total sample from][]{Kluge2020aa}.
Therefore, we explore three options for selecting the BCG:
\begin{enumerate}
    \item located at the X-ray center,
    \item located at the optical cluster center,
    \item being the brightest member.
\end{enumerate}

Figure \ref{fig:zspec_example} shows an illustrative example. 
The zoomed image in the bottom-left panel is centered on the galaxy, which visually fulfills best all commonly used selection criteria for BCGs. That galaxy is one of the most luminous and extended cluster members, and it is well-centered within the X-ray contours (white).
The location of the initial X-ray detection in the catalog (before refined modeling, \citealt{Bulbul2023}) is shown by the yellow cross. The nearest galaxy is marked by an orange circle. Due to its compactness, it is clearly not a good choice for the BCG. The BCG chosen by \eromapper (the brightest member galaxy) is marked by the red circle. In this example and in 70\% of the eRASS1 clusters, it coincides with the optical cluster center (green cross). However, selecting it as the BCG is inconsistent with our visual preference in this example.

The optical center is often, but not always well-centered on the X-ray peak. \cite{Seppi2023} found in 31\% of their analyzed clusters good agreement with only very small offsets of $\Delta_{\rm{X-O}} < 0.05\times R_{\rm{500c}}$, where $R_{\rm 500c}$ is the radius enclosing 500 times the critical density of the Universe. Visual inspection suggested no general preference for either choice. Either the optical center or the brightest member galaxy agreed with our visual preference in $\sim$85\% of the cases. Hence, we decided to follow the simplest approach and define the BCG in this work as the brightest member galaxy within $R_\lambda$ (see Equation (\ref{eq:clusterradius})).

To further evaluate the automatic BCG choice at low redshift $z<0.1$, we ran \eromapper on the cluster catalog of \cite{Kluge2020aa}. This catalog is mostly a subset of the ACO catalog with each cluster carefully centered on the BCG. The BCG was chosen as the largest and most central galaxy based on visual inspection of deep images that revealed visual features on levels 27--28 $g'$ mag arcsec$^{-2}$.
Of the 154 cataloged clusters that are within the LS footprint, we identified 130 cases with consistent redshift ($|z_{\rm best}-z_{\rm lit}|<0.01$). The lost fraction is attributed to the growing incompleteness at low redshifts (see Section \ref{sec:lowzcompleteness}). The median redshift of the sample from \cite{Kluge2020aa} is $z=0.06$.
Of the 130 clusters, our automatic BCG selection is in 85 cases consistent with the choice made by Kluge et al. 
In 23 cases, the BCG is not a member galaxy according to \eromapper, even though its redshift is consistent with the cluster redshift. These BCGs are typically marginal outliers in color space.
By only considering the remaining 107 clusters, the agreement with the BCG choice by \cite{Kluge2020aa} is 80\%, precisely what we expected (see beginning of this section). 
Larger color uncertainties at intermediate or higher redshift likely keep the BCG colors consistent with the red-sequence colors.
Therefore, we expect for the full \erass and reanalyzed literature samples that the selected BCG is a good choice in $\sim$80\% of the clusters.

\section{Number densities and mass limits} \label{sec:numberdensities}

The \erass cluster catalog's lower mass limit increases with redshift due to a combination of the lower total flux received and the effect of cosmic dimming of the X-ray surface brightness in more distant clusters. Only the most luminous clusters can be detected at high redshifts. This behavior is quantified in our X-ray selection function \citep{Clerc2023}.

An additional inhomogeneous optical selection function can alter the cluster catalog further. We examined its effect using the catalogs created with \eromapper in blind mode. This mode does not use positional priors. Hence, the obtained catalogs allow us to investigate purely optical selection effects. Our analysis utilized the full LS DR9 north and LS DR10 south surveys, making the blind-mode cluster catalog the most extensive to date (see Figure \ref{fig:zscanhisto} and Table \ref{tab:scancat}). The crucial disadvantage is, however, that the selection function is not well known. This knowledge is imperative to perform precise cosmological studies, for which \erosita was designed.

Number densities are computed by dividing the number of clusters per redshift bin by the survey area or comoving volume. The usable survey area depends on redshift because of the non-uniform depth of the LS. Therefore, we used the dashed curves in Figure \ref{fig:surveyarea} to calculate the redshift-dependent survey area and comoving volume for the \eromapper blind-mode catalogs.

\begin{figure*}
    \includegraphics[width=0.49\linewidth]{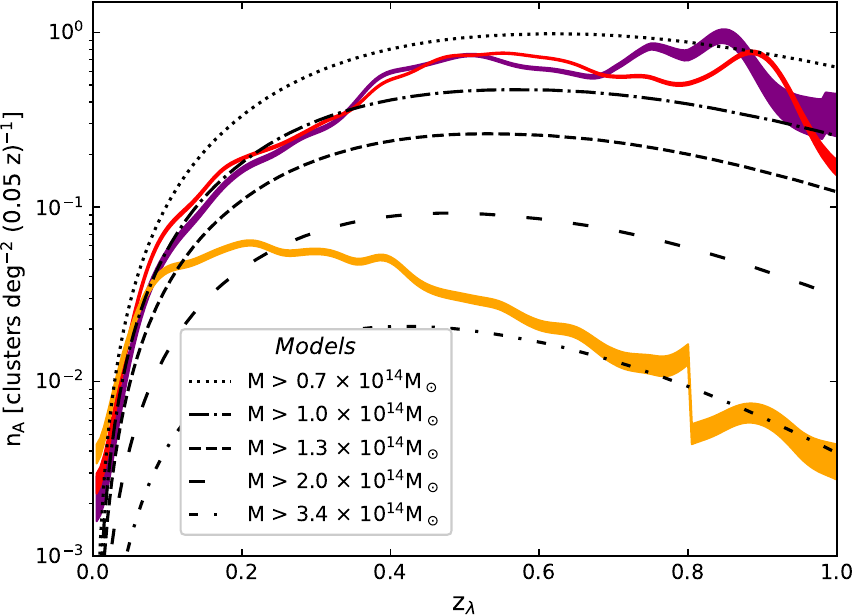}
    \includegraphics[width=0.49\linewidth]{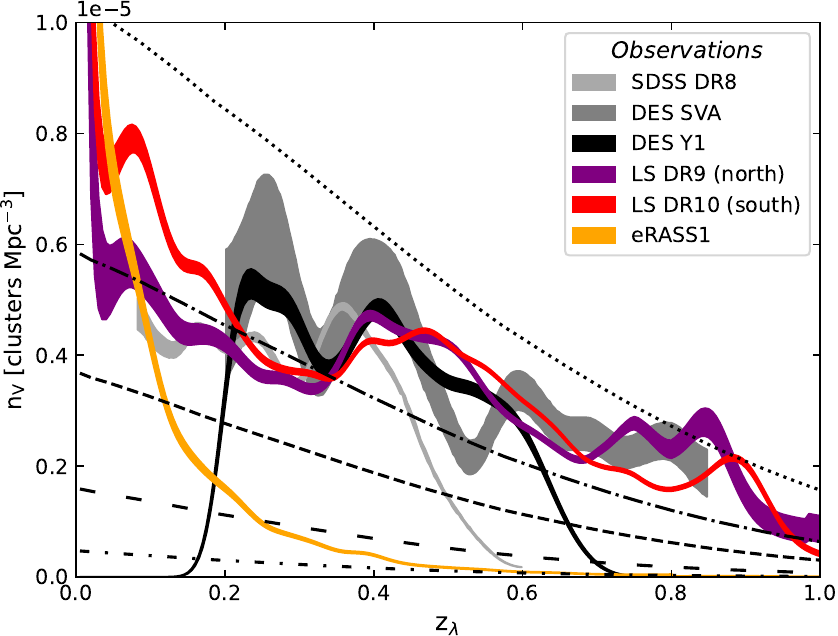}
    \caption{Cluster number density per sky area (left) and comoving volume (right) against redshift. Red and purple lines refer to clusters detected using the \eromapper blind mode on the LS DR10 south and LS DR9 north, respectively, in the $grz$ filter band combination. We restricted the samples to a reliable redshift range $z_\lambda<z_{\rm vlim}$. The orange line refers to the \erass clusters. A discontinuity at $z_\lambda=0.8$ arises from the merging of the low-$z$-optimized and high-$z$-optimized \eromapper runs (see Section \ref{sec:merging}).    
    A richness cut of $\lambda_{\rm norm}>16$ was applied only to the LS DR9 north and LS DR10 south catalogs. Volume densities for the SDSS DR8 and DES SVA catalogs were adopted from \cite{Rykoff2016aa} and for DES year 1, calculated assuming a survey area of 1437 deg$^2$ with uniform depth (see Table \ref{tab:scancat}). The curves were normalized for a bin size of $\Delta z=0.05$ and smoothed using a Gaussian kernel with a standard deviation of $\Delta z=0.02$. The width of the lines corresponds to the Poissonian error for the same bin size, consistently as in \cite{Rykoff2016aa}.
    \label{fig:cludens}}
\end{figure*}

\subsection{Construction of blind-mode catalogs} \label{sec:blind_construction}

As described in Section \ref{sec:merging}, we ran \eromapper in blind mode on six independent filter band and survey combinations. We did not merge the resulting catalogs because the clusters were not detected at the exact same locations in the different runs.
Table \ref{tab:scancat} lists the total number of detected clusters for each run. We restricted the blind-mode samples to a reliable redshift range $z_\lambda<z_{\rm vlim}$. Moreover, we removed galaxy groups to increase the purity. Therefore, we were restricting the richness to $\lambda_{\rm norm}>16$, which corresponds to a halo mass restriction of $M_{\rm 200m}\gtrsim10^{14}$\,M$_\odot$ according to the best-fit scaling relation for DES year 1 data \citep[][see also Section \ref{sec:richness}]{McClintock2019aa}.

The resulting blind-mode catalog for the $grz$ filter bands contains 140\,125 sources, 112\,609 + 27\,516 from the LS DR10 south and LS DR9 north, respectively. Another blind-mode catalog optimized for high redshifts using the $grzw1$ filter bands contains 366\,377 sources (273\,150 + 93\,227). More details are given in Table \ref{tab:scancat}.

\subsection{Resulting catalogs}

The number of blind-mode clusters against redshift is shown by the blue lines in Figure \ref{fig:zscanhisto}, right panel. The continuous (dashed) line corresponds to the $grz$ ($grzw1$) catalog for the combined LS DR9 and LS DR10 survey areas. 
For the $grzw1$ runs, we only considered brighter galaxies $L>0.4$\,L$_*$ (compared to $L>0.2$\,L$_*$ for the $grz$ runs), which increases the usable survey volume (see Appendix \ref{sec:zvlim}).
Below $z_{\rm best}<0.4$, the $grz$ and $grzw1$ runs have the same effective survey area (see Figure \ref{fig:surveyarea}). 
If there was no contamination and perfect completeness, the continuous and dashed lines should agree.
The fact that they disagree indicates a higher level of contamination in the $grzw1$ run compared to the $grz$ run. 
This is expected because the higher galaxy luminosity threshold allows for less secure cluster detections.
At high redshift ($z_{\rm best}>0.6$), the lines begin to deviate much more from each other. The reason is the larger effective survey area for the $grzw1$ runs (see Figure \ref{fig:surveyarea}).
Number densities per area ($n_{\rm A}$, left panel) and volume ($n_{\rm V}$, right panel) are shown in Figure \ref{fig:cludens}. The red (purple) curve corresponds to the LS DR10 south (DR9 north) $grz$ catalog. We find that the number density profiles for the clusters detected in the LS DR9 north and LS DR10 south are consistent within the amplitude of the systematic oscillations. The oscillations are larger than the statistical uncertainties and arise from the redshift-dependent redshift bias. We perform a detailed analysis in Section \ref{sec:redshift_bias}.

\subsection{Comparison to literature}

Table \ref{tab:scancat} lists the total number of clusters for various catalogs used in this work. The sample size of the \eromapper blind-mode catalogs surpasses the DES year 1 sample size by a factor of 10 in the overlapping redshift range $0.2<z_{\rm best}<0.65$ (see Figure \ref{fig:zscanhisto}). The main reason for this improvement is the $\sim$10 times larger survey area of the LS compared to the DES year 1 area as listed in Table \ref{tab:scancat} and shown in the last two panels in Figure \ref{fig:footprints}.

We compare our obtained volume number densities with results for the DES SVA (dark gray, $z_\lambda>0.2$) and SDSS DR8 (light gray, $0.08<z_\lambda<0.33$) by adopting the volume number densities from \cite{Rykoff2016aa}. Additionally, we calculated the number densities for the public DES year 1 cluster catalog (black, $0.2<z_\lambda<0.65$) assuming a redshift-independent survey area of 1437\,deg$^2$ (see Table \ref{tab:scancat}). All curves are consistent with each other within the amplitude of the systematic oscillations.

\subsection{Comparison to theoretical halo masses}

For the LS cluster catalogs obtained in \eromapper blind mode, we applied a richness cut. As richness correlates with halo mass, we compared the number densities with halo model predictions. We computed theoretical number densities by integrating the halo mass function \citep{Tinker2008aa,Diemer2018ab} above five different mass limits. 

The results are overlaid on Figure \ref{fig:cludens} with black lines.
The mass limit $M_{500\rm{c}}\gtrsim10^{14}$\,M$_\odot$ (densely dash-dotted line) corresponds to our richness cut of $\lambda_{\rm norm}>16$ that is applied to the blind-mode catalogs (red and purple lines). The curves agree with the amplitude of the systematic oscillations.
We conclude from the comparison that the impact of the optical selection function is small above $M_{500\rm{c}}\gtrsim10^{14}$\,M$_\odot$ because the observed number densities are consistent with the prediction.

We note that the predicted and observed number densities depend on cosmology. Applying the \erass cosmology \citep[$H_0=67.77$ km s$^{-1}$ Mpc$^{-1}$, $\Omega_{\rm m}=0.29$, and $\sigma_8=0.88$,][]{Ghirardini2023} shifts the predicted curves slightly upward.

\subsection{Conclusions for \erass}

We overplotted the number densities for the \erass clusters in Figure \ref{fig:cludens} in orange. They agree with the blind-mode sample below $z_\lambda\lesssim0.1$ and the predictions for a mass limit $M_{500\rm{c}}\lesssim10^{14}$\,M$_\odot$ at $z_\lambda<0.1$.
Beyond that, the number densities of the \erass sample are lower, as expected. The sensitivity of \erosita decreases for groups at higher redshift because of their low X-ray flux. 
The \erass mass limit corresponds to $M_{500\rm{c}}\gtrsim2\times10^{14}$\,M$_\odot$ at $z_\lambda=0.2$ and $M_{500\rm{c}}\gtrsim3\times10^{14}$\,M$_\odot$ at $z_\lambda=0.8$.

The intercepts between the \erass number densities and the model predictions roughly agree with the redshift-dependent mass limits of the \erass sample, obtained from the scaling relation between X-ray count rate and weak lensing shear \citep{Ghirardini2023}. The detection likelihood decreases toward higher redshifts and lower masses. This behavior is accounted for in the X-ray-selection function \citep{Clerc2023}.

At $z_\lambda=0.8$, we notice a discontinuity in the \erass cluster number surface density. It arises from merging the catalogs obtained for different filter band combinations (see Section \ref{sec:merging}).

\begin{figure*}
    \centering
    \includegraphics[width=\linewidth]{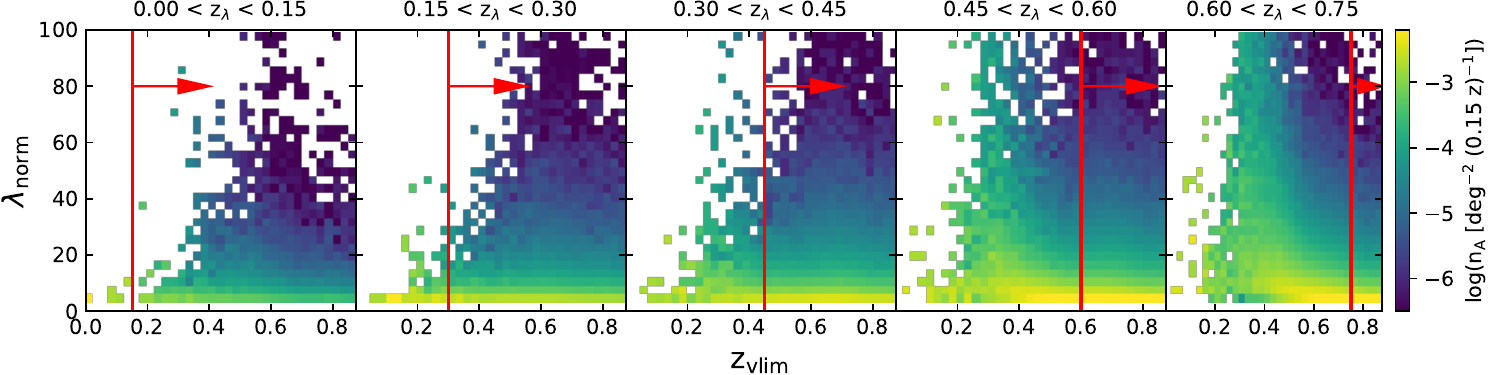}
    \caption{Dependency of the cluster number density on survey depth. The cluster number density per sky area $n_{\rm A}$ was calculated for the clusters detected using the \eromapper blind mode in the LS DR10 south using the $grz$ filter bands. The data was split into five bins in photometric cluster redshift $z_\lambda$. For each redshift bin, the densities were calculated for bins in limiting redshift $z_{\rm vlim}$ and richness $\lambda_{\rm norm}$. We consider the cluster properties reliable when local survey depth is sufficiently deep, that is, $z_\lambda<z_{\rm vlim}$ (see Appendix \ref{sec:zvlim} for details). Those clusters are located to the right of the red line where the number density does not depend on $z_{\rm vlim}$. The number density left of the red line increases, which indicates a growing number of false-positive detections. \label{fig:richness_cludens}}
\end{figure*}

\subsection{Dependency on LS survey depth -- richness bias} \label{sec:blind}

We utilized the blind-mode catalog obtained in the LS DR10 south ($grz$) further to demonstrate that the survey depth has no significant effect on the cluster number densities as long as only clusters with $z_\lambda<z_{\rm vlim}$ are selected, where $z_{\rm vlim}$ is the spatially dependent limiting redshift of the LS. Figure \ref{fig:richness_cludens} shows the blind-mode cluster numbers per sky area binned in intervals of limiting redshift $z_{\rm vlim}$ and richness $\lambda_{\rm norm}$ for five cluster redshift bins. The measured cluster number density increases where the LS is too shallow for the considered redshift (see Appendix \ref{sec:zvlim}). This effect must arise from measurement errors because the intrinsic cluster number density should not directly depend on the survey properties. A plausible explanation is an increasing number of false-positive member galaxies in shallower survey areas. This can occur due to the negative low-mass slope of the galaxy luminosity function \citep[e.g.,][]{Blanton2005}, which can lead to an increasing Eddington bias. Moreover, higher uncertainties in color can lead to an increased misidentification of interloping galaxies, or even stars, as cluster members. This can statistically inflate cluster richnesses, consequently increasing the observed cluster number densities. The affected clusters can be avoided by selecting only systems with photometric redshift lower than the local limiting redshift $z_\lambda<z_{\rm vlim}$. The red line in Figure \ref{fig:richness_cludens} marks the required minimum depth for each cluster redshift bin. The red arrow indicates the direction to where the reliable clusters are located.

\section{Purity of the \erass cluster catalog} \label{sec:pcont}

The \erass catalog was optimized for high completeness by selecting cluster candidates above a low extent likelihood threshold $\mathcal{L}_{\rm ext}>3$. The trade-off is high contamination. Using this selection, we expect from simulations \citep{Seppi2022} that only $\sim$50\% of the extended sources are genuine galaxy clusters above the approximate \erass flux limit of $F_{500}\approx4\times10^{-14}$ ergs s$^{-1}$ cm$^{-2}$ \citep{Bulbul2023}. Most other sources are misclassified point sources, such as AGNs (and stars). A smaller contribution comes from random background fluctuations \citep{Seppi2022}.

During the optical identification of these contaminants, it can either happen that no overdensity of red-sequence galaxies is found and the cluster candidate is discarded (2497 \erass extended sources in the LS footprint, see Section \ref{subsec:CTPs})\footnote{There is also the possibility that the X-ray signal originates from a real cluster whose redshift is much higher than the local limiting redshift of the LS.}, or the contaminant can be associated with an overdensity of red galaxies.
In the latter case, these are either (a) line-of-sight projections of galaxies that are not bound to one cluster or (b) real overdensities of red galaxies around AGN as galaxies tend to cluster. We considered these sources as contaminants because the X-ray signal is not associated with the ICM.

To quantify the remaining contamination in the \erass cluster catalog after our optical identification procedure, we introduce a mixture model in \cite{Ghirardini2023} \citep[see also][]{Bulbul2023}. It compares the properties of contaminants to those of clusters and attributes a probability $0\leq P_{\rm cont}\leq 1$, quantifying how likely each individual cluster is to be a contaminant. A detailed mathematical description of this mixture model can be found in \cite{Ghirardini2023}. In this work, we provide details on the optical properties of the contaminating sources and how they were used to estimate the contamination probability of the \erass clusters.

\begin{figure*}
    \includegraphics[width=\linewidth]{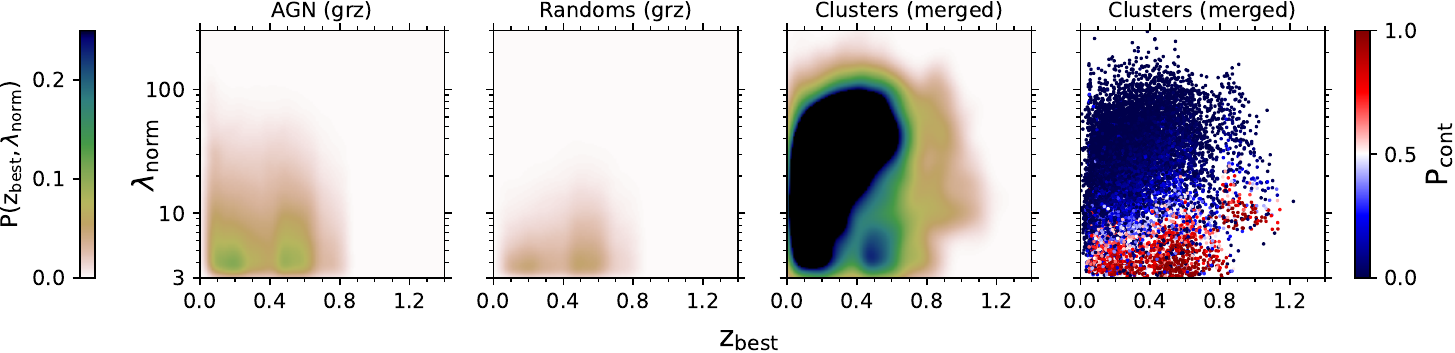}
    \caption{Kernel density estimates of AGNs, random points, and clusters in redshift ($z_{\rm best}$)--richness ($\lambda_{\rm norm}$) space (first three panels). The right panel shows the individual cluster data points color-coded by their probability of being a contaminant.}
    \label{fig:mixture_model}
\end{figure*}

To estimate the contribution of both types of contaminants, we ran \eromapper on 1\,250\,795 \erass point sources \citep{Merloni2023} and 1\,000\,000 random positions in the sky. Overdensities of red-sequence galaxies were identified for $\sim$850\,000 and $\sim$490\,000 sources, respectively. The exact numbers depend on the filter band combination.
Random points were positioned at least five cluster radii $R_\lambda$ away from extended \erass sources.
We analyzed the distribution of the randoms and \erass point sources in redshift and richness space (see Figure \ref{fig:mixture_model}, first two panels). The majority of the contaminants lie at low richnesses, as expected. 
We identified two peaks around $z_{\rm best}\approx0.2$ and $z_{\rm best}\approx0.5$. They are similarly pronounced for the point sources, while for the randoms, the second peak is slightly more dominant. The second peak is also apparent in the cluster sample (third panel). It indicates that there is indeed contamination left in the \erass cluster catalog after the optical cleaning was performed in this work.

\begin{figure*}
    \includegraphics[width=0.47\linewidth]{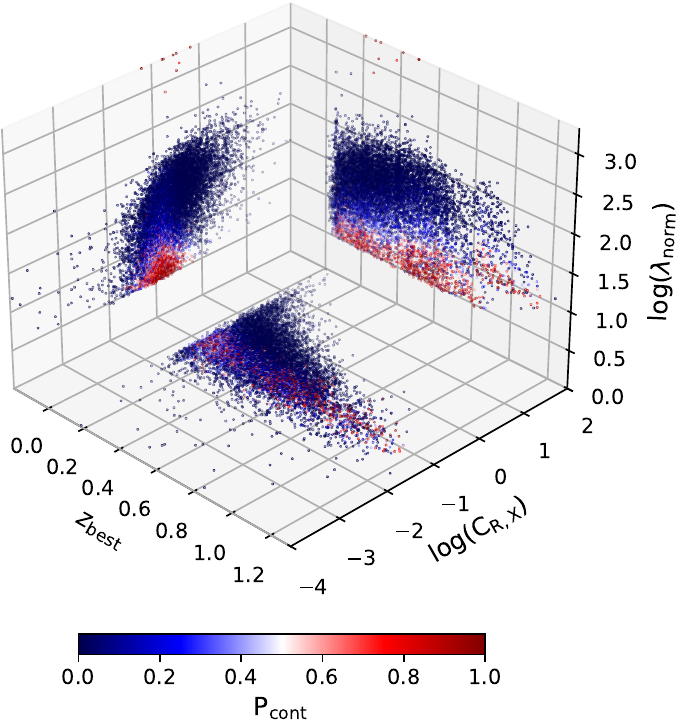}~~~~~~~~~~~~
    \includegraphics[width=0.47\linewidth]{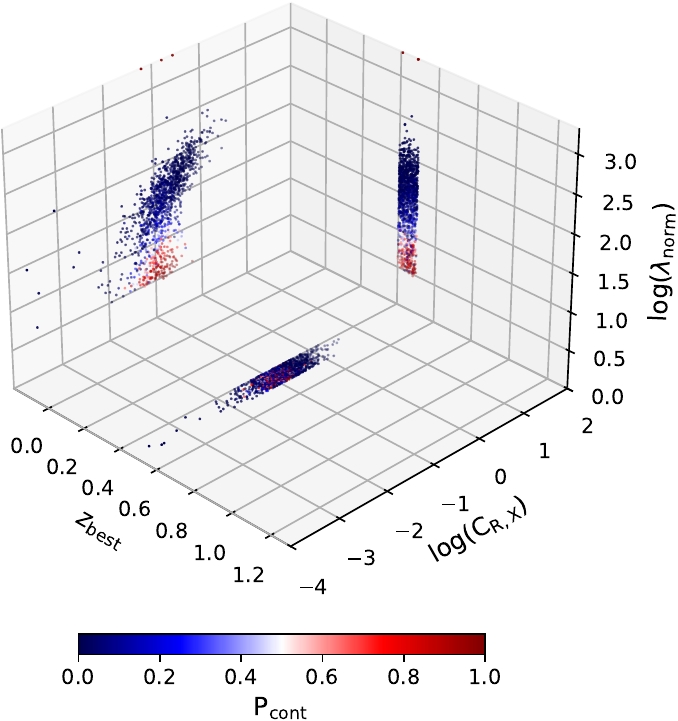}
    \caption{Distribution of clusters in redshift ($z_{\rm best}$)--richness ($\lambda_{\rm norm}$)--X-ray count rate ($C_{\rm R,X}$) space. The left panel shows the full distribution, and the right panel shows a slice in redshift $0.4<z_{\rm best}<0.5$ to enhance the dependence on the $\lambda_{\rm norm}$--$C_{\rm R,X}$ relation. The data points are projected on each side of the canvas. The probability of being a contaminant $P_{\rm cont}$ is color-coded. The top-right projection is equivalent to Figure \ref{fig:mixture_model}, right panel.}
    \label{fig:pcont}
\end{figure*}

The global amplitudes of the kernel density estimates for both types of contaminants were fitted in the mixture model \citep{Ghirardini2023}. For the primary \erass cluster sample with $\mathcal{L}_{\rm ext}>3$, we obtained a three times higher amplitude for the point sources $f_{\rm AGN}=0.0768$ than for the randoms $f_{\rm rand}=0.0246$. This aligns with the expectation that most of the contamination in the \erass cluster candidate catalog arises from X-ray point sources \citep{Seppi2022}.

Even though the distributions for the point sources and random points are similar in $z_{\rm best}$--$\lambda_{\rm norm}$ space, we can disentangle their contributions using the X-ray count rate $C_{\rm R,X}$. 
Figure \ref{fig:pcont} shows three projections of the 3-D distribution of the \erass clusters in $z_{\rm best}$--$\lambda_{\rm norm}$--$C_{\rm R,X}$ space. The contamination probability $P_{\rm cont}$ is color-coded.
AGNs typically have a higher X-ray flux than random background fluctuations. We modeled the $\lambda_{\rm norm}$--$C_{\rm R,X}$ distribution empirically using a redshift-dependent Schechter function \citep{Schechter1976aa}. Thereby, we assumed the power-law component to be the relation for clusters because both properties depend on the halo mass. Given this assumption, the exponential drop-off is due to the contamination. These objects are prominent in the right panel of Figure \ref{fig:pcont} as red points at low richnesses and high X-ray flux. It shows a selected redshift slice between $0.4<z_{\rm best}<0.5$.
By averaging the $P_{\rm cont}$ values of all $N=12\,247$ \erass clusters, we calculated the residual contamination of the \erass catalog to be

\begin{equation}
    \sum_i{P_{\rm cont,i}} / N = 14\%.
\end{equation}

Using this result, we also computed the purity of the \erass cluster candidate catalog (before optical identification). There are 16\,336 cluster candidates in the LS footprint and $N \cdot \sum_i{(1-P_{\rm cont,i})}\approx10\,581$ sources were optically identified after statistically removing the remaining contamination. The ratio of the two numbers gives a purity of 65\%, which is slightly higher than the predicted purity of 50\% \citep{Seppi2022}.

The advantage of our contamination estimator compared to traditional richness and redshift cuts \citep{Klein2021aa,Klein2023}, is that we keep lower-richness groups in our sample \citep{Bahar2024}.
We discuss the limitations of our contamination estimator, in particular for low-redshift groups, in Appendix \ref{sec:limitation_pcont}.

\section{Photometric redshift accuracy} \label{sec:redshift_accuracy}

In this section, we quantify the accuracy of the photometric redshifts derived from \eromapper. We characterize this to first order by bias and uncertainty.
The bias $\Delta z_\lambda$ is the systematic offset from the true redshift $z$ and is the floor reached when averaging many photometric redshifts. It arises primarily from red-sequence calibration errors (see Appendix \ref{sec:redseq}). 
The uncertainty $\delta z_\lambda$ is the statistical 1$\sigma$ error that arises from effects such as galaxy color measurement errors and interloping galaxies (e.g., see Figure \ref{fig:zspec_example}, bottom panel). 
The uncertainty decreases with increasing richness, thanks to ensemble averaging. 

\begin{figure*}
    \centering
    \includegraphics[width=0.455\linewidth]{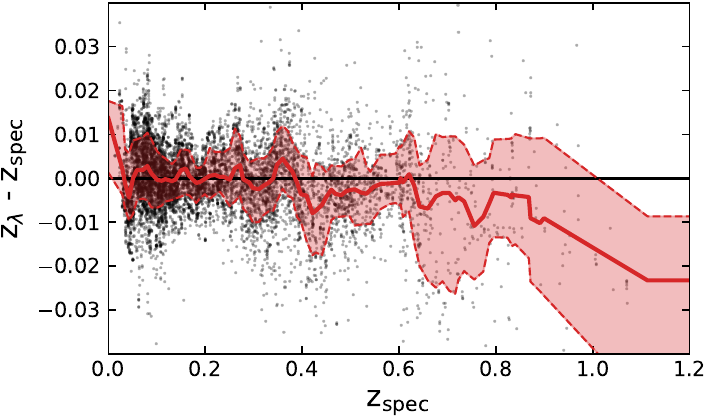}~~~~
    \includegraphics[width=0.535\linewidth]{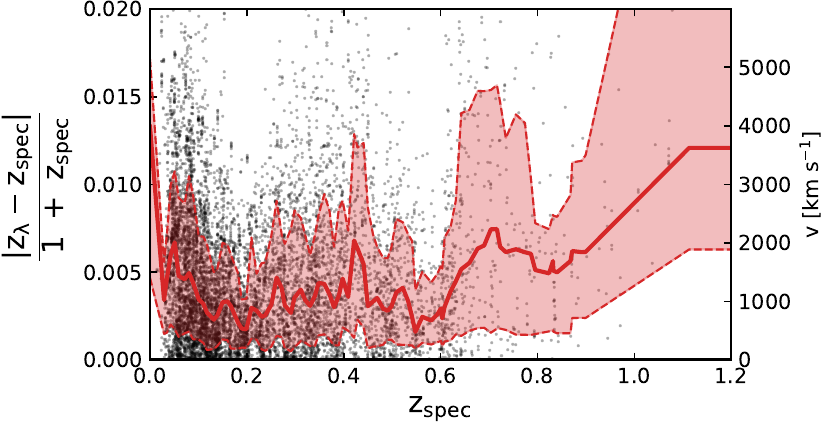}
    \caption{Bias and uncertainty of the photometric cluster redshifts $z_\lambda$ estimated by comparing to high-quality spectroscopic cluster redshifts $z_{\rm spec}$. Black points refer to 9522 clusters with high-quality spectroscopic redshifts (see Section \ref{sec:redshift_accuracy}). The catalogs were merged as described in Section \ref{sec:merging}. The red line is the running median, and the red shaded regions give the uncertainty $\delta z_\lambda$ taken as the 16$^{\rm th}$ and 84$^{\rm th}$ percentiles of the distribution.
    In the left panel, the continuous red line is the redshift bias $\Delta z_\lambda$, and the dashed lines give the redshift uncertainty $\delta z_\lambda$.
    The right panel shows the absolute errors normalized by $1+z_{\rm spec}$, which is proportional to the line-of-sight velocity $v$ (Equation (\ref{eq:proper_v})).
    \label{fig:zspec_zphot}}
\end{figure*}

Both bias and uncertainty depend on the true redshift $z$: $\Delta z_\lambda(z)$ and $\delta z_\lambda(z)$. 
We assumed negligible spectroscopic redshift uncertainties $\delta z_{\rm spec} \ll \delta z_\lambda$ and zero spectroscopic bias $\Delta z_{\rm spec}=0$, that is, we used $z_{\rm spec}$ as a proxy for the true redshift. 
The spectroscopic cluster redshifts have, on average, 1$\sim$7 (5$\sim$15) times lower uncertainties than the photometric cluster redshifts when less than 5 (more than 20) spectroscopic members are known.

We estimated the bias and uncertainty by comparing the photometric redshifts $z_\lambda$ to the high-quality spectroscopic redshifts $z_{\rm spec}$. 
By high quality, we mean a minimum number of spectroscopic members to $N_{\rm members}\geq10$ for $z_{\rm spec}<0.3$, $N_{\rm members}\geq7$ for $0.3<z_{\rm spec}<0.6$, and $N_{\rm members}\geq3$ for $z_{\rm spec}>0.6$
To have a sufficient sample size at high redshift, we ran \eromapper on a large set of test clusters compiled from the works listed in Table \ref{tab:litz_match_clusters} and \cite{Bohringer2000aa,Clerc2012aa,Balogh2014aa,Paterno-Mahler2017aa,Streblyanska2019aa,Huang2020ab,Huang2020aa}. 
The test cluster compilation comprises 151\,781 clusters distributed across the entire LS area. For a subset of 22\,450 clusters, we measured spectroscopic redshifts. 
We obtained 9522 clusters with high-quality spectroscopic redshifts to compare with the photometric redshifts (see black dots in Figure \ref{fig:zspec_zphot}).

We mention as a caveat to these bias and uncertainty estimates that the high-quality spectroscopic sample does not fairly sample the parent cluster sample. 
There could remain subtle biases that we can only uncover with more spectroscopy or narrow-band photometry. This task can be reassessed once the datasets from SDSS-V (10\,000 clusters), 4MOST (40\,000 clusters), DESI \citep{DESI2023}, and \textit{Euclid} \citep{Laureijs2011} become available.

The cosmological analysis in a companion paper \citep{Ghirardini2023} includes the empirical redshift uncertainties derived in this paper. The bias, however, was not included because it has negligible impact on the inferred cosmological parameters \citep{Ghirardini2023}. The redshift uncertainties were also accounted for in the study of superclusters and large-scale structures presented in another paper \citep{Liu2024}, which is sensitive to distance measurements.

\begin{figure}
    \centering
    \includegraphics[width=\linewidth]{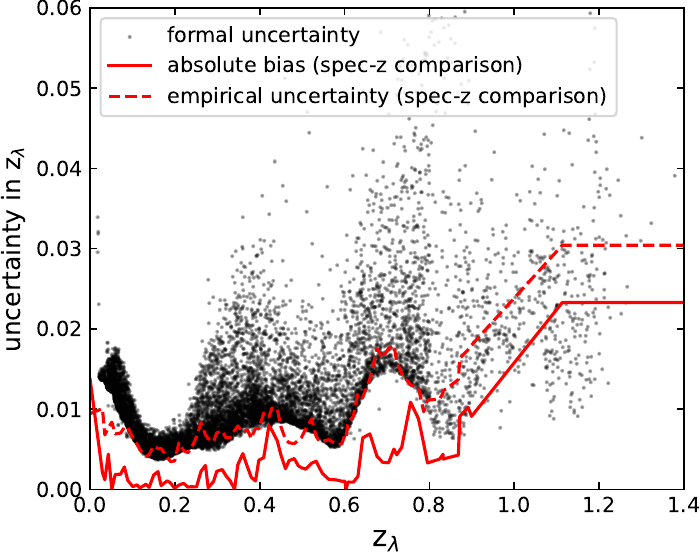}
    \caption{Uncertainties of the \erass photometric cluster redshifts. The black dots are the formal uncertainties $\delta z_\lambda^{\rm fml}$ calculated by \eromapper as the standard deviation of the redshift posterior $P(z)$. We adopted them as the final redshift uncertainties $z_{\lambda,{\rm err}}$. The dashed red line is the mean of the upper and lower empirical uncertainties $\delta z_\lambda^{\rm spec}$ derived from the comparison to the spectroscopic cluster redshifts. The continuous red line is the absolute redshift bias $|\Delta z_\lambda|$. 
    \label{fig:zlambda_error}}
\end{figure}

\subsection{Redshift bias} \label{sec:redshift_bias}

We estimated the redshift bias by applying a running median filter to the points shown in Figure \ref{fig:zspec_zphot}:

\begin{equation}
    \Delta z_\lambda (z) = \underset{z_{\rm spec}\in [z_{\rm min},z_{\rm max})}{\rm median}( z_\lambda - z_{\rm spec} ), \label{eq:zbias}
\end{equation}
where the window interval $[z_{\rm min}, z_{\rm max})$ was adjusted such that it had a minimum width of $\Delta z=0.02$ and encompassed a minimum of 100 clusters. Each result $\Delta z_\lambda$ was assigned to a redshift $z$ at the median $z_{\rm spec}$ in each interval. For the next step, we moved the window by $\Delta z=0.01$ from the previous initial redshift $z$.
Finally, the data points $\Delta z_\lambda(z)$ were linearly interpolated. 
The result is shown by the continuous red line in Figure \ref{fig:zspec_zphot}. It scatters around zero for $0.05 < z_{\rm spec} < 0.4$ and becomes slightly negative beyond that. 

\begin{align}
    \Delta z_\lambda && (     &z_{\rm spec}<0.05) & = +0.0040 \pm 0.0056, \nonumber \\
    \Delta z_\lambda && (0.05<&z_{\rm spec}<0.4 ) & = +0.0004 \pm 0.0016, \nonumber \\ 
    \Delta z_\lambda && ( 0.4<&z_{\rm spec}<0.8 ) & = -0.0041 \pm 0.0026, \nonumber \\ 
    \Delta z_\lambda && ( 0.8<&z_{\rm spec}     ) & = -0.0178 \pm 0.0242.
\end{align}

We notice stronger deviations around $z_{\rm spec}=0.4$ \citep[see also][]{Rykoff2014aa}. Here, the 4000\,\AA~break gets redshifted to the filter transition between the $g$ and $r$ bands. Consequently, neither the $g-r$ nor the $r-z$ color is sensitive to this redshift (see Appendix \ref{sec:redseq}). A positive bias just before $z_{\rm spec}<0.4$ and a negative bias just after $z_{\rm spec}>0.4$ creates an artificial clumping of clusters in redshift space at this location (see Figure \ref{fig:cludens}). 
A similar feature is visible at $z_{\rm spec}=0.7$. Here, the 4000\,\AA~break gets redshifted to the wavelength region between the $r$ and $z$ filters. Nevertheless, the bias remains below $|z_{\lambda}-z_{\rm spec}|<0.01$ for intermediate redshifts $0.6<z_{\rm spec}<0.9$. 
Beyond $z_{\rm spec}>0.9$, the bias increases to $|z_{\lambda}-z_{\rm spec}|>0.02$. Here, we reach the limiting redshift of the LS (see Figure \ref{fig:surveyarea}). The colors become less sensitive to redshift (see Appendix \ref{sec:redseq}), even though we included the $i$- and $w1$-band magnitudes for the red sequence colors beyond $z_\lambda>0.8$.
At very low redshifts $z_{\rm spec}<0.05$, the bias increases steeply because the red-sequence model was not calibrated (see Appendix \ref{sec:redseq}).

The right panel of Figure \ref{fig:zspec_zphot} shows the absolute deviations $|z_{\lambda}-z_{\rm spec}|$ normalized by $1+z_{\rm spec}$. We find
\begin{align}
    \frac{|z_{\lambda}-z_{\rm spec}|}{1+z_{\rm spec}} \lesssim 0.005 && \text{for~~} z_{\rm spec}\lesssim0.9.
\end{align}
The running median is below or around 0.5\% for $z_{\rm spec}\lesssim0.9$.
The previously discussed bias peaks around $z_{\rm spec}=0.4$ and $z_{\rm spec}=0.7$ are apparent here, too.

\subsubsection{Small-scale patterns in $\Delta z_\lambda$}

There are small-scale oscillations in photometric redshift bias with an amplitude of $z_{\lambda}-z_{\rm spec}\approx0.003$ and a period of $\Delta z_{\rm spec}\approx0.05-0.10$. This is visible as the systematic variations of the red line in Figure \ref{fig:zspec_zphot}, left panel. The effect is especially prominent around $z_{\rm spec}\approx0.4$, where neither the $g-r$ nor the $r-z$ colors are very sensitive to redshift (see Appendix \ref{sec:redseq}). The origin of these residuals can be traced back to the red-sequence model calibration. Our tests showed that the amplitude of these oscillations increased when we included the $i$-band in the $z<0.8$ \eromapper runs. Even though the statistical uncertainties decreased mildly because more information was being used, the systematic uncertainties increased because of color variations on small scales in redshift space. For this reason, we gave higher priority to the $grz$ and $grzw1$ \eromapper runs over the runs that include the $i$ band (see Section \ref{sec:merging}). In the future, we plan to improve this using finer-spaced spline fitting of the red sequence colors that include the $i$ band.

\subsubsection{Average bias correction}

The photometric redshifts can be corrected for the bias. To do so, we calculated $\Delta_z (z_\lambda)$ instead of $\Delta_z (z_{\rm spec})$ because $z_\lambda$ is known for every cluster and $z_{\rm spec}$ is only available for a subset. The procedure is equivalent to Equation (\ref{eq:zbias}). We computed $\Delta_z (z_\lambda)$ for the subset of clusters with known $z_{\rm spec}$ and subtracted it from $z_\lambda$ to obtain a bias-corrected photometric redshift $z_{\lambda,{\rm corr}}$:

\begin{equation}
    z_{\lambda,{\rm corr}} = z_\lambda - \Delta z_\lambda(z_\lambda). \label{eq:zlambdacorr}
\end{equation}

The corrected redshifts $z_{\lambda,{\rm corr}}$ are provided in the catalog, although all optical cluster properties were calculated at the uncorrected redshifts.
As stated in Section \ref{sec:redshift_accuracy}, the redshift bias has a negligible impact on the inferred cosmological parameters in \cite{Ghirardini2023}.

\subsection{Redshift uncertainties} \label{sec:redshift_uncertainties}

The formal redshift uncertainty $\delta z_\lambda^{\rm fml}$ was calculated by \eromapper for each cluster using the standard deviation of the redshift posterior $P(z)$ (see Section \ref{sec:eromapper_singleclusters}). We compared it to an empirical uncertainty $\delta z_\lambda^{\rm emp}$ derived from the comparison to spectroscopic cluster redshifts. The empirical uncertainties were calculated as the running 16$^{\rm th}$ and 84$^{\rm th}$ percentiles of the $z_{\rm spec}-z_\lambda$ distribution. 
It is analogous to the procedure described in Section \ref{sec:redshift_bias}, but we replaced the median with the percentiles and subtracted the redshift bias to isolate the contribution from the statistical uncertainties. The results for the upper and lower values of $\delta z_\lambda^{\rm emp}$ are

\begin{align}
    \delta z_\lambda^{\rm emp} && (     &z_{\rm spec}<0.05) & = +0.0093,-0.0078, \nonumber \\
    \delta z_\lambda^{\rm emp} && (0.05<&z_{\rm spec}<0.4 ) & = +0.0062,-0.0061, \nonumber \\ 
    \delta z_\lambda^{\rm emp} && (0.4< &z_{\rm spec}<0.8 ) & = +0.0095,-0.0110, \nonumber \\ 
    \delta z_\lambda^{\rm emp} && (0.8< &z_{\rm spec}     ) & = +0.0153,-0.0242.
\end{align}

These uncertainties are relatively small and almost symmetric at low to intermediate redshifts $0.05<z_{\rm spec}<0.4$. Below $z_{\rm spec}=0.05$, the 4000\,\AA~break reaches the blue limit of the $g$ band filter transmission curve, making the $g-r$ color less sensitive to redshift (see Appendix \ref{sec:redseq}).
In the future, $u$-band data from LSST \citep{Ivezic2019} will solve this issue.
Above $z_{\rm spec}>0.6$, the uncertainties increase because of higher photometric uncertainties for the fainter sources.

The running upper and lower percentiles are also shown in Figure \ref{fig:zspec_zphot}, left panel, by the dashed red lines. They confirm the mostly symmetric shape and the increase of $\delta z_\lambda^{\rm emp}$ around $z_{\rm spec}\approx0.7$. That increase happens because the 4000\,\AA~break is redshifted between the $g$ and $r$ filters.

We compare both types of uncertainties in Figure \ref{fig:zlambda_error}. The black points show $\delta z_\lambda^{\rm fml}$ for the individual clusters, and the red dashed line is the average of the upper and lower percentiles for $\delta z_\lambda^{\rm emp}(z_\lambda)$. In the range $0.1<z_\lambda<0.8$, the empirical uncertainties are consistent with the lower boundary of the formal uncertainties, but individual values for $\delta z_\lambda^{\rm fml}$ scatter upward. The redshift uncertainty is anti-correlated with richness, so as expected, the upscattering happens for poorer clusters with $\lambda\lesssim10$. 
Around $z_\lambda\approx0.4$ and $z_\lambda\approx0.7$, both uncertainties increase due to filter transitions. 
Above $z>0.8$, the formal uncertainties underestimate the empirical uncertainties by a factor of $\approx2$.

The continuous red line in Figure \ref{fig:zlambda_error} is the absolute value of the redshift bias evaluated on a grid for $z_\lambda$. It is lower than the uncertainties. That is, for an individual cluster, the statistical uncertainty dominates the error budget over the systematic error. We adopted the formal uncertainties, which were calculated by \eromapper as the standard deviation of the redshift posterior $P(z)$, as the final photometric redshift uncertainties $z_{\lambda,{\rm err}}=\delta z_\lambda^{\rm fml}$ because they also capture the richness dependence. Hence, they are a better estimate of an individual cluster's redshift uncertainty than the averaged empirical uncertainty for the ensemble.

\section{The richness -- velocity dispersion scaling relation} \label{sec:richness_vdisp}

We investigate the scaling relation between two cluster properties that correlate with cluster mass. The line-of-sight velocity dispersion $\sigma$ traces the dynamical mass $M_{\rm dyn}$ of a cluster 

\begin{equation}
    M_{\rm dyn}(<r) = c \frac{\sigma^2 r}{G},
\end{equation}
where $G$ is the gravitational constant, $r$ is the cluster radius, and $c$ corrects effects from 2D projection and orbit anisotropy \citep{Binney2008}. The richness $\lambda_{\rm norm}$ also correlates with total cluster mass \citep[e.g.,][]{McClintock2019aa} because deeper gravitational potential wells attract more baryonic matter.

To correct for redshift evolution, we multiplied $\lambda_{\rm norm}$ with $E(z)=\sqrt{\Omega_{\rm m}(1+z)^3 + \Omega_\lambda}$ \citep{Damsted2023}, although the effect is small. Most spectroscopic clusters are at low redshift (see Figure \ref{fig:bestztype}).
We explore the $\sigma-\lambda_{\rm norm}$ relation for the \erass sample augmented by the literature samples. 
The scaling relation between X-ray, weak lensing, and optical observables are detailed in companion papers (\citealt{Ghirardini2023,Grandis2023}; Pacaud et al. in prep.).

\subsection{Cluster selection}

\begin{figure}
    \centering
    \includegraphics[width=\linewidth]{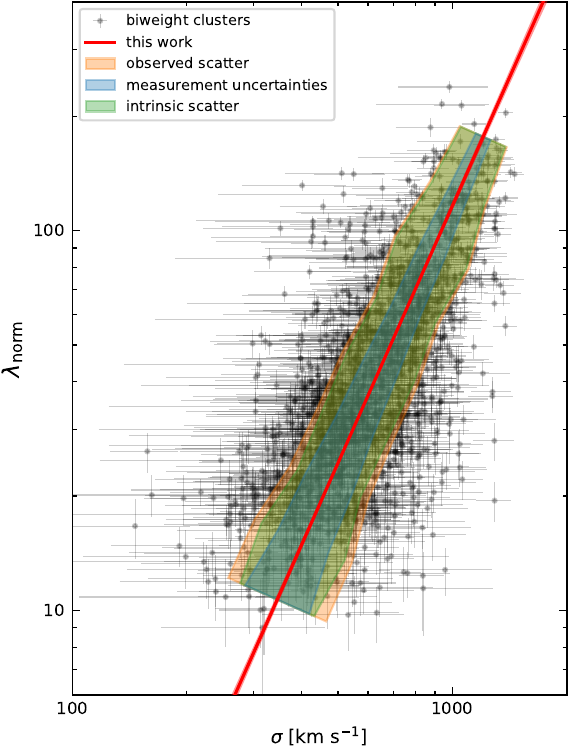}
    \caption{Relation for cluster line-of-sight velocity dispersion $\sigma$ and richness $\lambda_{\rm norm}$. The black data points are for all clusters in our analyzed catalogs (see Table \ref{tab:scancat}) apart from the blind-mode runs. In total, 1699 clusters with $N_{\rm members}\geq15$ spectroscopic members were selected. The red line is the best fit to the data, and the light red shades show the uncertainty of the relation. The observed scatter $\delta_{\rm obs,orth}$ (orange) was measured orthogonal to the best-fit relation in 9 bins. The median orthogonal component of the measurement uncertainties $\delta_{\rm m,orth}$ is shown for each bin in blue. By subtracting $\delta_{\rm m,orth}$ from $\delta_{\rm obs,orth}$ in quadrature, we estimated the intrinsic scatter $\delta_{\rm in}$ for each bin. It is shown in green.}
    \label{fig:lambda_vdisp}
\end{figure}

For the analysis, we considered only clusters with high-quality (unflagged) velocity dispersions determined using the bi-weight scale estimator (see Section \ref{sec:vdisp}).
As we required a large number of $\geq15$ spectroscopic members for this method, these clusters also have a high richness and a low relative uncertainty: $\delta\sigma/\sigma\sim20\%$. 
We show how these clusters populate the $\sigma-\lambda_{\rm norm}$ plane in Figure \ref{fig:lambda_vdisp}. 
The Figure contains data for 1699 clusters with a low masking fraction of $<20\%$. Of them, 309 are \erass clusters.

We assumed that sample selection effects are unimportant for this analysis. For \erass, the X-ray count rate is the primary selection variable. For a thin redshift slice, the velocity dispersion and the richness correlate with the X-ray count rate. Hence, the net selection effect is along the relation and, therefore, unimportant when applying orthogonal distance regression.

\subsection{Relation fitting}

We fitted a linear relation with free parameters $\alpha$ and $\beta$ to the logarithm of both variables;

\begin{equation}
    \log(\lambda_{\rm norm}) = \alpha \times \log(\sigma [{\rm km\,s^{-1}}]) + \beta.
\end{equation}
Our preferred fitting method \citep[ODR,][]{Boggs1989} minimizes the orthogonal residuals from the relation. Additionally, we weighed the individual data points by their measurement uncertainties in both variables and included the non-negligible intrinsic scatter. The method is described in detail in \cite{Kluge2023}. An advantage is that the method is insensitive to switching variables.

The best-fit values are

\begin{align}
\alpha &= \phantom{-}2.401\pm0.055,\\
\beta  &= -5.074\pm0.156,
\end{align}
with a covariance of COV$(\log\lambda_{\rm norm},\log\sigma)=-8.6\times10^{-3}$. 
Our best-fit slope agrees within 1.5$\sigma$ confidence with $\alpha=1/(0.323\pm0.057)=3.10^{+0.66}_{-0.46}$ obtained by \cite{Kirkpatrick2021aa}. The agreement improves to 0.8$\sigma$ ($\alpha=2.734\pm0.082$) confidence when we use the same fitting method \citep[BCES,][]{Akritas1996,Nemmen2012}.

Moreover, our best-fit slope also agrees within 1.6$\sigma$ confidence with $\alpha=1/(0.365\pm0.029)=2.74^{+0.23}_{-0.20}$ obtained by \cite{Damsted2023}. However, their applied fitting method \texttt{linmix} \citep{Kelly2007aa} is sensitive to switching variables. Using that method, we got consistent results only for $(\sigma|\lambda_{\rm norm})$: $1/\alpha=0.392\pm0.010$. For $(\lambda_{\rm norm}|\sigma)$, we got $\alpha=1.753\pm0.045$, a 9$\sigma$ deviation.

The statistical uncertainties of the best-fit slopes improved by a factor of 4--10 compared to previous works. This is attributed to the large gain in sample size of clusters with $N_{\rm members}\geq15$ spectroscopic members. Our sample includes 1699 clusters compared to 530 clusters \citep{Damsted2023}. The sample of \citep{Kirkpatrick2021aa} includes 2740 clusters but with a lower median of 10 spectroscopic members.

\subsection{Intrinsic scatter}

We estimated the intrinsic scatter following the procedure in \cite{Kluge2023}.
First, we calculated the observed orthogonal scatter $\delta_{\rm obs,orth}$ and the median orthogonal component of the measurement uncertainties $\delta_{\rm m,orth}$. Then, the intrinsic scatter $\delta_{\rm in}$ was estimated by

\begin{align}
    \delta_{\rm in} = \sqrt{\delta^2_{\rm obs,orth} - \delta^2_{\rm m,orth}}.
\end{align}
To capture a possible variation with $\sigma$ or $\lambda_{\rm norm}$, we split the sample into 9 equal-sized bins perpendicular to the best-fit line. In each bin, we calculated $\delta_{\rm obs,orth}$, $\delta_{\rm obs,orth}$, and $\delta_{\rm in}$, and show results in Figure \ref{fig:lambda_vdisp} by the orange, blue, and green shades, respectively. Notably, not only do the measurement uncertainties increase for the lower $\sigma$ or $\lambda_{\rm norm}$, as expected, but also the intrinsic scatter increases. 

We report the obtained values for two larger bins, split at $\sigma=620$\,km\,s$^{-2}$ and $\lambda_{\rm norm}=43$. For the subsample at high $\sigma$ and $\lambda_{\rm norm}$, we obtained $\delta_{\rm obs,orth}=0.100$ dex, $\delta_{\rm m,orth}=0.039\pm0.026$ dex, and $\delta_{\rm in}=0.092\pm0.008$ dex. For the subsample at low $\sigma$ and $\lambda_{\rm norm}$, we obtained $\delta_{\rm obs,orth}=0.118$ dex, $\delta_{\rm m,orth}=0.058\pm0.029$ dex, and $\delta_{\rm in}=0.102\pm0.012$ dex. This is slightly lower than $\delta_{\rm in}=0.138\pm0.017$ obtained by \cite{Damsted2023}, possibly because we applied different fitting and scatter-estimation methods.
We conclude that the $\sigma-\lambda_{\rm norm}$ relation obtained in this work is dominated by intrinsic scatter and not by measurement uncertainties.

\section{Summary} \label{sec:summary}

The first \erosita All-Sky Survey (\erass) provides the largest sample of ICM-selected galaxy clusters and groups in the western Galactic hemisphere to date.
In this work, we identify 12\,247 candidates by associating them with a coincident overdensity of red-sequence galaxies or, in 247 cases, by matching them with known clusters from the literature.
We ran the red-sequence-based cluster finder \eromapper on optical and near-infrared data from DESI Legacy Imaging Surveys DR9 (at Decl. $>32.375\degr$) and DR10 (at Decl. $<32.375\degr$) on 13\,116\,deg$^2$ of the sky. The \erass sample of cluster and group candidates is designed to be as complete as possible. After discarding candidates without optical identification, we estimated the purity using a mixture model to be 86\%.

We provide optical (redshifts, richnesses, optical centers, BCG position) and spectroscopic (redshifts and velocity dispersions) properties.
The photometric redshifts have an excellent accuracy of $\Delta z/(1+z)\lesssim0.005$ for $0.05<z<0.9$. For individual clusters, the accuracy is dominated by statistical uncertainties.

Spectroscopic redshifts were calculated for a subsample of 3210 \erass clusters and groups. Of those, 1759 have at least three spectroscopic members. The spectroscopic galaxy redshifts were compiled from public catalogs from the literature and dedicated follow-up programs and matched to the photometrically selected member galaxies. The spectroscopic cluster redshifts are, on average, 1$\sim$7 (5$\sim$15) times more accurate than the photometric cluster redshifts when less than 5 (more than 20) spectroscopic members are known.
This improvement in accuracy makes the \erass clusters align well with the large-scale structure as traced by individual spectroscopic galaxy redshifts.
Velocity dispersions were measured for 1906 clusters, of which 1499 are robust, and 358 have high quality with relative uncertainty $\delta\sigma/\sigma\approx20\%$.

In addition to the \erass catalog, we consistently remeasured the redshifts on known clusters from the literature. We recovered consistent redshifts in the range $0.05\lesssim z\lesssim1.1$ for $>95\%$ of the clusters detected with the high-$z$-sensitive Sunyaev-Zeldovich effect or low-$z$-sensitive optical selection. The outliers are explained by multiple structures coinciding along the line of sight.

Moreover, we detected galaxy clusters without positional priors in the full footprint of the Legacy Surveys (24\,069\,deg$^2$). The number density of the found clusters is independent of the optical survey depth below the local limiting redshift $z_{\rm vlim}$.
The depth of the LS limits the high-quality subsample at around $z_{\rm vlim}\approx0.7$ for a minimum galaxy luminosity of 0.2\,L$_*$. By increasing the luminosity threshold to $L>0.4$\,L$_*$ and including the $i$ and $w1$ filter bands, we increased the limiting depth to $z_{\rm vlim}\approx1.0$ and even higher locally.

Combining all analyzed cluster catalogs, we obtained a sample of 1699 clusters with high-quality velocity dispersion measurements. We correlated it with the richness and found a best-fit relation $\log(\lambda_{\rm norm}) = 2.401 (\pm0.055) \times \log(\sigma [{\rm km\,s^{-1}}]) - 5.074 (\pm0.156)$, which agrees with previous results from other studies.
The scatter in the relation is dominated by an intrinsic scatter $\delta_{\rm in}$ for the full range $300\lesssim\sigma\lesssim1000$\,km\,s$^{-1}$ and $10\lesssim\lambda\lesssim200$. It increases only slightly from $\delta_{\rm in}=0.092\pm0.008$\,dex to $\delta_{\rm in}=0.102\pm0.012$\,dex toward lower $\sigma$ and $\lambda$.

In the future, the next-generation \erosita cluster catalogs will be supported by a wealth of spectroscopic input from SDSS-V (10\,000 clusters), 4MOST (40\,000 clusters), LSST, and \textit{Euclid}. These data will expand bias-free redshift measurements to higher redshifts.
A complete sampling of the nodes of the cosmic web will yield exquisite constraints on cosmological models and primordial non-Gaussianity \citep{Stopyra2021}.

\section*{Data availability}

We make available the optical properties of the \erass galaxy clusters and groups as a catalog in electronic form. It can be accessed at the CDS via anonymous ftp to \url{cdsarc.u-strasbg.fr} (\href{ftp://130.79.128.5/}{130.79.128.5}) or via \url{https://cdsarc.cds.unistra.fr/viz-bin/cat/J/A+A/688/A210}. Alternatively, the catalog is available via the following webpage: \url{https://erosita.mpe.mpg.de/dr1/AllSkySurveyData_dr1/Catalogues_dr1/}. Moreover, the measured optical properties of the clusters from literature catalogs and \eromapper runs without positional priors are made available, too. They can be accessed and explored using a visual inspection tool on the following webpage: \url{https://erass-cluster-inspector.com}. Reasonable requests for the member galaxy catalogs can be made to the corresponding author.

\section*{Acknowledgement}

The authors thank the referee for helpful and constructive
comments on the draft.
This work is based on data from \erosita, the soft X-ray instrument aboard SRG, a joint Russian-German science mission supported by the Russian Space Agency (Roskosmos), in the interests of the Russian Academy of Sciences represented by its Space Research Institute (IKI), and the Deutsches Zentrum f{\"{u}}r Luft und Raumfahrt (DLR). The SRG spacecraft was built by Lavochkin Association (NPOL) and its subcontractors and is operated by NPOL with support from the Max Planck Institute for Extraterrestrial Physics (MPE).

The development and construction of the \erosita X-ray instrument was led by MPE, with contributions from the Dr. Karl Remeis Observatory Bamberg \& ECAP (FAU Erlangen-Nuernberg), the University of Hamburg Observatory, the Leibniz Institute for Astrophysics Potsdam (AIP), and the Institute for Astronomy and Astrophysics of the University of T{\"{u}}bingen, with the support of DLR and the Max Planck Society. The Argelander Institute for Astronomy of the University of Bonn and the Ludwig Maximilians Universit{\"{a}}t Munich also participated in the science preparation for \erosita.

The \erosita data shown here were processed using the \esass software system developed by the German \erosita consortium.
V. Ghirardini, E. Bulbul, A. Liu, C. Garrel, E. Artis, M. Kluge, and X. Zhang acknowledge financial support from the European Research Council (ERC) Consolidator Grant under the European Union’s Horizon 2020 research and innovation program (grant agreement CoG DarkQuest No 101002585). N. Clerc was financially supported by CNES. T. Schrabback and F. Kleinebreil acknowledge support from the German Federal
Ministry for Economic Affairs and Energy (BMWi) provided
through DLR under projects 50OR2002, 50OR2106, and 50OR2302, as well as the support provided by the Deutsche Forschungsgemeinschaft (DFG, German Research Foundation) under grant 415537506.

The Legacy Surveys consist of three individual and complementary projects: the Dark Energy Camera Legacy Survey (DECaLS; Proposal ID \#2014B-0404; PIs: David Schlegel and Arjun Dey), the Beijing-Arizona Sky Survey (BASS; NOAO Prop. ID \#2015A-0801; PIs: Zhou Xu and Xiaohui Fan), and the Mayall z-band Legacy Survey (MzLS; Prop. ID \#2016A-0453; PI: Arjun Dey). DECaLS, BASS and MzLS together include data obtained, respectively, at the Blanco telescope, Cerro Tololo Inter-American Observatory, NSF’s NOIRLab; the Bok telescope, Steward Observatory, University of Arizona; and the Mayall telescope, Kitt Peak National Observatory, NOIRLab. Pipeline processing and analyses of the data were supported by NOIRLab and the Lawrence Berkeley National Laboratory (LBNL). The Legacy Surveys project is honored to be permitted to conduct astronomical research on Iolkam Du’ag (Kitt Peak), a mountain with particular significance to the Tohono O’odham Nation.

NOIRLab is operated by the Association of Universities for Research in Astronomy (AURA) under a cooperative agreement with the National Science Foundation. LBNL is managed by the Regents of the University of California under contract to the U.S. Department of Energy.

This project used data obtained with the Dark Energy Camera (DECam), which was constructed by the Dark Energy Survey (DES) collaboration. Funding for the DES Projects has been provided by the U.S. Department of Energy, the U.S. National Science Foundation, the Ministry of Science and Education of Spain, the Science and Technology Facilities Council of the United Kingdom, the Higher Education Funding Council for England, the National Center for Supercomputing Applications at the University of Illinois at Urbana-Champaign, the Kavli Institute of Cosmological Physics at the University of Chicago, Center for Cosmology and Astro-Particle Physics at the Ohio State University, the Mitchell Institute for Fundamental Physics and Astronomy at Texas A\&M University, Financiadora de Estudos e Projetos, Fundacao Carlos Chagas Filho de Amparo, Financiadora de Estudos e Projetos, Fundacao Carlos Chagas Filho de Amparo a Pesquisa do Estado do Rio de Janeiro, Conselho Nacional de Desenvolvimento Cientifico e Tecnologico and the Ministerio da Ciencia, Tecnologia e Inovacao, the Deutsche Forschungsgemeinschaft and the Collaborating Institutions in the Dark Energy Survey. The Collaborating Institutions are Argonne National Laboratory, the University of California at Santa Cruz, the University of Cambridge, Centro de Investigaciones Energeticas, Medioambientales y Tecnologicas-Madrid, the University of Chicago, University College London, the DES-Brazil Consortium, the University of Edinburgh, the Eidgenossische Technische Hochschule (ETH) Zurich, Fermi National Accelerator Laboratory, the University of Illinois at Urbana-Champaign, the Institut de Ciencies de l’Espai (IEEC/CSIC), the Institut de Fisica d’Altes Energies, Lawrence Berkeley National Laboratory, the Ludwig Maximilians Universitat Munchen and the associated Excellence Cluster Universe, the University of Michigan, NSF’s NOIRLab, the University of Nottingham, the Ohio State University, the University of Pennsylvania, the University of Portsmouth, SLAC National Accelerator Laboratory, Stanford University, the University of Sussex, and Texas A\&M University.

BASS is a key project of the Telescope Access Program (TAP), which has been funded by the National Astronomical Observatories of China, the Chinese Academy of Sciences (the Strategic Priority Research Program “The Emergence of Cosmological Structures” Grant \# XDB09000000), and the Special Fund for Astronomy from the Ministry of Finance. The BASS is also supported by the External Cooperation Program of Chinese Academy of Sciences (Grant \# 114A11KYSB20160057), and Chinese National Natural Science Foundation (Grant \# 12120101003, \# 11433005).

The Legacy Survey team makes use of data products from the Near-Earth Object Wide-field Infrared Survey Explorer (NEOWISE), which is a project of the Jet Propulsion Laboratory/California Institute of Technology. NEOWISE is funded by the National Aeronautics and Space Administration.

The Legacy Surveys imaging of the DESI footprint is supported by the Director, Office of Science, Office of High Energy Physics of the U.S. Department of Energy under Contract No. DE-AC02-05CH1123, by the National Energy Research Scientific Computing Center, a DOE Office of Science User Facility under the same contract; and by the U.S. National Science Foundation, Division of Astronomical Sciences under Contract No. AST-0950945 to NOAO.

The \textit{Hobby-Eberly Telescope} (HET) is a joint project of the University of Texas at Austin, the Pennsylvania State University, Ludwig-Maximillians-Universit\"at M\"unchen, and Georg-August Universit\"at G\"ottingen. The HET is named in honor of its principal benefactors, William P. Hobby and Robert E. Eberly.

VIRUS is a joint project of the University of Texas at Austin, Leibniz-Institut f\"ur Astrophysik Potsdam (AIP), Texas A\&M University (TAMU), Max-Planck-Institut f\"ur Extraterrestrische Physik (MPE), Ludwig-Maximilians-Universit\"at Muenchen, Pennsylvania State University, Institut f\"ur Astrophysik G\"ottingen, University of Oxford, and the Max-Planck-Institut f\"ur Astrophysik (MPA). In addition to Institutional support, VIRUS was partially funded by the National Science Foundation, the State of Texas, and generous support from private individuals and foundations.

\bibliography{ref,ref2}{}

\appendix

\section{Limitations \& caveats} \label{sec:limitations}

This section gives a concise overview of the limitations of the \erass cluster and group catalog.

\subsection{Completeness}

The optical completeness decreases at low redshift $z<0.05$ because we discarded galaxies that overlap with large galaxies (see Sections \ref{sec:catprocessing} and \ref{sec:lowzcompleteness}). This affects especially clusters or groups that host a cD galaxy. At high redshifts, the catalog is almost complete for high cluster masses and is mostly limited by the depth of the LS. It is quantified by the parameter $z_{\rm vlim}$ (see Appendix \ref{sec:zvlim}). If a cluster was found at a higher redshift $z_\lambda > z_{\rm vlim}$, the probability increases that the detection is a contaminant (see Figure \ref{fig:richness_cludens}). Avoiding these clusters using the boolean catalog parameter IN\_ZVLIM is advisable for selecting a clean sample.

Our cluster finder is based on the colors of galaxies, which follow the red sequence. Hence, star-forming late-type galaxies cannot be part of our member galaxy sample. To confirm this, we ran \eromapper on the Hickson compact group catalog \citep{Hickson1982aa}, which consists mostly of low-$z$ late-type galaxies. We recovered 17/100 groups that consist of mostly early-type galaxies at $z_\lambda<0.1$. The remaining 83/100 groups were not detected by \eromapper, as expected.

The intrinsic proportion of blue star-forming galaxies increases not only for lower masses but also for higher redshifts, known as the Butcher--Oemler effect \citep{ButcherOemler1978}. In the densest clusters, the fraction of quiescent galaxies over all galaxies decreases from 75\% at $z=0$ to 45\% at $z=0.7$ \citep{Hahn2015}.
Assessing the precise impact on the optical completeness of the \erass catalog would require running \eromapper on simulated galaxy catalogs \citep[e.g.,][]{To2024}. No such dedicated analysis exists to our knowledge, yet.
Nevertheless, we can infer a plausible range for the optical completeness based on existing comparisons for different types of cluster finders. A red-sequence-based cluster finder similar to \eromapper yields a relative optical completeness of $\sim$100\% ($\sim$70\%) at $z<0.6$ ($z<1.0$) compared to a redshift-based cluster finder that is more sensitive to blue star-forming galaxies (see Section \ref{sec:rm} for more details). 
However, the cluster masses of $M>10^{14}$\,M$_\odot$ in that analysis are lower than the mass limit of the \erass catalog at $z>0.6$ with $M\approx3\times10^{14}$\,M$_\odot$ (see Figure \ref{fig:cludens}). 
Given the observed decrease in the proportion of blue, star-forming galaxies with increasing cluster mass \citep{Hahn2015}, we can regard the 70\% optical completeness as a lower limit. 
Moreover, the \erass mass limit at $z>0.6$ is comparable to the mass limit in the ACT survey with which we estimated a high optical completeness for the \erass identified catalog of $>95\%$ (see Section \ref{sec:highzcompleteness}).
Finally, the good agreement of the cluster number densities obtained with the \eromapper run in blind mode with the theoretical prediction for the halo mass function (see Figure \ref{fig:cludens}) underlines that the optical completeness remains high out to high redshift.

\subsection{Contamination estimates for galaxy groups} \label{sec:limitation_pcont}

Our contamination estimator (see Section \ref{sec:pcont}) performs well at assigning a high contamination probability $P_{\rm cont}$ to possible AGN for high X-ray count rates and low richnesses (see Figure \ref{fig:pcont}, right panel, top left projection). However, sources with low X-ray count rates always received $P_{\rm cont}\approx0$. These count rates occurred during the multi-component fitting using MBproj2D \citep[see][]{Bulbul2023}. They are not part of the expected count rates for random sky points. Hence, the kernel density estimate for this region in parameter space is almost zero for the contaminants. Consequently, our contamination estimator underestimated the true contamination on the galaxy group scale.

In contrast, visual inspection of real groups ($\lambda_{\rm norm}\lesssim10$) at low redshift ($z_{\rm best}\lesssim0.1$) revealed that many of them received a high $P_{\rm cont}$ value. In principle, lowering the member luminosity threshold below $L<0.2$\,L$_*$ can help distinguish them better from AGN. This would increase the measured richnesses of real groups while we expect the richnesses of AGN to remain low.

\subsection{Cluster properties}

The cluster parameters were always calculated at the photometric redshift $z_\lambda$. This concerns the galaxy memberships, richnesses $\lambda$ and $\lambda_{\rm norm}$, optical centers, and velocity dispersions $\sigma$. In cases where the best redshift type was changed to a literature redshift, the cluster parameters are not reliable because they refer to a foreground or background cluster.

\subsection{Masked clusters} \label{sec:infootprint}

Few clusters had a large mask applied, which covered roughly one-half of their extent. One example is 1eRASS J022553.0-415457 with a masking fraction of 40\%. We noticed that we detected member galaxies only on the other side of this cluster when we ran \eromapper on a location within the initial mask. That means, the mask was not created based on flagged photometry in the LS. We suspect that this effect occurs when a cluster is located near the edge of a heal pixel. Nevertheless, the richnesses were corrected for the masking fraction (see Section \ref{sec:rm}), and, hence, they are noisier but unbiased. Also, we have no reason to assume that the photometric redshifts would be biased in the described cases. Moreover, only a small fraction of the \erass clusters is affected. Four percent of them have a masking fraction of $>40\%$ and for most of them, a saturated nearby star is the reason for the large mask.

\subsection{Velocity dispersion}

We provide cluster velocity dispersions even when the velocity clipping diverged in some of the bootstrapped samples (see Section \ref{sec:vdisp}). This affects 407 out of 1906 cases that can be identified by the flag VDISP\_FLAG\_BOOT>0. Two examples with unrealistically high velocity dispersions $\sigma>10\,000$\,km\,s$^{-1}$ are 1eRASS J104344.4+240537 and 1eRASS J025101.7-353415. In both cases, the velocity dispersion is high because one wrongly measured redshift around $z_{\rm spec}=0$ was not clipped. This error is taken into consideration in the uncertainty $\delta\sigma>10\,000$\,km\,s$^{-1}$, which is almost equal to $\sigma$ itself. The spectroscopic redshifts are less affected by these outliers because the biweight location for the spectroscopic cluster redshifts happens to be more robust than the gapper estimator for the velocity dispersion.

\subsection{Bias-corrected photometric redshifts}

Red-sequence model calibration errors (see Section \ref{sec:redseq}) propagate to biases in the photometric cluster redshifts (see Section \ref{sec:redshift_bias}). We have measured this bias and provide corrected photometric redshifts $z_{\lambda,{\rm corr}}$ (see Equation (\ref{eq:zlambdacorr})) and corresponding uncertainties. These corrections are small at low redshifts but become increasingly important at $z\gtrsim0.8$. In particular, the formal uncertainties underestimate the real uncertainties significantly at $z\gtrsim0.8$.

\section{Limiting depth and limiting redshift} \label{sec:zvlim}

\begin{figure*}
    \centering
    LS DR10 south\\~\\
    \includegraphics[width=0.19\linewidth]{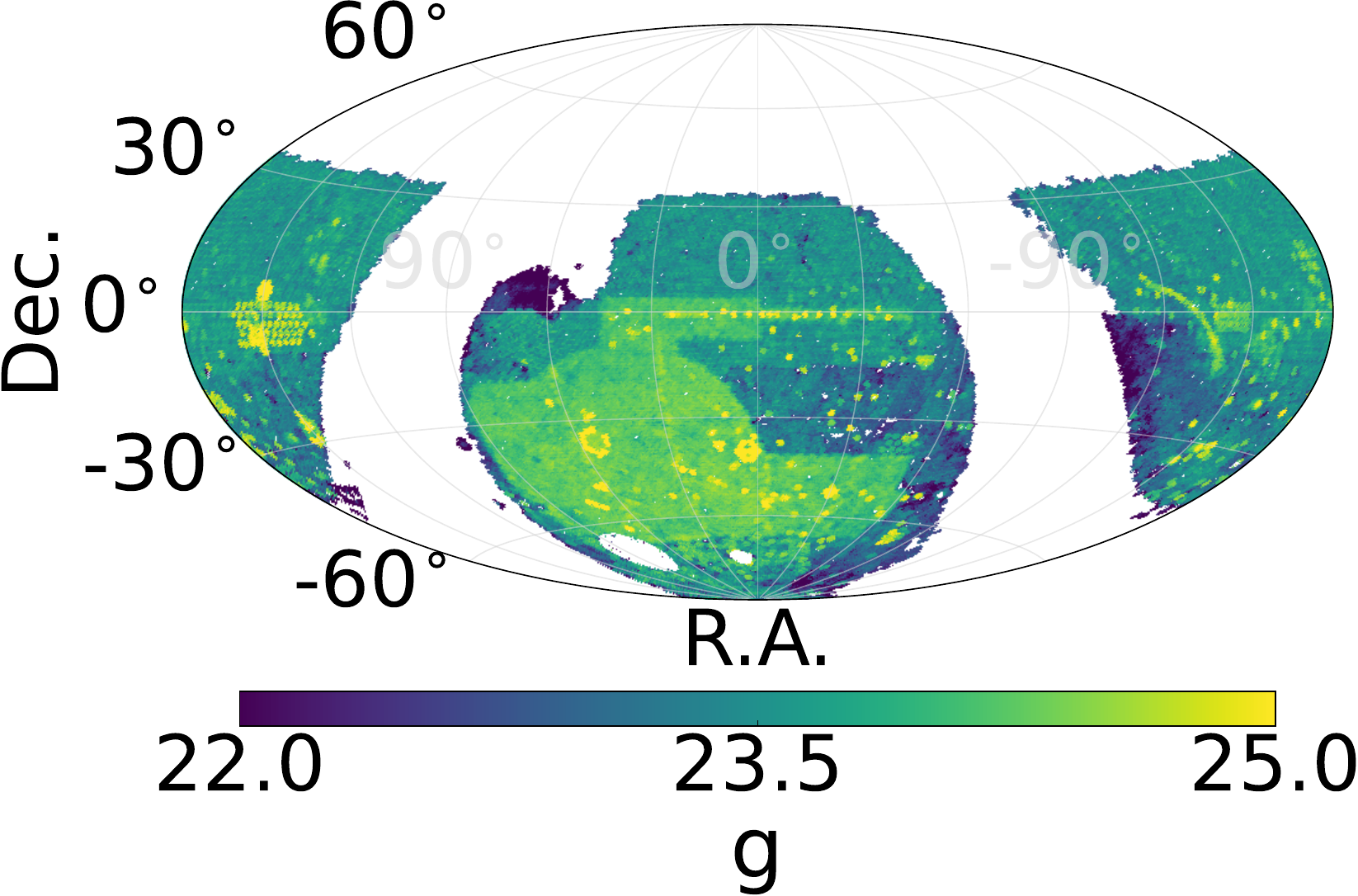}
    \includegraphics[width=0.19\linewidth]{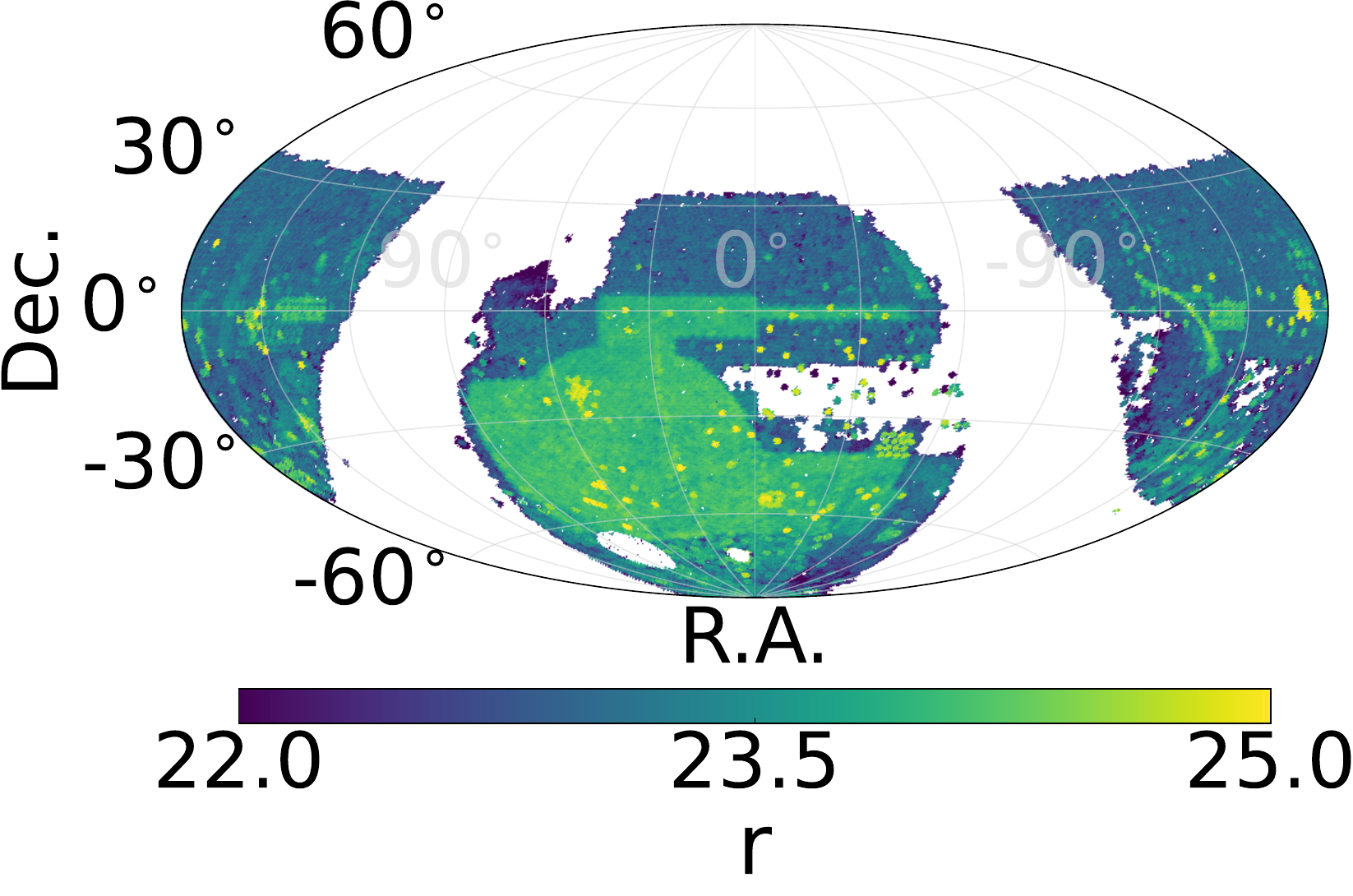}
    \includegraphics[width=0.19\linewidth]{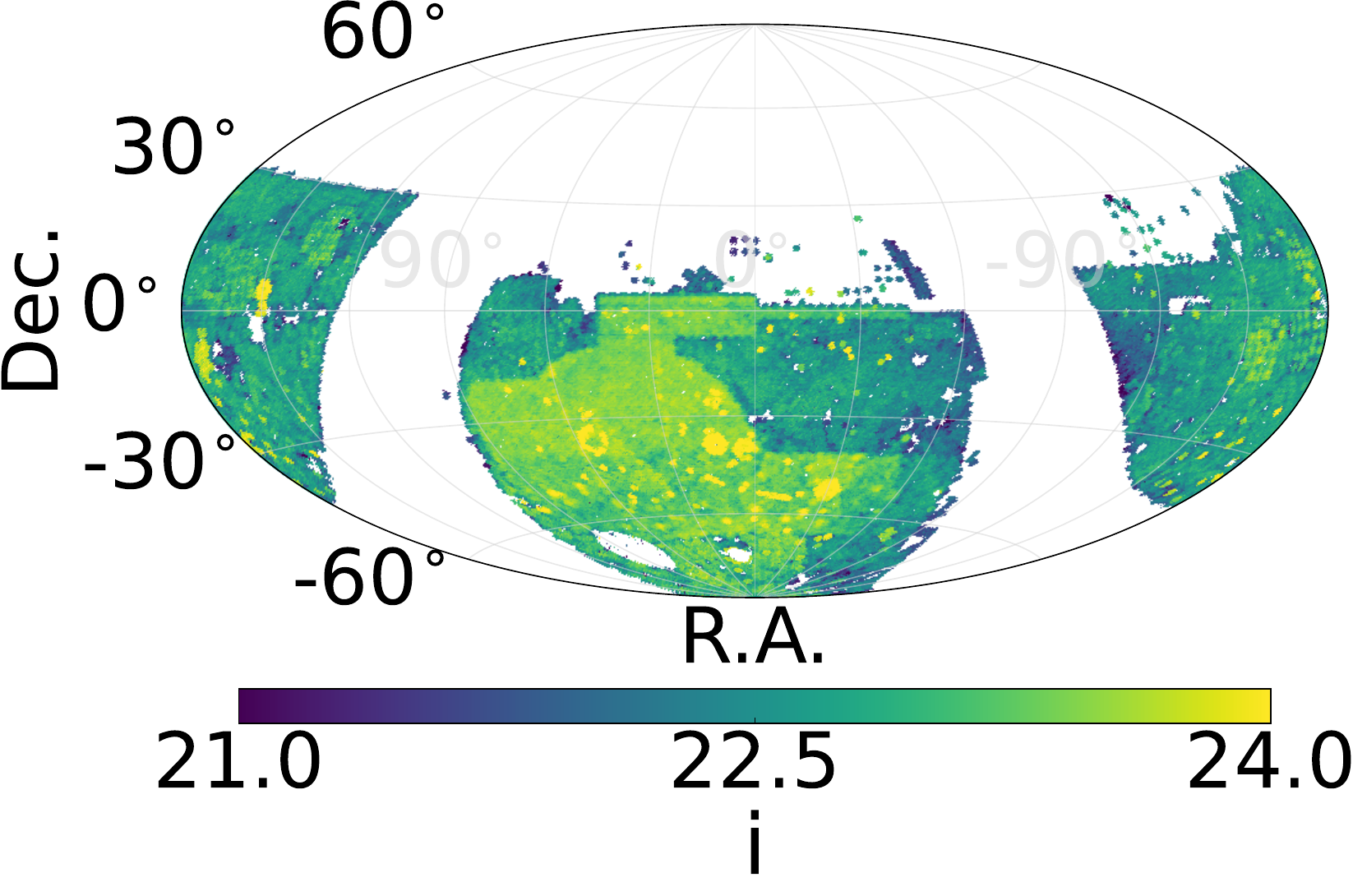}
    \includegraphics[width=0.19\linewidth]{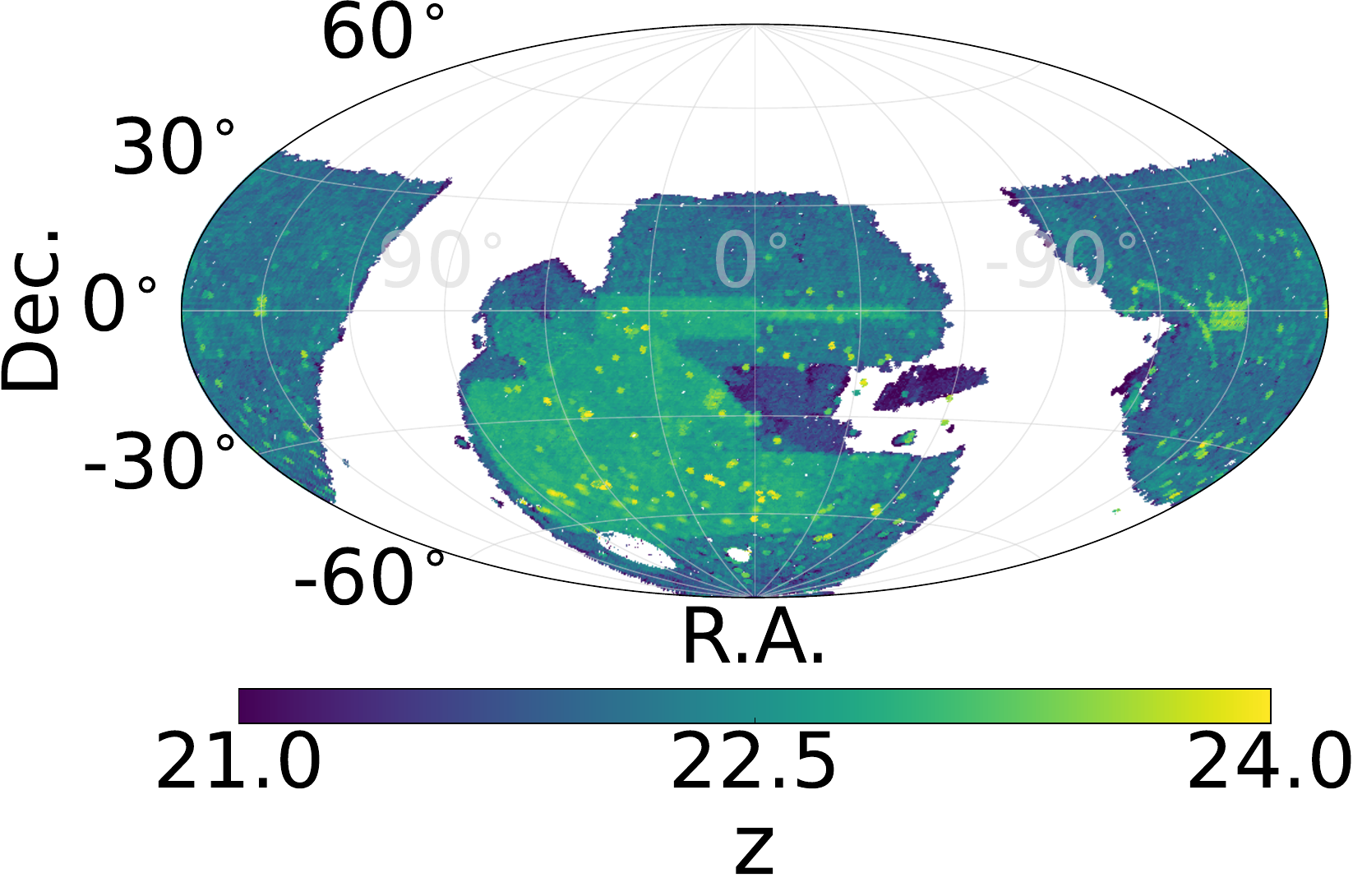}
    \includegraphics[width=0.19\linewidth]{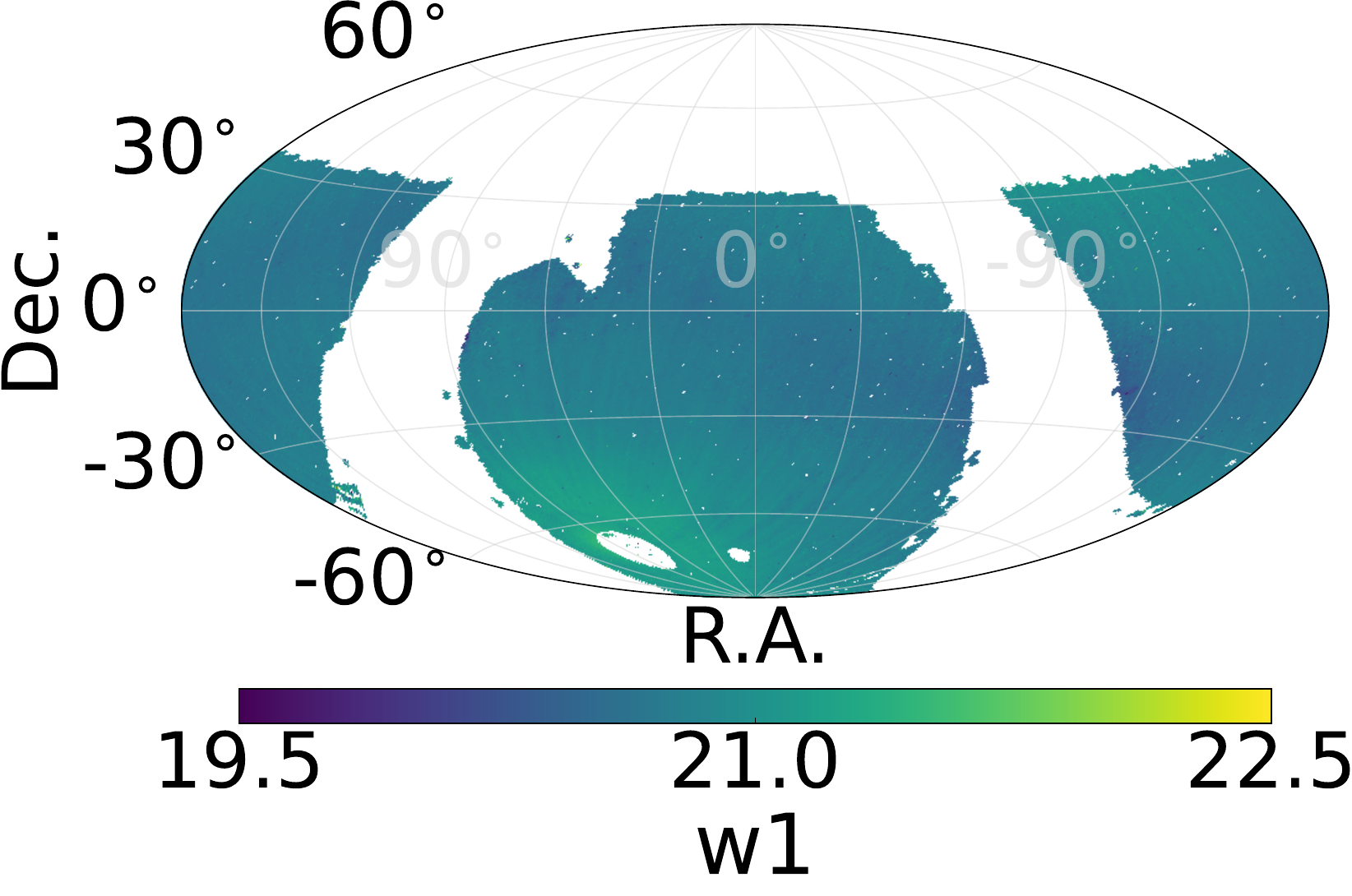}\\~\\
    LS DR9 north\\~\\
    \includegraphics[width=0.19\linewidth]{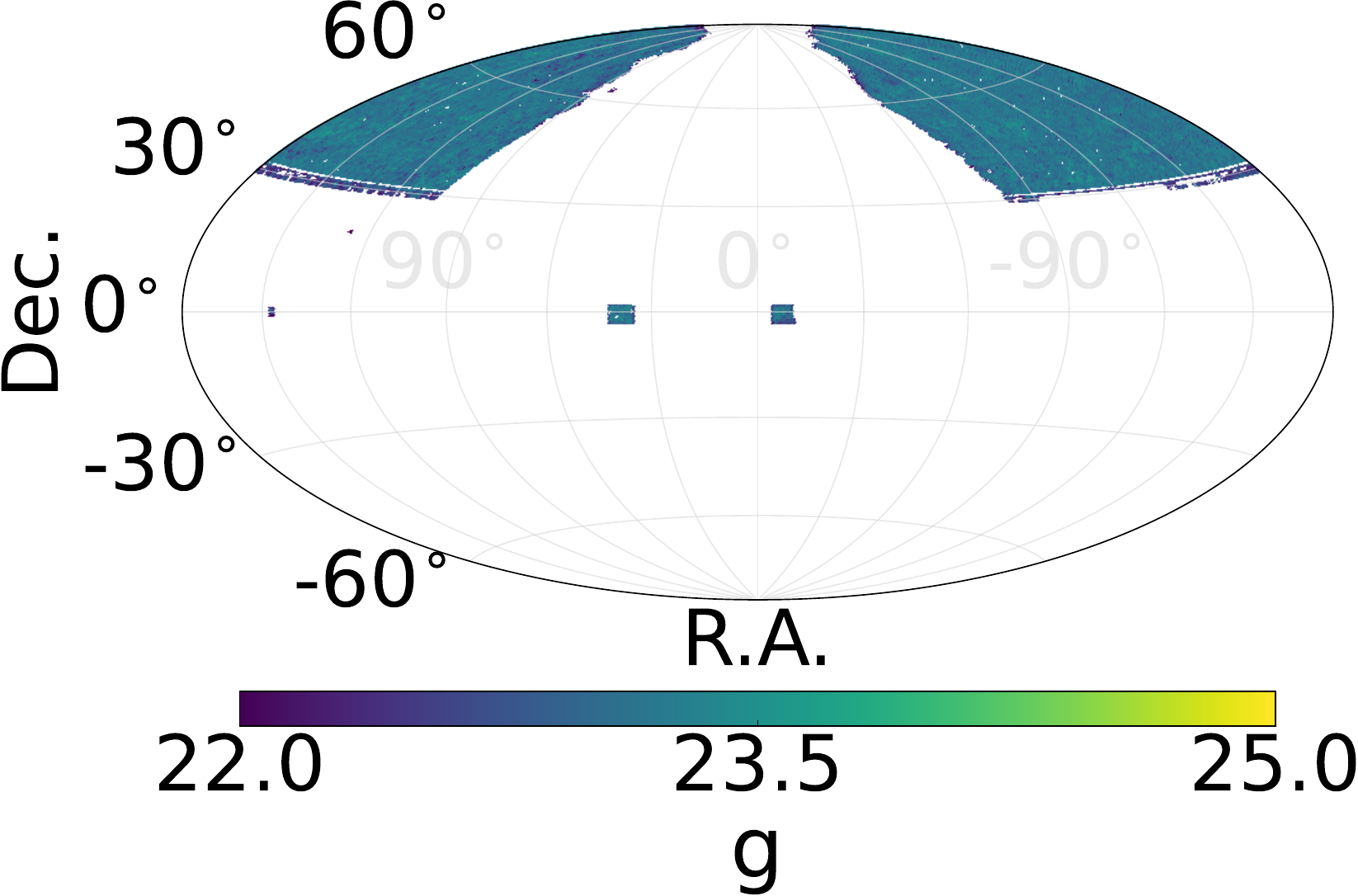}
    \includegraphics[width=0.19\linewidth]{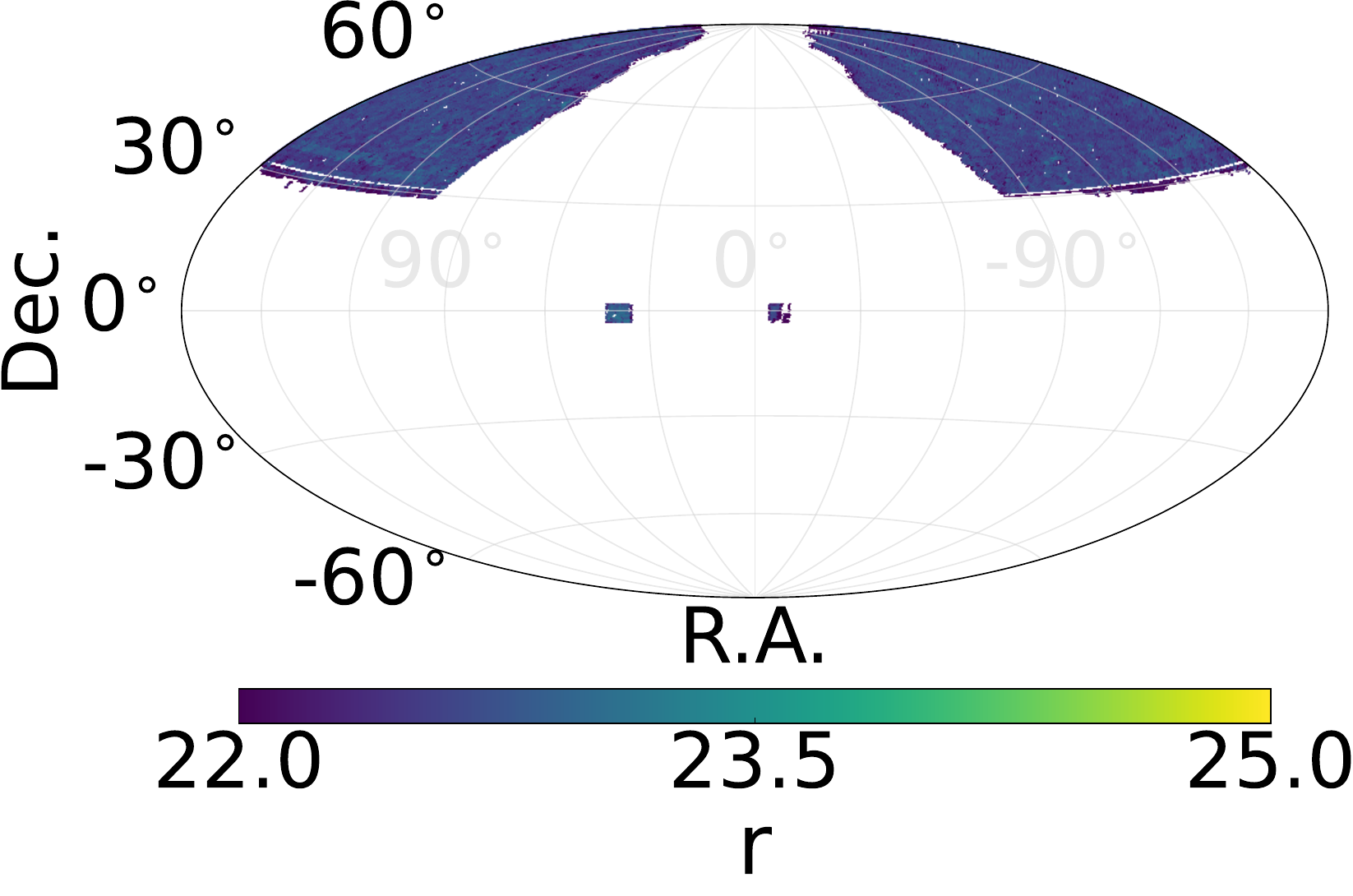}\hspace{3.58cm}
    \includegraphics[width=0.19\linewidth]{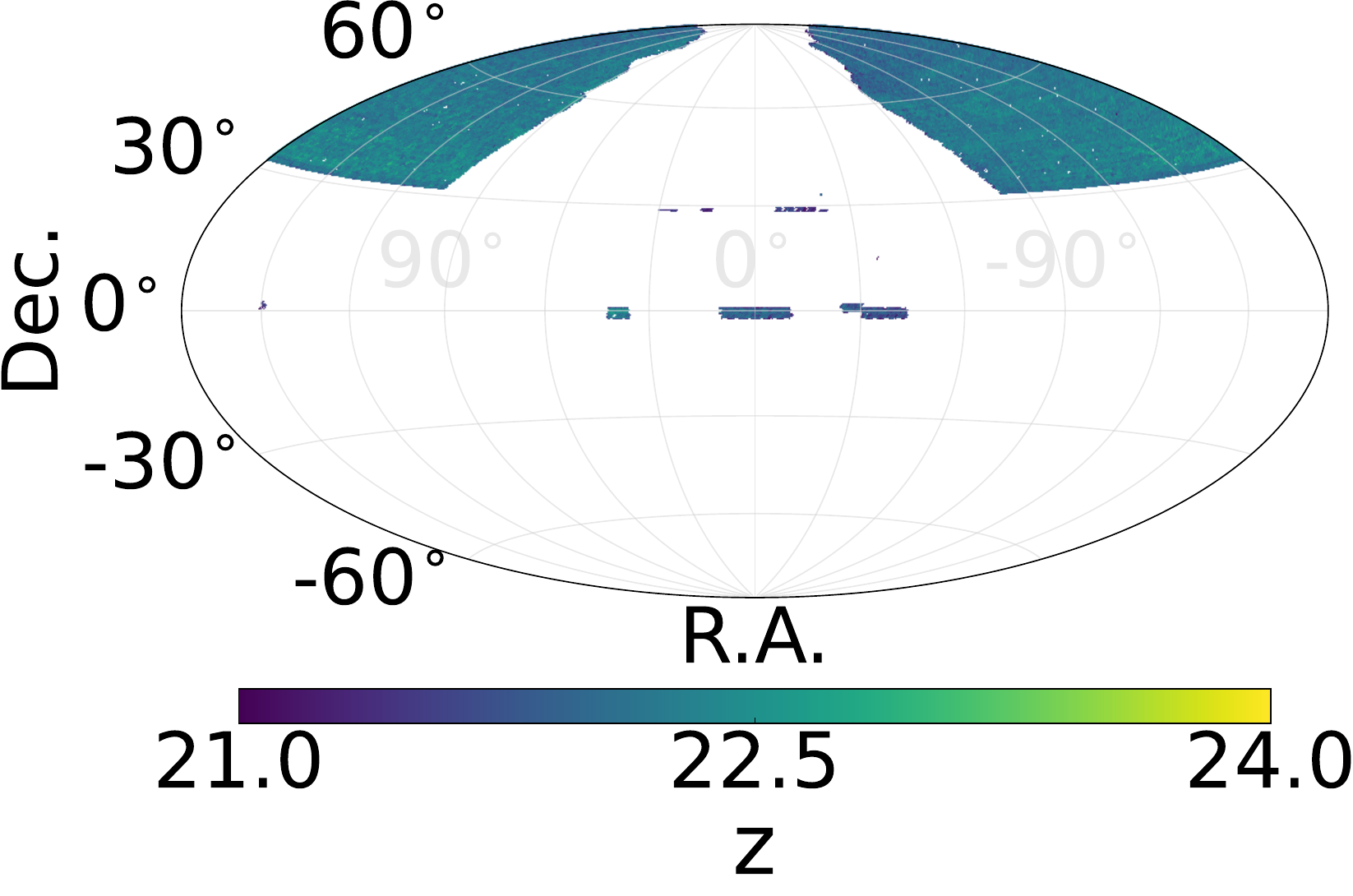}
    \includegraphics[width=0.19\linewidth]{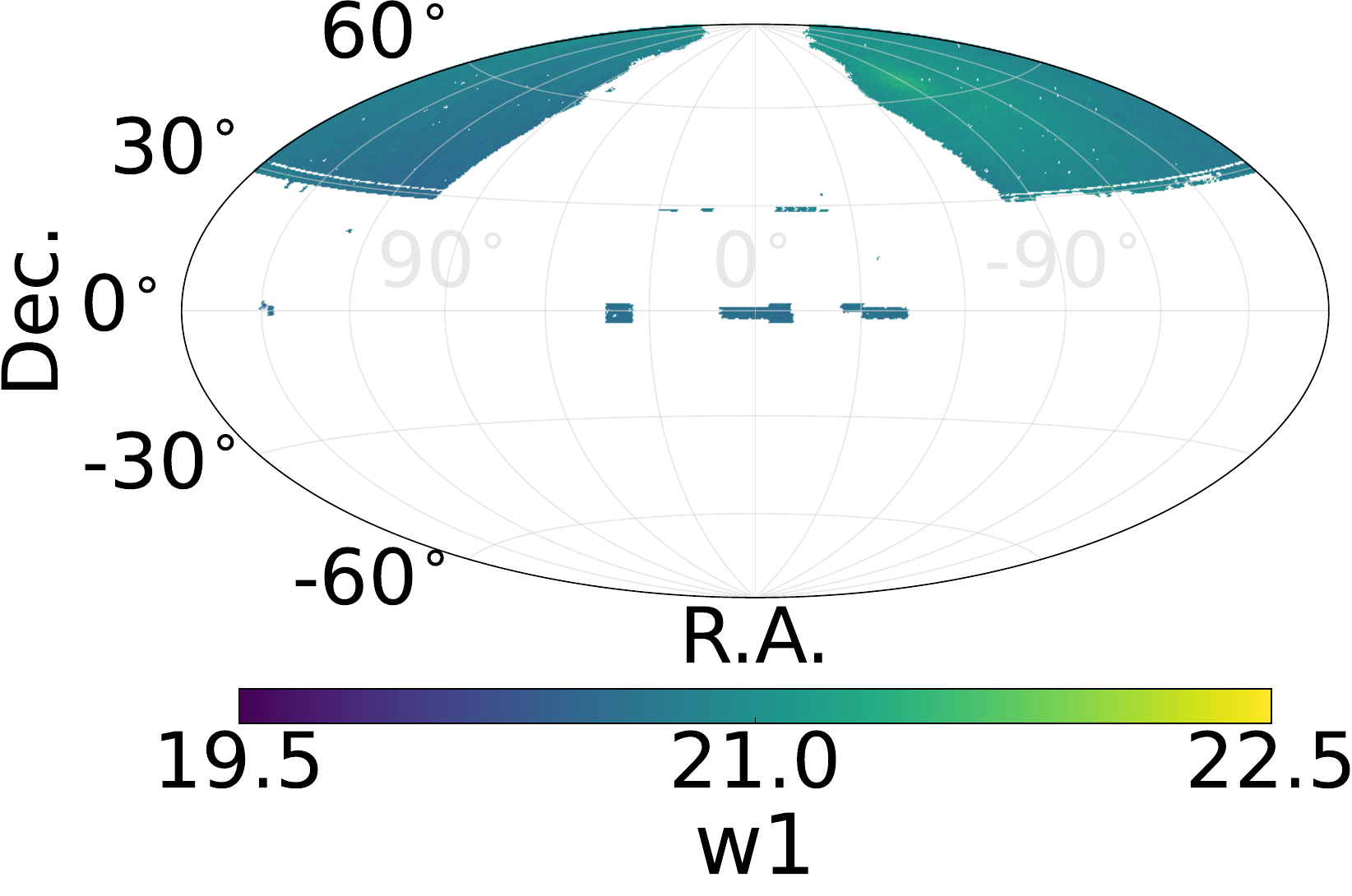}
    \caption{Depth maps showing the limiting magnitude for the Legacy Surveys DR10 south (top panels) and DR9 north (lower panels) in different filter bands. The maps present the full coverage in each band independently rather than combining the footprints as in Figure \ref{fig:footprint}. The limiting magnitude is defined as the brightness of a galaxy whose flux has a signal-to-noise ratio of 10.}
    \label{fig:depthmaps}
\end{figure*}

To minimize the number of spurious sources, it is necessary to know whether the local survey depth is sufficiently deep. This information is stored in depth maps. The sky position-dependent depth was estimated via a parametric model \citep{Rykoff2015aa} that uses the galaxy magnitudes and uncertainties in the LS. The limiting magnitude is defined by that of a galaxy with a detection confidence of 10$\sigma$, that is, its measured flux has a signal-to-noise ratio of 10. These limits vary across the sky and for different filter bands. Depth maps store this information in the HEALPix format \citep{Gorski2005aa}. We used a spatial resolution of \texttt{NSIDE}=4096, which corresponds to an area of $~\sim1\arcmin\times1\arcmin$ per pixel. If a heal pixel contained an insufficient number of galaxies for the model to converge, the depth was approximated by recursively expanding out to the next largest pixel in the nested scheme until it did.

One depth map was created for each filter band and for each of the LS DR10 south and LS DR9 north survey parts (see Figure \ref{fig:depthmaps}). The spatial coverage for the different filter bands is similar but not identical. Therefore, we set all heal pixels to empty which were also empty in at least one of the depth maps used in each \eromapper run. The resulting $z$-band depth map for the $grz$ filter band combination is shown in Figure \ref{fig:footprint}. The top panel refers to the LS DR10 south and the bottom panel refers to the LS DR9 north.

\begin{figure}
    \centering
    \includegraphics[width=\linewidth]{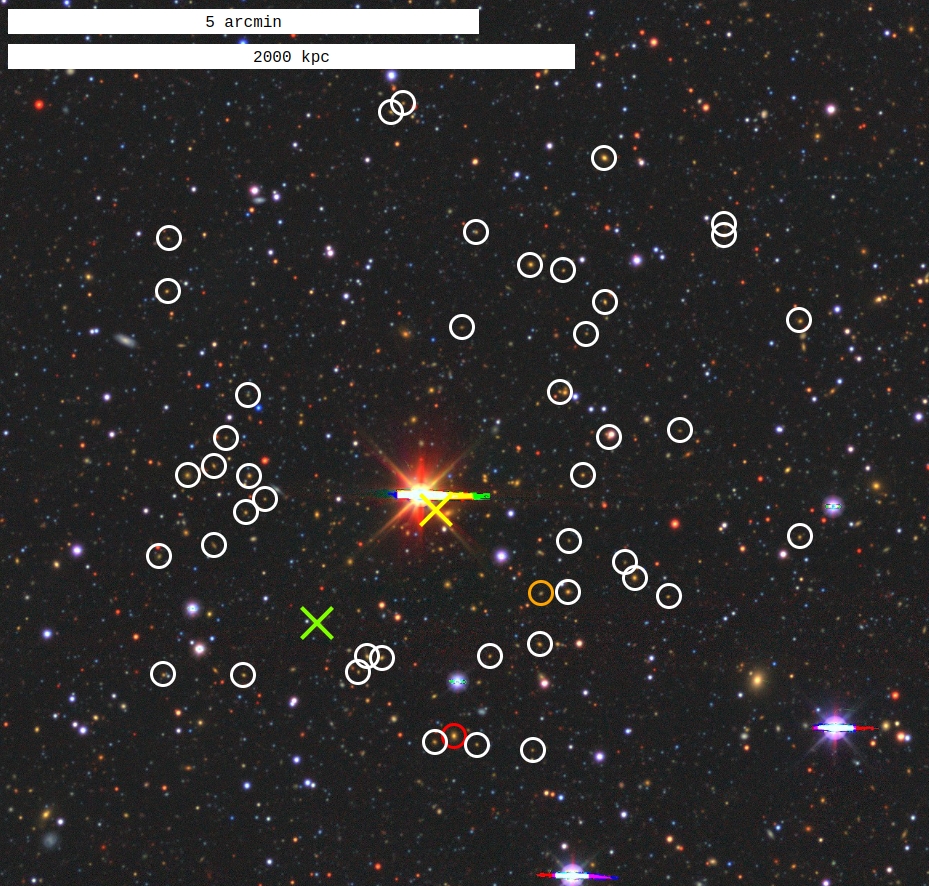}
    \caption{Example of a cluster 1eRASS J010123.5-764801 located in a masked area of the footprint map. The saturated foreground star near the X-ray detection (yellow cross) was responsible for the mask. Cluster member galaxies are marked by circles (see Figure \ref{fig:zspec_example} for further descriptions of the labels). Only the members in the outskirts of the cluster were identified. North is up, east is left.}
    \label{fig:outfootprint_example}
\end{figure}

Saturated sources leave small holes in the depth maps. If a cluster center is located within such a hole, we describe it as formally outside the footprint and flagged it in the catalog as {\tt IN\_FOOTPRINT=False}. Part of the cluster can still be detected when the cluster radius $R_\lambda$ is larger than the size of the masked region. This affects 363 \erass clusters. We show one example in Figure \ref{fig:outfootprint_example}. A bright foreground star overlaps with a cluster in the background. The galaxies in projected proximity have flagged photometry in the LS. They were rejected by \eromapper (see Section \ref{sec:rm}). However, cluster members in the outskirts of the cluster were still detected and marked by the circles.
As expected, the masking fraction is high with {\tt MASKFRAC=0.51}, and therefore, the richness $\lambda$ was upscaled by a factor of {\tt SCALEVAL=2.03}. We consider the photometric redshift $z_\lambda=0.400\pm0.012$ reliable because 45 member galaxies were detected despite the high masking fraction. For this reason, we kept these clusters in our sample.

\begin{figure}
    \includegraphics[width=\linewidth]{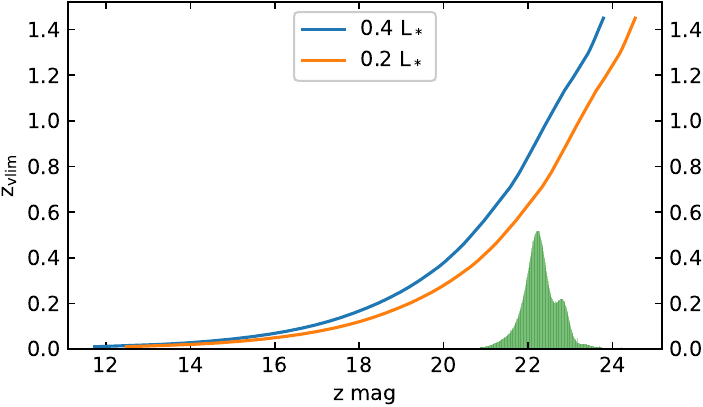}
    \caption{Conversion from limiting $z$-band magnitude to limiting redshift $z_{\rm vlim}$ for galaxies with a luminosity $L=0.2$\,L$_*$ (orange) and $L=0.4$\,L$_*$ (blue). Restricting the sample to brighter members allows for probing higher redshifts. For reference, the green histogram shows the distribution of limiting magnitudes in the $z$ band for the combined full LS DR10 south and LS DR9 north areas.}
    \label{fig:zmag_vs_zvlim}
\end{figure}

The limiting magnitude in our reference band, the $z$ band, can be converted to a limiting redshift $z_{\rm vlim}$ for non-empty heal pixels. It is defined as the redshift, at which a member galaxy at the low-luminosity limit $L_{\rm thresh}$ can still be detected with a 10$\sigma$ significance. To calculate this limit, we predicted the brightness in the DECam $z$ band of a galaxy with luminosity $L_{\rm thresh}$ for all redshifts $0.01<z<1.45$ using the tool {\texttt EzGal} \citep{Mancone2013aa}. Thereby, we assumed that it behaves according to the prediction for a passively evolving single stellar population \citep{Bruzual2003aa}. We set the low-luminosity limit to $L_{\rm thresh}=0.2$\,L$_*$ for the \eromapper runs using the $grz$ and $griz$ filter bands, and $L_{\rm thresh}=0.4$\,L$_*$ for the runs that include the $w1$ band (see Section \ref{sec:merging} and Figure \ref{fig:zmag_vs_zvlim}). Hereby, L$_*$ is the luminosity at the break of the Schechter function \citep{Schechter1976aa}. The magnitude of an L$_*$ galaxy was normalized to 17.85 $i$ mag in the Sloan photometric system at redshift $z=0.2$, consistently to \cite{Rykoff2016aa}.

Our choice to base the largest reliable redshift purely on the $z$ band deviates from \cite{Rykoff2016aa} who used a combination of the limiting magnitudes in every filter band to calculate a maximum redshift $z_{\rm max}$. We have shown that the cluster number density remains unbiased for clusters $z_\lambda<z_{\rm vlim}$ (see Figure \ref{fig:richness_cludens}). As $z_{\rm max} < z_{\rm vlim}$, this allowed us to probe even higher redshifts. We emphasize that we did not discard clusters with higher than the limiting redshift $z_\lambda>z_{\rm vlim}$. They were kept in the catalogs with a flag {\tt IN\_ZVLIM=False}.

\section{Red sequence models} \label{sec:redseq}

\begin{figure*}
    \includegraphics[width=\linewidth]{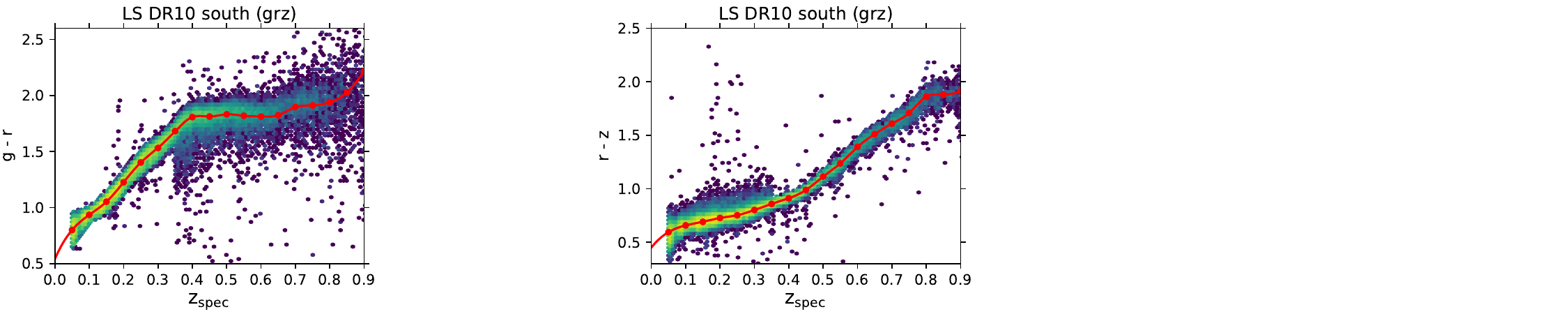}\\
    \includegraphics[width=\linewidth]{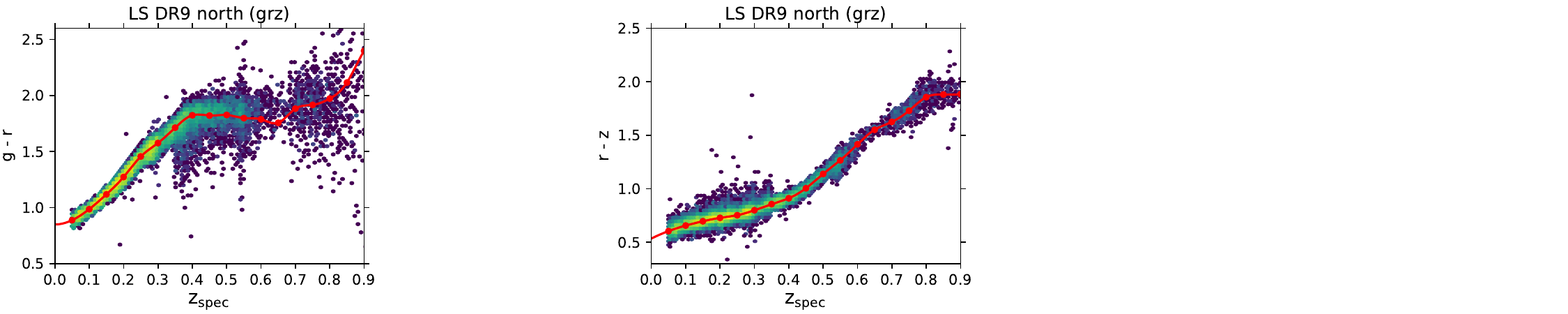}\\
    \includegraphics[width=\linewidth]{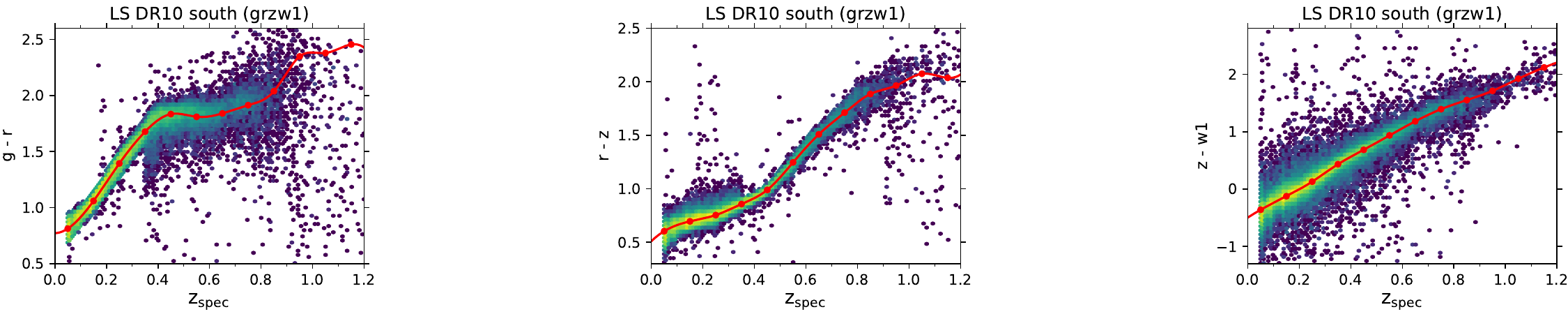}\\
    \includegraphics[width=\linewidth]{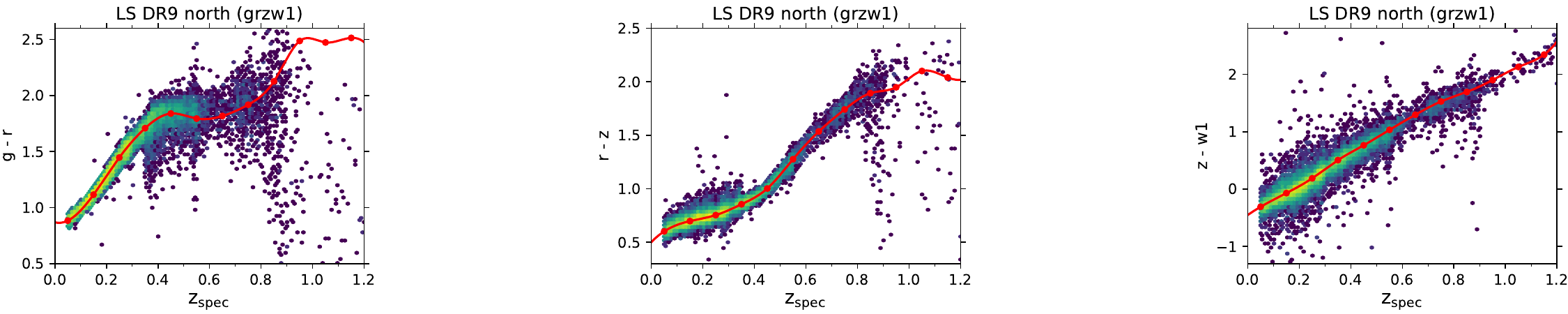}\\
    \includegraphics[width=\linewidth]{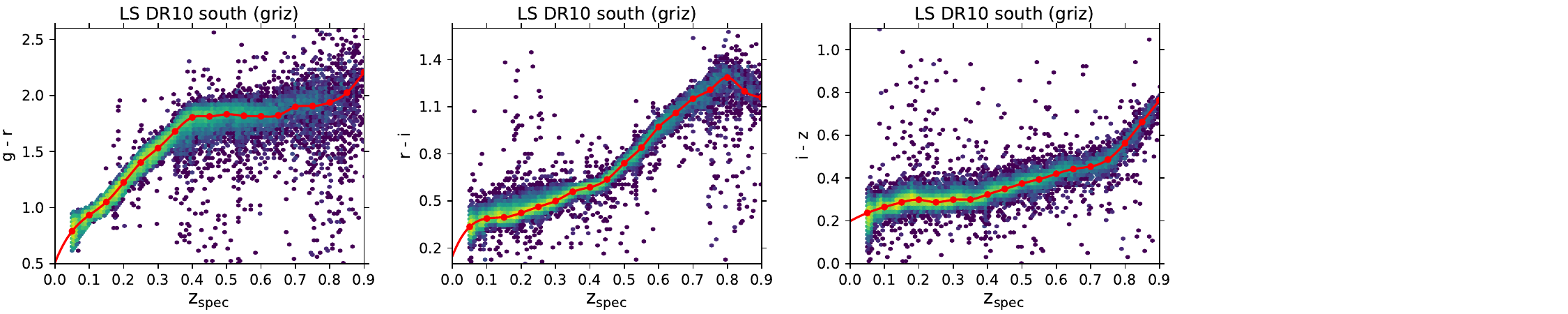}\\
    \includegraphics[width=\linewidth]{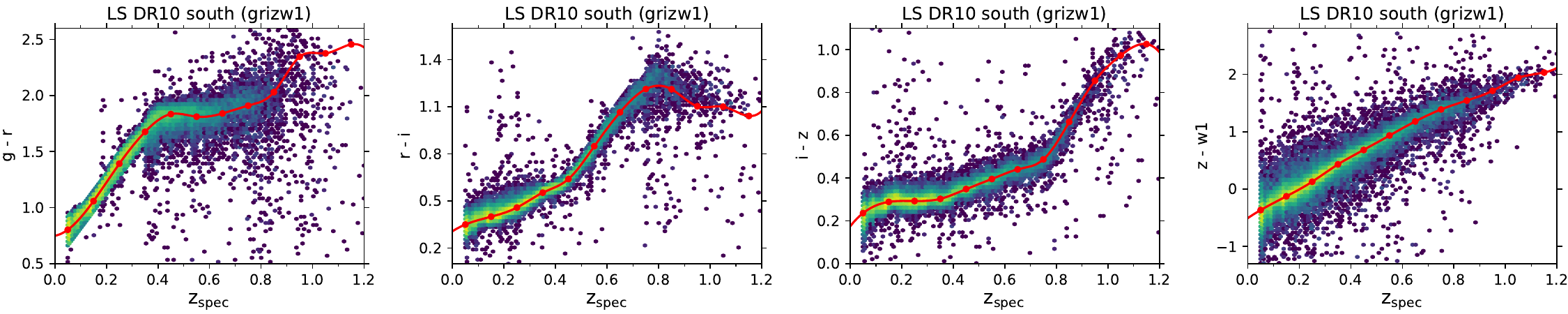}
    \caption{Red-sequence models for the six different calibrations (rows). Each filter band combination and LS parts are labeled in the titles of the subpanels. The combination of $n$ filter bands results in red-sequence models for $n-1$ colors (columns). The red-sequence models are shown by the red splines and red points mark the spline nodes. The training galaxies are color-coded according to their point density. Their total numbers are, after clipping, $\sim$93\,000 for LS DR10 ($grz$ and $grzw1$), $\sim$72\,000 for LS DR10 ($griz$ and $grizw1$), and $\sim$36\,000 for LS DR9 north ($grz$ and $grzw1$).
    \label{fig:redseq}}
\end{figure*}

Photometric redshifts were calculated by comparing the galaxy colors to a spline-based red-sequence model \citep{Rykoff2014aa}. The color--redshift relations for the red-sequence model must be calibrated with the colors of red galaxies with known spectroscopic redshifts. For this, we utilized our spectroscopic galaxy redshift compilation (see Section \ref{sec:spec_compilation} and Appendix \ref{sec:spec_compilation_refs}). As explained in Section \ref{sec:merging}, we ran \eromapper six times using different filter band combinations. The red-sequence model was calibrated for each run independently, corresponding to the six rows in Figure \ref{fig:redseq}.

We began the model fitting procedure by assuming an initial red-sequence model derived from the predictions of a passively evolving single stellar population \citep{Bruzual2003aa}. We then selected red galaxies whose colors are consistent with the initial model. Figure \ref{fig:redseq} shows the color-coded point density of the final sample for each photometric color.

The spline nodes of the red-sequence model (red points) were then fitted to the colors in redshift bins. Outliers were iteratively clipped. We chose a small node spacing of $\Delta z_{\rm spec}=0.05$ for the $grz$ and $griz$ filter band combinations to capture high-frequency color variations. If not fitted well, systematic residuals lead to a higher local redshift bias (see Section \ref{sec:redshift_bias}), which in turn leads to local clumping of the cluster number densities (see Figure \ref{fig:cludens}). A larger node spacing of $\Delta z_{\rm spec}=0.10$ was applied for the $grizw1$ and $grzw1$ filter band combinations. This was better suited for the red-sequence model calibration for the high-redshift ($z_\lambda>0.8$) \eromapper runs because the number of training galaxies became low.

The minimum redshift of the model is $z_{\rm spec}=0.05$. The maximum redshift is $z_{\rm spec}=0.9$ for the $grz$ and $griz$ filter band combinations and $z_{\rm spec}=1.2$ for the $grzw1$ and $grizw1$ combinations. Including the near-infrared $w1$ band helps to constrain better cluster redshifts at $z>0.8$ because the $z-w1$ colors have a non-negligible slope. However, it does not necessarily help for low$-z$ cluster redshifts because of the relatively large scatter of the $z-w1$ colors. For this reason and the more uncertain photometry in the $w1$ band (see Section \ref{subsec:data_legacy}), we did exclude it for the low$-z$ \eromapper runs.

As mentioned previously, photometric redshifts were calculated by comparing the galaxy colors to the red-sequence model. Thereby, all filter bands were taken into account at all redshifts. Figure \ref{fig:redseq} confirms that different filter bands are suited better or worse to constrain specific redshift intervals. The higher the slope of the color--redshift relation, the better the color constrains the redshift. Hence, each color was weighted by the local slope of its color--redshift relation.

\clearpage

\section{Details about the spectroscopic galaxy redshift compilation} \label{sec:spec_compilation_refs}

\begin{figure}[h]
    \includegraphics[width=\linewidth]{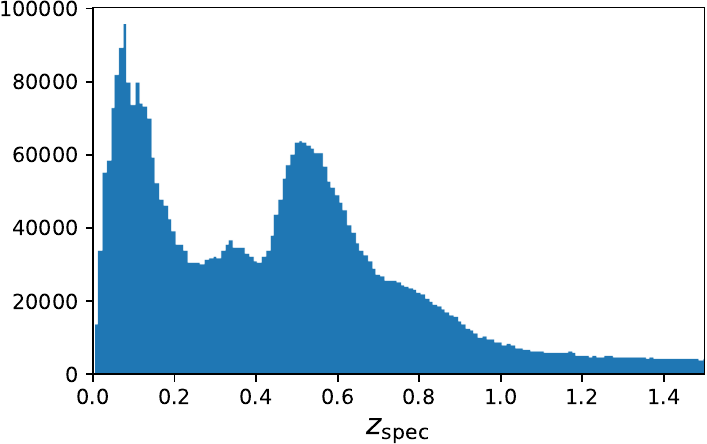}
    \caption{Redshift distribution of the spectroscopic galaxy redshift compilation.}
    \label{fig:speczcomp_histo}
\end{figure}

The spectroscopic galaxy compilation is used to calibrate the red-sequence models and to calculate spectroscopic cluster redshifts and velocity dispersions.
Figure \ref{fig:speczcomp_histo} shows the distribution of all redshifts.
Table \ref{tab:speczpost} lists the individual references in column (1), the number of unique galaxy redshifts in column (2), and the applied selection criteria in column (3). Please note that the \cite{Zou2019aa} source is a compilation of several spectroscopic surveys: 2dFGRS \citep{Colless2001aa}, 2SLAQ \citep{Cannon2006aa}, 6dFGS \citep{Jones2004aa, Jones2009aa}, CFRS \citep{Lilly1995aa}, CNOC2 \citep{Yee2000aa}, DEEP2 \citep{Davis2003aa, Newman2013aa}, SDSS DR14 \citep{Abolfathi2018aa}, VIPERS \citep{Garilli2014aa, Guzzo2014aa}, VVDS \citep{LeFevre2005aa,Garilli2008aa}, WiggleZ \citep{Drinkwater2010aa,Parkinson2012aa}, and zCOSMOS \citep{Lilly2007aa}.

\onecolumn

\begin{longtable}{lrl}

\caption{\label{tab:speczpost} Spectroscopic galaxy compilation.}\\
\hline\hline
Source & \# Galaxies & Query \\
\hline
\endfirsthead
\caption{continued.}\\
\hline
Source & \# Galaxies & Query \\
\hline
\endhead
\hline
\endfoot
\hline
\endlastfoot

\cite{Abolfathi2018aa} & 900 & in\_DESComplilation and des\_flags >= 3 \\
\cite{Aguado-Barahona2019aa} & 21 &  \\
\cite{Agulli2016aa} & 430 &  \\
\cite{Ahumada2020aa} & 450\,130 & zwarning == 0 and z > z\_err and z\_err > 0 \\
\cite{Alam2015aa} & 933\,644 & zWarning == 0 \\
\cite{Almeida2023aa} & 5034 & zwarning == 0 and z > z\_err and z\_err > 0 \\
&& and survey == "BHM" and plug\_ra != 0 \\
\cite{Amodeo2018aa} & 10 &  \\
\cite{Bacon2022aa} & 37 &  \\
\cite{Balogh2014aa} & 29 & q\_z >= 3 and Class == "p  " \\
\cite{Balogh2020aa} & 507 & member == 1 and Redshift\_Quality > 2 and Spec\_Flag < 2 \\
Balzer et al. (in prep) & 154 &  \\
\cite{Bayliss2011aa} & 26 &  \\
\cite{Bayliss2016aa} & 1655 &  \\
\cite{Bayliss2017aa} & 1578 &  \\
\cite{Bezanson2018aa} & 99 &  \\
\cite{Blake2016aa} & 53\,534 & in\_DESComplilation and des\_flags >= 3 \\
\cite{Bradshaw2013aa}; &&  \\
\cite{McLure2013aa} & 1296 &  \\
\cite{Caminha2017aa} & 3448 & q\_z > 2 \\
\cite{Caminha2019aa} & 451 & z > 0 and (q\_z == 2 or q\_z == 3) \\
\cite{Carrasco2017aa} & 730 &  \\
\cite{Castagne2012aa} & 92 &  \\
\cite{Castignani2020aa} & 5 &  \\
\cite{Cava2009aa} & 2686 & Mm == 1 \\
\cite{Childress2017aa} & 11\,264 & flag !=6 \\
\cite{Clerc2020aa} & 27\,783 & z < 0.6 and screen\_ismember\_w > 0.5 \\
\cite{Coe2012aa} & 165 & Mm == 1 \\
\cite{Coil2011aa}; && \\
\cite{Cool2013aa} & 168\,546 & ZQUALITY >= 3 and not a star \\
\cite{Colless2001aa} & 165\,911 & q\_z >= 3 \\
\cite{Connelly2012aa} & 4566 &  \\
\cite{Connor2019aa} & 26 & q\_z == 3 \\
\cite{Cooper2012aa} & 4366 & Q >= 3 \\
\cite{Crawford2011aa} & 585 & q\_z >= 2 \\
\cite{Czoske2001aa} & 382 & q\_z == "A" or q\_z == "B" \\
\cite{Davis2003aa}; && \\
\cite{Davis2007aa} & 4 & in\_DESComplilation and des\_flags >= 3 \\
\cite{Davis2003aa}; && \\
\cite{Newman2013aa} & 20\,022 & ZQUALITY >= 3 and not a star \\
\cite{Dawson2013ab} & 688 &  \\
\cite{Deger2018aa} & 167 & Set == "C" or Set == "G" \\
\cite{Demarco2005aa} & 24 &  \\
\cite{Demarco2007aa} & 33 &  \\
\cite{Demarco2010ab} & 30 &  \\
\cite{Deshev2017aa} & 189 & Memb == 1 \\
\cite{Dressler1999aa} & 144 & Mm == "c" \\
\cite{Drinkwater2010aa} & 21\,656 & Qop>=4 \\
\cite{Driver2011aa} & 48\,875 & in\_DESComplilation and des\_flags >= 3 \\
\cite{Ebeling2014aa} & 1589 &  \\
\cite{Ebeling2017aa} & 82 &  \\
\cite{Eisenhardt2008aa} & 76 &  \\
\cite{Ferrari2003aa} & 105 & q\_cz == 1 \\
\cite{Foex2017aa} & 738 & EZflag > 0 and zspec > 0 \\
\cite{Foley2011aa} & 12 &  \\
\cite{Fossati2019aa} & 71 & Redshift > 0 \\
\cite{LeFevre2004aa}; && \\
\cite{Garilli2008aa} & 359 & in\_DESComplilation and des\_flags >= 3 \\
\cite{Garilli2014aa} & 44\,103 & 2.6<flag<4.6, 22.6<flag<24.6 (secondary),\\
&& 12.6<flag<14.6 (broad-line AGN),\\
&& 212.6<flag<214.6 (broad-line AGN, secondary) \\
\cite{Garilli2021aa} & 146 & zflg == 3 or zflg == 4 \\
\cite{Geha2017aa} & 15\,313 & in\_DESComplilation and des\_flags >= 3 \\
\cite{George2011aa} & 2033 &  \\
\cite{Girardi2011aa} & 126 &  \\
\cite{Golovich2019aa} & 3061 &  \\
\cite{Gomez2012aa} & 40 &  \\
\cite{Gonzalez2019aa} & 1 &  \\
\cite{Gschwend2018aa} & 7156 & in\_DESComplilation and des\_flags >= 3 \\
\cite{Guennou2014aa} & 69 &  \\
\cite{Guglielmo2018aa} & 1888 & q\_z == 400 and DRr200\_1 > 0 \\
\cite{Haines2021aa} & 10\,895 & CL == 1 \\
\cite{Halliday2004aa} & 72 &  \\
\cite{Hansen2002aa} & 36 & Note == "b" and u\_z != ":" and u\_z != "::" \\
\cite{Hilton2020aa} & 726 & redshiftType == "spec" \\
\cite{Huchra2012aa} & 13\,375 & n\_cz != "S" and n\_cz != "6" \\
\cite{Hwang2014aa} & 4142 &  \\
\cite{Inami2017aa} & 346 & Confid > 1 \\
\cite{Jones2009aa} & 93\,437 & q\_z == 4 \\
\cite{Jorgensen2014aa} & 10 & Mm == 1 \\
\cite{Jorgensen2017aa} & 71 &  \\
\cite{Karman2015aa} & 31 &  \\
\cite{Khabibullina2009aa} & 1478 &  \\
\cite{Khullar2018aa} & 34 &  \\
\cite{Kirk2015aa} & 107 & Memb == "V" \\
\cite{Kluge2020aa} & 90 &  \\
\cite{Koranyi2002aa} & 225 &  \\
\cite{LeFevre2013aa} & 32\,288 & flag=3, 4, 22, 24 (2xs are secondary),\\
&& 13, 14 (broad-line AGN),\\
&& 213, or 214 (broad-line AGN, secondary) \\
\cite{Lee2019aa} & 50 & flag != "star" and Cluster\_ID != "-99\,999" \\
\cite{Lemze2013aa} & 1598 &  \\
\cite{Lidman2016aa} & 3066 & in\_DESComplilation and des\_flags >= 3 \\
\cite{Lidman2020aa} & 894 & Object\_types.str.contains("luster") \\
\cite{Lilly1995aa} & 381 & Class >= 3 \\
\cite{Lilly2009aa} & 12\,870 & Flag 4s, 3s, 2.5, 2.4, 1.5, 9.5, 9.4, 9.3 are considered secure \\
\cite{Liske2015aa} & 41\,261 & NQ>=4 \\
\cite{Liu2012aa} & 6 &  \\
\cite{Liu2018aa} & 47 & Mm == "Y" \\
\cite{Mao2012aa} & 524 & in\_DESComplilation and des\_flags >= 3 \\
\cite{Masters2017aa,Masters2019aa} & 3092 &  \\
\cite{Maturi2019aa} & 7470 & prob > 0.8 and has\_spec\_z \\
\cite{McClintock2019aa} & 654 & Z\_SPEC > 0 and P > 0.9 \\
\cite{McDonald2016aa} & 29 &  \\
\cite{McLure2017aa} & 85 & q\_zsp > 2 \\
\cite{Mirkazemi2015aa} & 3 &  \\
\cite{Momcheva2016aa} & 31\,043 & in\_DESComplilation and des\_flags >= 3 \\
\cite{Moran2005aa} & 87 &  \\
\cite{Moretti2017aa} & 7014 & Memb >= 1 \\
\cite{Morris2007aa} & 279 &  \\
\cite{Muzzin2009aa} & 25 &  \\
\cite{Muzzin2012aa} & 217 & in\_DESComplilation and des\_flags >= 3 \\
\cite{Nanayakkara2016aa} & 40 & in\_DESComplilation and des\_flags >= 3 \\
\cite{Nastasi2014aa} & 108 &  \\
\cite{NiloCastellon2014aa} & 79 &  \\
\cite{Oguri2018aa} & 753 & z\_spec > 0 \\
\cite{Olsen2003aa} & 128 & Memb == "m" \\
\cite{Olsen2005aa} & 266 &  \\
\cite{Owen1995aa} & 65 &  \\
\cite{Owers2011aa} & 2350 &  \\
\cite{Parkinson2012aa} & 8 & in\_DESComplilation and des\_flags >= 3 \\
\cite{Pierre1997aa} & 4 &  \\
\cite{Pierre2016aa} & 51 &  \\
\cite{Pimbblet2006aa} & 5015 &  \\
\cite{Postman2001aa} & 12 & (f\_\_PLO2001\_ == " *" or f\_\_PLO2001\_ == "* " or\\
&& f\_\_PLO2001\_ == "*" or f\_\_PLO2001\_ == "**")\\
&& and q\_z >= 2 \\
\cite{Pranger2014aa} & 139 &  \\
\cite{Psychogyios2020aa} & 23 & m == 1 \\
\cite{Rescigno2020aa} & 27 & QF > 1 \\
\cite{Rest2014aa}; && \\
\cite{Scolnic2014aa}; && \\
\cite{Kaiser2010aa} & 1128 & in\_DESComplilation and des\_flags >= 3 \\
\cite{Ricci2018aa} & 111 & flag == "s" \\
\cite{Richard2021aa} & 1271 & zconf > 1 \\
\cite{Rines2000aa} & 153 &  \\
\cite{Rines2016aa} & 2603 & Mm == 1 \\
\cite{Rines2018aa} & 2929 & Mm == 1 \\
\cite{Robotham2011aa} & 32\,361 & GroupID != 0 \\
\cite{Rudnick2017aa} & 104 & Env == "cluster" or Env == "group" \\
\cite{Ruel2014aa} & 948 &  \\
\cite{Runge2018aa} & 126 & zspec > 0 \\
\cite{Rykoff2012aa} & 9353 & zsp > 0 \\
\cite{Rykoff2016aa} & 28\,950 & zspec > 0 and PMem > 0.9 \\
\cite{Schirmer2011aa} & 36 & Mm == "b" \\
\cite{Shectman1996aa} & 8862 &  \\
\cite{Sifon2013aa} & 219 &  \\
\cite{Silverman2015aa} & 450 & flag >= 3 \\
\cite{Skelton2014aa}; && \\
\cite{Momcheva2016aa} & 6644 & z\_best\_s = 2 \\
\cite{Sluse2019aa} & 59 & Type == "E" \\
\cite{Smith2000aa} & 94 &  \\
\cite{Sohn2018aa} & 2013 & Mem == "Y" \\
\cite{Sohn2018ab} & 762 &  \\
\cite{Sohn2019aa} & 152 & Mem == "Y" and BCG == "Y" \\
\cite{Sohn2019ab} & 877 &  \\
\cite{Sohn2020aa} & 14\,388 &  \\
\cite{Sohn2020ab} & 131 &  \\
\cite{Song2017aa} & 48 & Mm == 1 \\
\cite{Stanford2014aa} & 171 &  \\
\cite{Stanford2021aa} & 568 &  \\
\cite{Stott2008aa} & 52 &  \\
\cite{Straatman2018aa} & 1195 & Use == 1 \\
\cite{Strait2018aa} & 15 & QF == "3" or QF == "4" \\
\cite{Streblyanska2019aa} & 17 &  \\
\cite{Sullivan2011aa} & 243 & in\_DESComplilation and des\_flags >= 3 \\
\cite{Tanaka2009aa} & 89 & q\_z == 0 \\
\cite{Tasca2017aa} & 139 & in\_DESComplilation and des\_flags >= 3 \\
\cite{Tran2007aa} & 41 & q\_z > 1 \\
\cite{Treu2015aa} & 603 & in\_DESComplilation and des\_flags >= 3 \\
\cite{Verdugo2008aa} & 143 & Memb != "field" \\
\cite{Wen2018aa} & 1927 &  \\
\cite{Willis2013aa} & 21 &  \\
\cite{Wilson2016aa} & 3005 &  \\
\cite{Yang2018aa} & 2993 &  \\
\cite{Zaznobin2020aa} & 27 &  \\
\cite{Zeimann2013aa} & 7 & abs(\_\_z\_ - z) < 0.001 \\
\cite{Zou2019aa} (compilation) & 2\,470\,093 &  \\
&&\\
\hline
&&\\
total & 4\,882\,137 &  \\
    \hline\hline
\end{longtable}

\clearpage

\twocolumn

\section{Catalog column descriptions} \label{sec:column_descriptions}

Table \ref{tab:coldesc} gives an overview of the column names, data types, and descriptions of the columns in the \erass identified cluster catalog. The data are available in electronic form at the CDS. More X-ray properties of the clusters can be found in \cite{Bulbul2023}.

\begin{table*}[b]
  \caption{Column names, data types, and descriptions of the \erass identified cluster catalog.}
  \centering
  \small
  \begin{tabular}{lll}
  \hline\hline
     Column Name & Type & Description \\
     \hline
    {\tt NAME}                    & 32A  & Cluster name \\
    {\tt DETUID}                  & 32A  & Unique eSASS detection ID \\
    {\tt RA}                      & E4   & Right Ascension J2000 of X-ray prior \\
    {\tt DEC}                     & E5   & Declination J2000 of X-ray prior \\
    {\tt RADEC\_ERR}              & E2   & Uncertainty of RA, DEC (arcsec) \\
    {\tt RA\_OPT}                 & E4   & Right Ascension J2000 of optical center \\
    {\tt DEC\_OPT}                & E5   & Declination J2000 of optical center \\
    {\tt RA\_BCG}                 & E4   & Right Ascension J2000 of the brightest member \\
    {\tt DEC\_BCG}                & E5   & Declination J2000 of the brightest member \\
    {\tt REFMAG\_BCG}             & E3   & Galactic extinction-corrected $z$-band AB magnitude of the brightest member \\
    {\tt BEST\_Z}                 & E5   & Best available cluster redshift \\
    {\tt BEST\_Z\_ERR}            & E5   & Error on best available cluster redshift \\
    {\tt BEST\_Z\_TYPE}           & 16A  & Type of best available cluster redshift \\
    {\tt Z\_LAMBDA}               & E5   & Uncorrected cluster photometric redshift \\
    {\tt Z\_LAMBDA\_ERR}          & E5   & Uncorrected cluster photometric redshift uncertainty \\
    {\tt Z\_LAMBDA\_CORR}         & E5   & Bias-corrected cluster photometric redshift \\
    {\tt Z\_LAMBDA\_CORR\_UERR}   & E5   & Bias-corrected cluster photometric redshift upper error \\
    {\tt Z\_LAMBDA\_CORR\_LERR}   & E5   & Bias-corrected cluster photometric redshift lower error \\
    {\tt Z\_LAMBDA\_SECOND}       & E5   & Uncorrected foreground/background cluster photometric redshift \\
    {\tt Z\_LAMBDA\_SECOND\_ERR}  & E5   & Uncorrected foreground/background cluster photometric redshift uncertainty \\
    {\tt LIT\_Z}                  & E5   & Literature cluster redshift \\
    {\tt LIT\_Z\_ERR}             & E5   & Literature cluster redshift error \\
    {\tt LIT\_Z\_SRC}             & 32A  & Reference for LIT\_Z \\
    {\tt SPEC\_Z\_BOOT}           & E5   & Bootstrap estimate of biweight spectroscopic redshift \\
    {\tt SPEC\_Z\_BOOT\_ERR}      & E5   & Bootstrap error on biweight spectroscopic redshift \\
    {\tt N\_MEMBERS}              & J    & Number of members used in biweight redshift and velocity dispersion estimate \\
    {\tt CG\_SPEC\_Z}             & E5   & Spectroscopic redshift of the central galaxy \\
    {\tt CG\_SPEC\_Z\_ERR}        & E5   & Error on spectroscopic redshift of the central galaxy \\
    {\tt BCG\_SPEC\_Z}            & E5   & Spectroscopic redshift of the brightest member \\
    {\tt VDISP\_BOOT}             & E2   & Bootstrap estimate of velocity dispersion (km\,s$^{-1}$) \\
    {\tt VDISP\_BOOT\_ERR}        & E2   & Error on bootstrap estimate of velocity dispersion (km\,s$^{-1}$) \\
    {\tt VDISP\_FLAG\_BOOT}       & E4   & Mean vdisp\_flag for bootstrap estimate \\
    {\tt VDISP\_TYPE}             & 8A   & Estimate used to determine vdisp. Gapper or Biweight depending on number of members \\
    {\tt LAMBDA\_NORM}            & E2   & Richness normalized to the definition of the $grz$ run \\
    {\tt LAMBDA\_NORM\_ERR}       & E2   & Richness error normalized to the definition of the $grz$ run \\
    {\tt LAMBDA\_OPT\_NORM}       & E2   & Richness (optical center) normalized to the definition of the $grz$ run \\
    {\tt LAMBDA\_OPT\_NORM\_ERR}  & E2   & Richness (optical center) error normalized to the definition of the $grz$ run \\
    {\tt SCALEVAL}                & E5   & Richness scale factor Eq. 2 of \cite{Rykoff2016aa} \\
    {\tt MASKFRAC}                & E3   & Fraction of cluster area which is masked \\
    {\tt RUN}                     & 128A & RedMaPPer calibration: survey\_bands\_refband\_iteration \\
    {\tt LMAX}                    & E4   & Optical maximum likelihood \\
    {\tt NHI}                     & E5   & HI Column Desnsity (10$^{21}$ cm$^{-2}$) from HI4PI. Nside=1024 \\
    {\tt EXT\_LIKE}               & E5   & Extent likelihood of the X-ray source \\
    {\tt CR500X\_NH0}             & D12  & X-ray count rate \\
    {\tt PCONT}                   & E4   & Probability that the cluster is a contaminant \\
    {\tt SHARED\_MEMBERS}         & L    & There is another cluster with $\geq70\%$ of the same members and higher LMAX \\
    {\tt IN\_FOOTPRINT}           & L    & Cluster is within the optical survey footprint \\
    {\tt LIMMAG\_G}               & E2   & Limiting $g$ galaxy magnitude from \cite{Rykoff2015aa} depth (map) model \\
    {\tt LIMMAG\_R}               & E2   & Limiting $r$ galaxy magnitude from \cite{Rykoff2015aa} depth (map) model \\
    {\tt LIMMAG\_I}               & E2   & Limiting $i$ galaxy magnitude from \cite{Rykoff2015aa} depth (map) model \\
    {\tt LIMMAG\_Z}               & E2   & Limiting $z$ galaxy magnitude from \cite{Rykoff2015aa} depth (map) model \\
    {\tt LIMMAG\_W1}              & E2   & Limiting $w1$ galaxy magnitude from \cite{Rykoff2015aa} depth (map) model \\
    {\tt IN\_ZVLIM}               & L    & Photometric redshift is smaller than the limiting optical survey redshift ZVLIM\_02 or ZVLIM\_04 \\
    {\tt ZVLIM\_02}               & E3   & Maximum observable galaxy redshift at 0.2\,L$_*$ volume limit \\
    {\tt ZVLIM\_04}               & E3   & Maximum observable galaxy redshift at 0.4\,L$_*$ volume limit \\
    {\tt GAIA\_10MAG\_30\_ARCSEC} & L    & A star in the \textit{Gaia} eDR3 catalog brighter than 10 $G$ mag is closer than 30 arcsec to the X-ray source \\
    {\tt NGC\_30\_ARCSEC}         & L    & A galaxy in the NGC catalog is closer than 30 arcsec to the X-ray source \\
    {\tt HECATE\_30\_ARCSEC}      & L    & A galaxy in the HECATE survey \citep{Kovlakas2021} is closer than 30 arcsec to the X-ray source \\
        \hline\hline

  \end{tabular}
  \label{tab:coldesc}
\end{table*}

\end{document}